\documentclass[apj, letterpaper]{emulateapj}  

\pdfoutput=1

\usepackage{amsmath}

\def\oslow{$\Omega_{\rm slow}$} 
\newcommand{\omed}{$\Omega_{\rm med}$}
\newcommand{\ofast}{$\Omega_{\rm fast}$}
\def\tt{}

\begin{document}

\slugcomment{\bf}
\slugcomment{ApJ, in press}

\title{Three-dimensional Atmospheric Circulation of Warm and Hot
  Jupiters: Effects of Orbital Distance, Rotation Period, and
  Non-Synchronous Rotation}

\shorttitle{Atmospheric circulation of warm and hot Jupiters}
\shortauthors{Showman et al.}

\author{Adam P. Showman\altaffilmark{1}, Nikole K. Lewis\altaffilmark{2,3,4},
Jonathan J. Fortney\altaffilmark{5}}

\altaffiltext{1}{Department of Planetary Sciences and Lunar and
  Planetary Laboratory, The University of Arizona, 1629 University
  Blvd., Tucson, AZ 85721 USA; showman@lpl.arizona.edu}
\altaffiltext{2}{Sagan Fellow}
\altaffiltext{3}{Department of Earth, Atmospheric, and Planetary Sciences,
Massachusetts Institute of Technology, Cambridge, MA 02139 USA}
\altaffiltext{4}{Space Telescope Science Institute, 3700 San Martin
Drive, Baltimore, MD 21218 USA}
\altaffiltext{5}{Department of Astronomy and Astrophysics, University
  of California, Santa Cruz, CA 95064, USA}
\begin{abstract}
\label{Abstract}
Efforts to characterize extrasolar giant planet (EGP) atmospheres have
so far emphasized planets within 0.05 AU of their stars. Despite this
focus, known EGPs populate a continuum of orbital separations from
canonical hot Jupiter values (0.03-0.05 AU) out to 1 AU and
beyond. Unlike typical hot Jupiters, these more distant EGPs will not
generally be synchronously rotating. In anticipation of observations
of this population, we here present three-dimensional atmospheric
circulation models exploring the dynamics that emerge over a broad
range of rotation rates and incident stellar fluxes appropriate for
warm and hot Jupiters. We find that the circulation resides in one of
two basic regimes. On typical hot Jupiters, the strong day-night
heating contrast leads to a broad, fast superrotating (eastward)
equatorial jet and large day-night temperature differences. At faster
rotation rates and lower incident fluxes, however, the day-night
heating gradient becomes less important, and baroclinic instabilities
emerge as a dominant player, leading to eastward jets in the
midlatitudes, minimal temperature variations in longitude, and, often,
weak winds at the equator. Our most rapidly rotating and least
irradiated models exhibit similarities to Jupiter and Saturn,
illuminating the dynamical continuum between hot Jupiters and the
weakly irradiated giant planets of our own Solar System. We present
infrared (IR) light curves and spectra of these models, which depend
significantly on incident flux and rotation rate. This provides a way
to identify the regime transition in future observations.  In some
cases, IR light curves can provide constraints on the rotation rate of
non-synchronously rotating planets.
\end{abstract}



\section{Introduction}
\label{Introduction}

Ever since their initial discovery, the atmospheric structure and
circulation of hot Jupiters has been a subject of intense focus.
Light curves and secondary eclipse measurements have now been obtained
for a variety of objects, constraining the three-dimensional
temperature structure, day-night heat transport, and circulation
regime.  This observational vanguard has motivated a growing body of
modeling studies of the three-dimensional atmospheric circulation of
hot Jupiters \citep[e.g.][]{showman-guillot-2002, cooper-showman-2005,
  dobbs-dixon-lin-2008, dobbs-dixon-agol-2013, showman-etal-2009,
  showman-etal-2013, menou-rauscher-2010, lewis-etal-2010,
  heng-etal-2011, thrastarson-cho-2010, perna-etal-2012,
  rauscher-kempton-2014, mayne-etal-2014}.  Most of these models have emphasized
synchronously rotating hot Jupiters in $\sim$2--4-day orbits with
properties similar to HD 189733b or HD 209458b.

Despite the focus of modeling efforts on HD 189733b-like and 209458b-like
planets, groundbased surveys and the CoRoT
and Kepler missions have greatly expanded the catalog of known
extrasolar giant planets (EGPs) to include many Jupiter-sized
objects outside the close-in hot-Jupiter population.
Known EGPs populate nearly a continuum of orbital
separations from canonical hot-Jupiter values ($\sim$0.03--0.05 AU)
out to 1 AU and beyond.   As we will show, for Jupiter-like tidal $Q$
values of $\sim$$10^5$, planets beyond $\sim$0.2 AU have
tidal spindown times comparable to typical system ages, implying
that these more distant EGPs will not in general be synchronously
rotating.  Depending on tidal $Q$ values, orbital histories, 
and other factors, even some planets inward of $\sim$0.1--0.2 AU
may rotate asychronously.  In anticipation of observations of
this wider population, there is thus a strong motivation to explore
the atmospheric circulation of hot and warm Jupiters over a wide range
of incident stellar fluxes and rotation rates.

To date, however, no such systematic investigation has been
carried out.  \citet{showman-etal-2009} and \citet{rauscher-kempton-2014}
explored the effects of factor-of-two deviations from synchronous
rotation in models of HD 189733b and/or HD 209458b.  \citet{lewis-etal-2014}
performed an analogous study for the eccentric hot Jupiter HAT-P-2b.
\citet{kataria-etal-2013} performed a thorough parameter study of the 
effect of eccentricity and stellar flux on the circulation of eccentric
hot Jupiters, considering both synchronous and pseudo-synchronous
rotation rates.  There has been no comparably thorough exploration
isolating how both widely varying stellar flux and rotation rate affect the
circulation for hot Jupiters on circular orbits.

Such a study can address fundamental questions on the mechanisms
controlling hot-Jupiter atmospheric circulation and their relationship
to the dominant circulation mechanisms of Solar-System planets such as
Earth and Jupiter.  Most models of synchronously rotating hot Jupiters
in $\sim$3-day orbits predict significant day-night temperature
contrasts and several broad zonal jets\footnote{Zonal and meridional
  denote the east-west and north-south directions, respectively.
  Zonal wind is the eastward wind and meridional wind is the northward
  wind; a zonal average is an average in longitude.}, including an
eastward (superrotating) jet at the equator, which in some cases
causes an eastward displacement of the hottest regions from the
substellar longitude \citep[e.g.,][]{showman-guillot-2002}.
\citet{showman-polvani-2011} showed that many features of this
circulation regime, including the equatorial superrotation, can be
understood in terms of the interaction of standing, planetary-scale
waves with the mean flow.  However, it is unknown whether this
circulation regime should apply across the range of hot and warm
Jupiters\footnote{We loosely refer to hot and warm Jupiters as EGPs 
with equilibrium temperatures greater or less than $1000\rm\,K$, 
respectively.} or whether it should give way to other circulation regimes
under greatly different conditions.  Earth, for example, exhibits
significant equator-to-pole temperature differences, only modest
variations of temperature in longitude, and zonal winds that peak in
midlatitudes, with westward zonal-mean flow in the equatorial
troposphere.  Such a regime, if it occurred on a close-in EGP, could
lead to very different lightcurves and spectra than would otherwise
occur.

Here, we present new three-dimensional circulation models over a broad
range of rotation rates and incident stellar fluxes comparable to and
less than that received by HD 189733b with the aim of understanding
the conditions under which transitions to different circulation
regimes occur, establishing the link to more Earth-like and
Jupiter-like regimes, and determining the implications for
observables.  Section~\ref{theory} presents theoretical arguments
anticipating a transition in the circulation regime for warm and hot Jupiters.
Section~\ref{model} presents our dynamical model used to test these ideas.  
Section~\ref{basic-results}
describes the basic circulation regimes of our model integrations,
including a diagnosis of the conditions under which regime transitions
occur.  Section~\ref{mechanisms} presents more detailed diagnostics
illuminating the dynamical mechanisms.  Section~\ref{observables}
presents light curves and spectra that our models would imply.  Our
summary and conclusions are in Section~\ref{discussion}.

\section{Background theory and prediction of a regime transition}
\label{theory}

\begin{figure*}
\centering
\includegraphics[scale=0.6, angle=0]{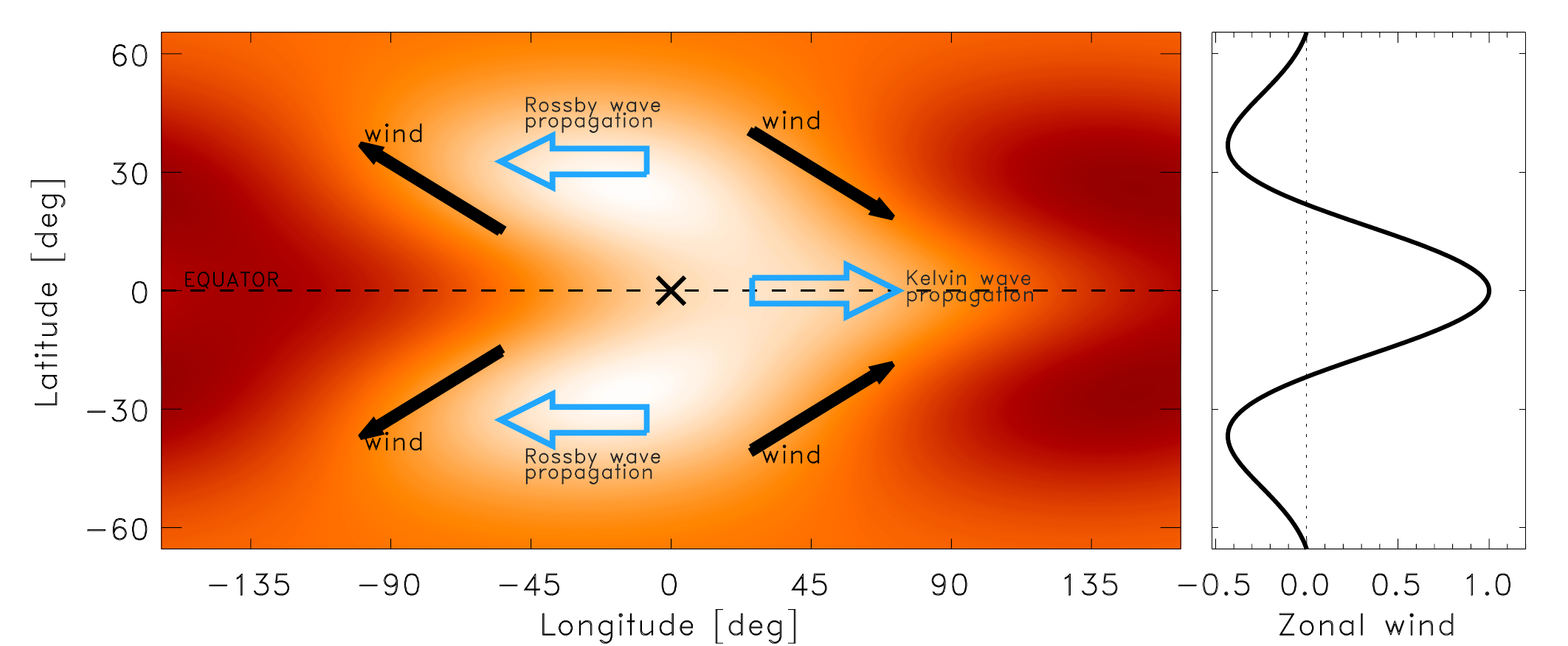}
\put(-330.,150.){(a)}\\
\includegraphics[scale=0.6, angle=0]{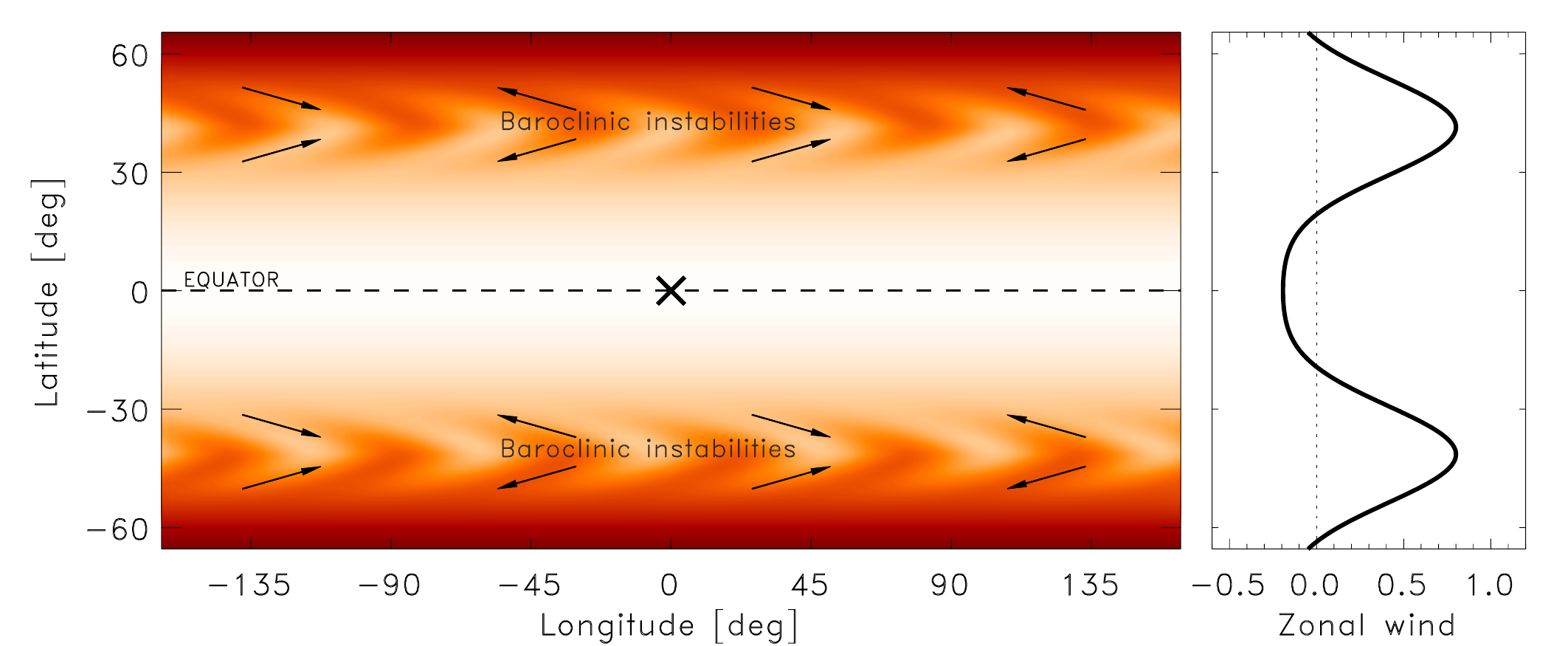}
\put(-330.,150.){(b)}
\caption{Two regimes of atmospheric circulation of warm/hot Jupiters
  as proposed in this study.  The top row, (a), represents the
  ``canonical'' hot Jupiter regime of slowly rotating, highly
  irradiated planets.  The strong day-night thermal forcing induces
  significant day-night temperature differences and planetary-scale
  Kelvin and Rossby wave modes, which exhibit phase tilts that
  cause momentum to be transported to the equator.  The result is a
  strong eastward (superrotating) jet at the equator.  The bottom row,
  (b), is a regime we hypothesize to occur on warm Jupiters when the
  rotation is sufficiently fast and/or the stellar irradiation is
  sufficiently weak.  In this case, the day-night forcing (i.e., the
  diurnal cycle) is relatively unimportant, and the circulation is
  instead driven by the zonal-mean, equator-to-pole contrast in
  stellar heating.  Longitudinal temperature variations are small, but
  latitudinal temperature gradients (from equator to pole) can be
  large.  In such a configuration, the midlatitudes can become
  baroclinically unstable, and the Rossby waves so generated cause
  momentum transport into the instability latitude, leading to the
  generation of ``eddy driven'' zonal jets.  Zonal winds peak in the
  midlatitudes rather than the equator. The cross marks the substellar
  point.}
\label{schematic}
\end{figure*}

Circulation models of ``typical'' hot Jupiters---defined here as those
in several-day orbits with effective temperatures of
1000--$1500\rm\,K$ such as HD 189733b and HD 209458b---generally
exhibit a circulation pattern near the infrared (IR) photosphere
dominated by a fast, broad eastward jet centered at the equator
(equatorial ``superrotation'') and large temperature differences
between the dayside and nightside \citep[e.g.][]{showman-guillot-2002,
  cooper-showman-2005, showman-etal-2008a, showman-etal-2009,
  showman-etal-2013, heng-etal-2011, heng-etal-2011b, perna-etal-2012,
  rauscher-menou-2010, rauscher-menou-2012b, rauscher-menou-2013,
  dobbs-dixon-lin-2008, dobbs-dixon-agol-2013, rauscher-kempton-2014}.
The day-night temperature differences in these models reach
$\sim$200--$1000\rm\,K$ depending on pressure and on how the day-night
thermal forcing is introduced.  In many cases, the equatorial jet
induces an eastward displacement of the hottest regions from the
substellar point.  Models that include radiative transfer show that IR
lightcurves exhibit large flux variations with orbital phase, with a
flux peak that often occurs before secondary eclipse
\citep{showman-etal-2009, heng-etal-2011b, rauscher-menou-2012b,
  perna-etal-2012, dobbs-dixon-agol-2013}.  Observational support for
this dynamical regime comes from the overall agreement between
observed and synthetic light curves and spectra---including inferences
of eastward hotspot offsets---for HD 189733b \citep{showman-etal-2009,
  knutson-etal-2012, dobbs-dixon-agol-2013}, HD 209458b
\citep{zellem-etal-2014}, WASP-43b \citep{stevenson-etal-2014,
  kataria-etal-2014b}, and HAT-P-2b \citep{lewis-etal-2013,
  lewis-etal-2014}.  Eastward offsets have also been inferred
on WASP-12b \citep{cowan-etal-2012} and Ups And b \citep{crossfield-etal-2010},
although 3D models of those planets have yet to be published.

\citet{showman-polvani-2011} showed that the key dynamical feature of
this circulation regime---the superrotating equatorial jet---results
from a wave-mean-flow interaction caused by the strong day-night
heating contrast.  Figure~\ref{schematic}a schematically illustrates
this mechanism.  The day-night heating contrast induces standing
planetary-scale waves.  In particular, a Kelvin wave\footnote{See
  \citet[][pp.~394-400, 429-432]{holton-2004} for an introduction to
  equatorially trapped Kelvin and Rossby waves.} is generated at low
latitudes; it is centered about and exhibits thermal maxima occurring
at the equator.  Kelvin waves exhibit (group) propagation to the east,
and in the context of a continuously forced and damped circulation,
this leads to a quasi-steady, eastward displacement of the thermal
pattern at the equator (Figure~\ref{schematic}a, left).  Moreover,
equatorially trapped Rossby waves are generated on the poleward flanks
of the Kelvin wave.  These waves exhibit group propagation to the
west, and in the presence of continuous thermal forcing and damping, a
quasi-steady westward displacement of the thermal pattern emerges at
mid-latitudes (Figure~\ref{schematic}a, left).  This latitudinally
varying zonal phase shift implies that the thermal and eddy wind
structure exhibits northwest-southeast tilts in the northern
hemisphere and southwest-northeast tilts in the southern hemisphere.
In turn, this eddy pattern induces a transport of eddy momentum from
the midlatitudes to the equator, leading to equatorial superrotation
(Figure~\ref{schematic}a, right).  See \citet{showman-polvani-2010,
  showman-polvani-2011}, \citet{showman-etal-2013}, and
\citet{tsai-etal-2014} for additional discussion; a review can be
found in \citet{showman-etal-2013b}.

But theory and simulations from the solar-system literature indicate
that, at sufficiently fast rotation and weak irradiation, the dynamics
should shift to a different regime.  When the irradiation is
sufficiently weak, the day-night thermal forcing (i.e., the diurnal
cycle) lessens in importance and the latitudinal variation of the {\it
  zonal-mean} heating becomes the dominant driver of the circulation.
This heating pattern leads to significant temperature gradients in
latitude but only small temperature gradients in longitude.  Dynamical
(e.g., thermal wind) balance requires that these meridional
temperature gradients will be accompanied by strong zonal flow.  A
large body of work shows that, when the rotation is sufficiently
fast, this configuration tends to be dynamically unstable: small
perturbations in an initially zonally symmetric flow will grow via
baroclinic instability, leading to baroclinic eddies that transport
thermal energy poleward (for reviews see
\citealt{pierrehumbert-swanson-1995} or \citealt[][Chapter
  6]{vallis-2006}). Although such instability on Earth and Mars is
enhanced by the existence of entropy gradients on the lower surface,
several studies have shown that baroclinic instabilities are possible
even on gas giants like Jupiter that lack such surfaces
\citep{conrath-etal-1981, read-1988, williams-2003a,
  lian-showman-2008, ogorman-schneider-2008, polichtchouk-cho-2012}.
Typically, such instabilities occur most readily at mid-to-high
latitudes, where meridional temperature gradients are large and
isentropes slope steeply.  The thermal pattern expected in this regime
is shown schematically in Figure~\ref{schematic}b (left panel).

On rotating planets, because the gradient of the Coriolis parameter
with northward distance, $\beta$, is nonzero, baroclinic instability
can transport momentum from surrounding latitudes into the instability
latitude, leading to the formation of so-called ``eddy-driven'' zonal
jet streams at the instability latitude.  Linear stability analyses
demonstrate this transport in an idealized setting
\citep[e.g.][]{conrath-etal-1981, held-andrews-1983, james-1987}, and
nonlinear studies of forced-equilibrium circulations show how it can
lead to the formation of zonal jets (\citealt{williams-1979,
  williams-2003a}; \citealt{panetta-1993};
\citealt{ogorman-schneider-2008}; \citealt{lian-showman-2008}, and
many others).  The process causing this momentum convergence is often
described phenomenologically in terms of the excitation of Rossby
waves at the instability latitude and their propagation to other
latitudes where they dissipate or break \citep[e.g.,][] {held-2000,
  vallis-2006}.  Rossby waves propagating northward from the
instability latitude exhibit northward group velocity, whereas those
propagating southward exhibit southward group velocity.  Because
Rossby waves with northward group velocity exhibit southward angular
momentum transport, whereas those with southward group velocity
exhibit northward angular momentum transport, the implication is that
angular momentum is transported into the instability latitude from
surrounding regions \citep{thompson-1971, held-1975, held-2000,
  vallis-2006, showman-etal-2013b}.  To the extent that wave
generation preferentially occurs at some latitudes, and wave
breaking/dissipation in others, eddy-driven zonal jets will emerge in
the midlatitudes, as shown schematically in Figure~\ref{schematic}b
(right panel).

Several eddy-mean-flow feedbacks promote the emergence of discrete
midlatitude zonal jets (as opposed to simply broad strips of eastward
wind) in this regime.  When the zonal wind speeds increase with
height, eastward jets tend to be more baroclinically unstable than
westward jets; the former support a greater number of possible
instability modes, and generally exhibit greater baroclinic
instability growth rates, than the latter \citep[e.g.,][]{wang-1990,
  polichtchouk-cho-2012}.  This may allow preferential Rossby wave
generation at eastward jets, thereby promoting preferential angular
{\tt momentum} transport into them.  Perhaps more importantly, the
breaking of Rossby waves tends to occur more readily in westward jets,
where the meridional potential vorticity (PV)\footnote{\tt Potential
  vorticity is essentially the local vorticity (including
  contributions both from winds and the planetary rotation) divided by
  a measure of the vertical spacing between isentropes; it is a
  materially conserved quantity in adiabatic, frictionless, stratified
  flow.  In the 3D system, it is defined as
  $\rho^{-1}(\nabla\times{\bf u} + 2{\bf \Omega})\cdot \nabla\theta$,
  where $\rho$ is density, ${\bf u}$ is the 3D velocity vector, ${\bf
    \Omega}$ is the planetary rotation vector, and $\theta$ is
  potential temperature.  See \citet{holton-2004} or
  \citet{vallis-2006} for a description of its uses in atmospheric
  dynamics.} gradient is small, than it does in
eastward jets, where the PV gradient is large.  The meridional mixing
caused by Rossby-wave breaking decreases still further the (already
weak) PV gradient in westward jets, promoting even stronger Rossby
wave breaking there.  This is a positive feedback, which implies that,
even if Rossby wave generation occurs randomly everywhere, the wave
breaking---and associated PV mixing---will self-organize.  The result
tends to be series of zonal strips of nearly constant PV, with sharp
PV jumps in between, which corresponds to the spontaneous emergence of
zonal jets when the Rossby number is sufficiently small.  See
\citet{dritschel-mcintyre-2008} or \citet{showman-etal-2013b} for
reviews.

Assembling these arguments, we thus predict a regime transition for
warm/hot Jupiters as a function of stellar irradiation and rotation
rate.  When stellar irradiation is strong and the rotation rate is
modest---as for synchronously rotating planets on several-day
orbits---we expect large day-night temperature differences and strong
equatorial superrotation---as shown in Figure~\ref{schematic}a. When
stellar irradiation is weak or the rotation rate is fast, we expect
minimal longitudinal temperature differences, significant
equator-pole temperature differences, midlatitude baroclinic eddies,
and zonal-mean zonal winds that reach a maximum at the midlatitudes
rather than the equator.  These two regimes will have very different
predictions for light curves, IR spectra, and Doppler signatures that
may be detectable in observations.

What should be the criterion for the transition?
\citet{perez-becker-showman-2013} presented an analytic theory for the
day-night temperature difference for synchronously rotating planets;
however, no such theory yet exists for the more general case of
non-sychronously rotating planets, especially when baroclinic
instabilities become important.  Nevertheless, a reasonable hypothesis
is that the transition between these regimes should occur when the
radiative time constant (e.g., at the photosphere) approximately
equals the solar day.   {\tt Here, the radiative time constant is
the characteristic time for the atmosphere to gain or lose energy
by radiation, and the solar day is the characteristic time (say) between
two successive sunrises at a given point on the planet. }
When the radiative time constant is shorter than
the solar day, the day-night (diurnal) forcing is strong, leading to
wave-driven equatorial superrotation as shown in
Figure~\ref{schematic}a.  When the radiative time constant is longer
than the solar day, the diurnal cycle becomes dynamically less
important than the meridional gradient in zonal-mean heating, and one
obtains the regime of mid-latitude jets shown in
Figure~\ref{schematic}b.  When winds are a significant fraction of the
planetary rotation speed, the ``solar day'' should be generalized to
include the zonal advection by winds.\footnote{The criterion can thus
  be viewed as a comparison of the radiative time constant with a
  generalized horizontal advection time that includes both rotation
  and winds.  In certain cases, horizontal wave timescales and/or
  vertical advection timescales could alter the transition; evaluating
  this idea will require an extension of the theory of
  \citet{perez-becker-showman-2013} to non-synchronously rotating
  planets.}

Let us quantify this criterion.  The solar day is
$P_{\rm solar} = 2\pi/|\Omega - n|$, where $n$ is the orbital mean
motion (i.e., the mean orbital angular velocity).  Thus,
\begin{equation}
P_{\rm solar} = {1\over  |{1\over P_{\rm rot}} - {1\over P_{\rm orb}}|}
= {1\over |{1\over P_{\rm rot}} - {1\over k a_{\rm orb}^{3/2}}|}
\label{psolar}
\end{equation}
where $P_{\rm rot}=2\pi/\Omega$ is the (siderial) rotation period of
the planet, $P_{\rm orb}$ is the orbital period, $a_{\rm orb}$ is the
orbital semimajor axis, and in the second expression we have used
Kepler's 3rd law.  Here, $k=2\pi/\sqrt{G(M_* + M_p)}=3.46\times 10^7 \rm
\,s \,AU^{-3/2}$, where $G$ is the gravitational constant, $M_*$ and
$M_p$ are the stellar and planetary masses, and we have evaluated the
expression for HD 189733.  To order of magnitude, the radiative time
constant near the photosphere is \citep{showman-guillot-2002}
\begin{equation}
\tau_{\rm rad} = {p c_p\over 4 g \sigma T_e^3}
\label{taurad}
\end{equation}
where $p$ is the pressure of the heated part of the atmosphere, $c_p$
is the specific heat at constant pressure, $g$ is gravity, $\sigma$ is
the Stefan-Boltzmann constant, and $T_e$ is the planet's equilibrium
temperature.  Assuming zero albedo, the latter can be expressed as
$T_e = 2^{-1/2} (R_*/a_{\rm orb})^{1/2} T_*$, where $R_*$ and $T_*$
are the stellar radius and temperature, respectively.  Thus,
\begin{equation}
\tau_{\rm rad} \approx {p c_p\over g \sigma T_*^3 } \left({a_{\rm orb}
\over R_*}\right)^{3/2}
\label{taurad2}
\end{equation}
Equating this expression to the solar day, we obtain---as a function
of of orbital semimajor axis and stellar properties---the planetary
rotation period for which the solar day equals the radiative time
constant
\begin{equation}
P_{\rm rot} \approx {1\over {1\over k a_{\rm orb}^{3/2}} +
{g \sigma T_*^3\over p c_p}\left({R_*\over a_{\rm orb}}\right)^{3/2}}.
\label{transition}
\end{equation}
This expression assumes the rotation is prograde and faster than the
orbital motion.\footnote{In other words, defining $P_{\rm orb}$
  positive, Equation~(\ref{transition}) is valid for $P_{\rm rot}^{-1}
  > P_{\rm orb}^{-1}$.  The case $P_{\rm rot}^{-1} < P_{\rm
    orb}^{-1}$---corresponding to any magnitude of retrograde
  rotation, or to prograde rotation with a rotation period longer than
  the orbital period---would be obtained by replacing the positive
  sign in the denominator of (\ref{transition}) with a negative sign.
  In this paper we consider only prograde rotation; given our
  expression for the radiative time constant, the solar day equals the
  radiative time constant for a prograde-rotating planet only when
  $P_{\rm rot}^{-1} > P_{\rm orb}^{-1}$, implying that
  Equation~(\ref{transition}) is the most relevant case.}
Generalizing to include a constant equatorial zonal-wind speed $U$,
the time between successive sunrises for a zonally circulating air
parcel would become $P_{\rm solar}= 2\pi/|\Omega + U/a - n|$, where
$a$ is the planetary radius.  Equating this expression to the
radiative time constant yields the following generalized version of
Equation~(\ref{transition}):
\begin{equation}
P_{\rm rot} \approx {1\over {1\over k a_{\rm orb}^{3/2}} - {U\over 2\pi a}
+ {g \sigma T_*^3\over p c_p}\left({R_*\over a_{\rm orb}}\right)^{3/2}}.
\label{transition2}
\end{equation}

\begin{figure}
\includegraphics[scale=0.5, angle=0]{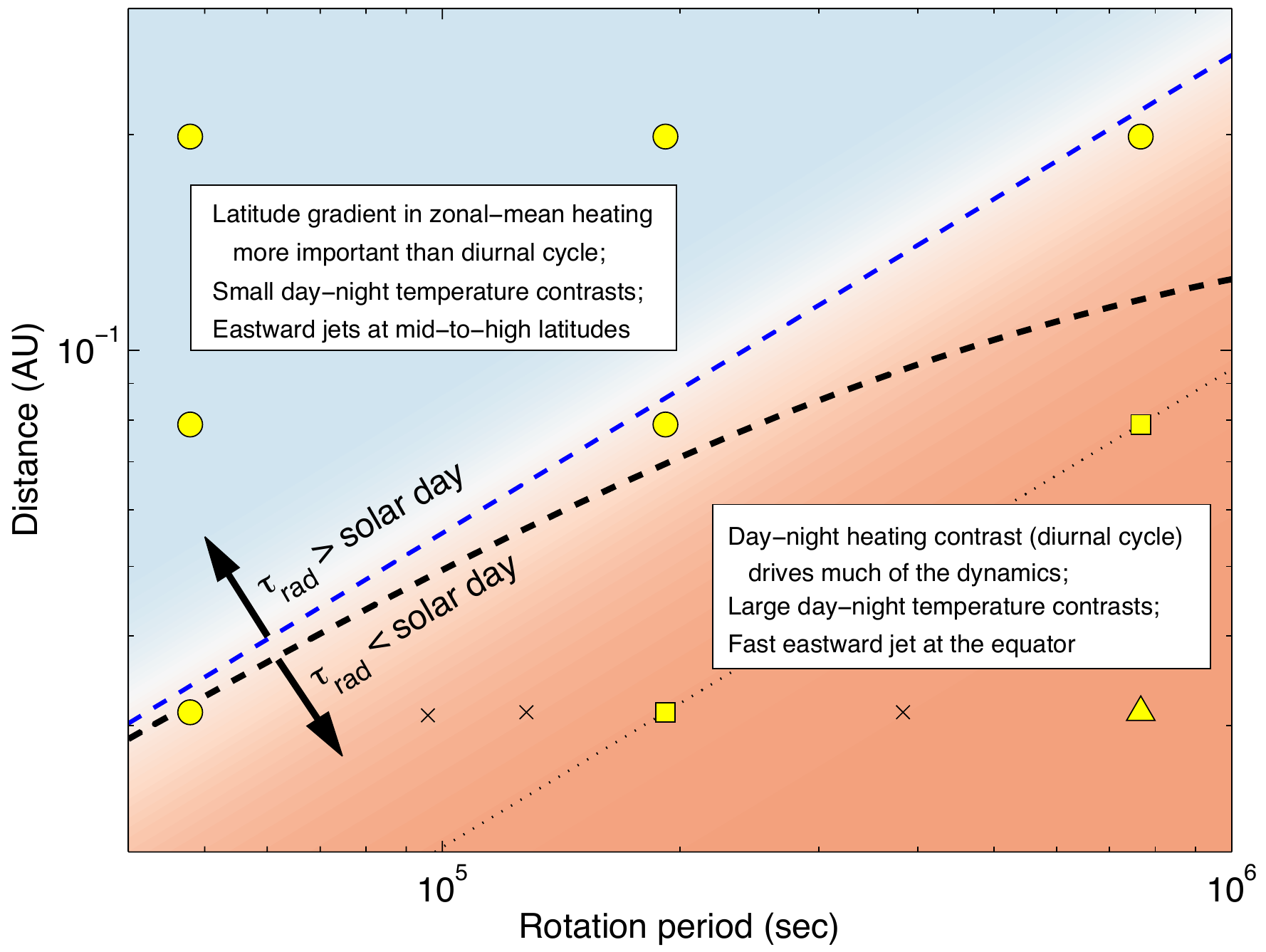}
\caption{Predicted phase diagram of dynamical regimes.  Blue dashed
  line shows the planetary rotation rates and orbital semimajor axes
  at which the planet's solar day equals its photospheric radiative
  time constant (Equation~\ref{transition}); black dashed line shows
  the same criterion generalized to include a constant eastward zonal
  wind of $U=1\rm\,km\,s^{-1}$ (Equation~\ref{transition2}). Regions
  below the line (orange) have $\tau_{\rm rad}<P_{\rm solar}$ and
  should exhibit large-day night temperature differences with a fast
  eastward equatorial jet.  Regions above the line (blue) exhibit
  $\tau_{\rm rad}>P_{\rm solar}$ and should have a circulation
  exhibiting mid-to-high latitude zonal jets with little zonal
  temperature variation.  The equation is evaluated using
  $T_*=4980\rm\,K$ and $R_*=0.788 R_\odot = 5.5\times10^8\rm\,m$
  (appropriate to HD 189733), and
  $c_p=1.3\times10^4\rm\,J\,kg^{-1}\,K^{-1}$ and $g=21\rm\, m\,s^{-2}$
  appropriate to the planet.  We adopt $p\approx 0.25\rm\,bar$,
  equivalent to that adopted by \citet{perez-becker-showman-2013}, who
  found a good fit between their model and lightcurve observations for
  hot Jupiters.  Overplotted yellow symbols show the parameter
  combinations for GCM simulations presented in this paper; squares
  are synchronously rotating models, whereas circles (triangles)
  denote models whose rotation periods are shorter (longer) than their
  orbital periods.  Crosses show non-synchronous models presented in
  \citet{showman-etal-2009}. Thin dotted line denotes synchronous
  rotation.}
\label{parameter-space}
\end{figure}

We plot these expressions as a function of $P_{\rm rot}$ and $a_{\rm
  orb}$ in Figure~\ref{parameter-space}, using an eastward wind
$U=1\rm\,km\,s^{-1}$.\footnote{Adopting eastward wind (positive $U$) is most
relevant because the atmospheric winds tend to be broadly eastward
throughout the atmosphere (see Figures~\ref{zonal-winds}--\ref{temp-winds}).
Westward wind, though less relevant, would be represented as negative $U$
and would imply the transition would occur {\it above} the dashed blue
line in Figure~\ref{parameter-space}.} Regions below the dashed
line have $\tau_{\rm rad}<P_{\rm solar}$ and should exhibit the regime
of large day-night temperature differences and equatorial
superrotation shown in Figure~\ref{schematic}a.  Regions above the
dashed line have $\tau_{\rm rad}>P_{\rm solar}$ and should exhibit the
regime of mid-to-high-latitude jet and small zonal temperature
variations of Figure~\ref{schematic}b. Our simulations serve as a test
of this prediction and will clarify the dynamics operating in each
regime.

Although we focus here on this basic regime shift, it is worth
emphasizing that the dynamics are complex and can exhibit richer
behavior.  In particular, each of the regimes (orange and blue)
identified in Figure~\ref{parameter-space} could themselves split into
several sub-regimes.  For example, in the orange region, models in
which the radiative time constant becomes particularly short---and/or
the rotation rate becomes particularly long---could transition from a
regime of fast equatorial superrotation to a regime where day-night
flow dominates the circulation \citep[e.g.][]{showman-etal-2013}.  In
the blue region, sufficiently rapid rotation should lead to a tropical
zone\footnote{We here define tropics and extratropics following
  \citet{showman-etal-2013b}: tropics are regions where the Rossby
  number $Ro \gtrsim 1$, whereas the extratropics are the regions
  where $Ro\ll 1$.}  confined near the equator, with a broad
extratropical zone at mid-to-high latitudes in which baroclinic
instabilities occur as depicted in Figure~\ref{schematic}a.  However,
if the rotation rate is sufficiently slow (while still maintaining
$\tau_{\rm rad}>P_{\rm solar}$), the planet may become an ``all
tropics'' world with a global Hadley circulation in which baroclinic
instabilities play a minimal role.  While such a planet would still
exhibit small zonal temperature variations and zonal winds peaking in
midlatitudes, the dynamics controlling those jets may differ from the
rapidly rotating case.  Titan and Venus provide the best solar system
analogues.  This regime has been well studied for terrestrial planets
\citep[e.g.,][]{delgenio-suozzo-1987, delgenio-zhou-1996,
  mitchell-etal-2006, mitchell-vallis-2010, kaspi-showman-2014} but
has yet to receive attention for gas giants.  Moreover, our
considerations neglect any role for strong frictional drag, strong
magnetic coupling, or strong internal convection, all of which may
influence the dynamical regime.

\section{Model}
\label{model}

We solved the coupled hydrodynamics and radiative transfer equations
using the Substellar and Planetary Atmospheric Radiation and
Circulation (SPARC/MITgcm) model of \citet{showman-etal-2009}.  This
model solves the global, three-dimensionsal primitive equations in
spherical geometry, with pressure as a vertical coordinate, using the
MITgcm \citep{adcroft-etal-2004}.  The radiative transfer is solved
using the two-stream variant of the multi-stream, non-grey radiative
transfer model of \citet{marley-mckay-1999}.  Opacities are treated
using the correlated-$k$ method, which retains most of the accuracy of
full line-by-line calculations but with dramatically reduced
computational overhead. Opacities are treated statistically in each of
11 wavelength bins (see \citealt{kataria-etal-2013}), allowing the
inclusion of $10^5$ to $10^6$ individual opacity points within each
bin; this is far more accurate than grey or multi-band approaches that
adopt a single, mean opacity in each of a small number of wavelength
bins (e.g., \citealt{rauscher-menou-2012}, \citealt{heng-etal-2011b},
\citealt{dobbs-dixon-agol-2013}; see \citealt{amundsen-etal-2014} for
a detailed discussion).  To date, the SPARC model is the only general
circulation model (GCM) that has been used to model the 3D circulation
of hot Jupiters including a realistic representation of non-grey
radiative transfer, as necessary for accurate assessment of the
opacities, heating rates, and temperature structure under particular
assumptions about the atmospheric composition.  Gaseous opacities are
determined assuming local chemical equilibrium (accounting for rainout
of condensates) at a specified atmospheric metallicity.  We have
explored a variety of metallicities in prior work
\citep{showman-etal-2009, lewis-etal-2010}, but as our emphasis here
is on the dependence of the circulation regime on irradiation and
rotation rate, we adopt solar metallicity for the current work. We
neglect any opacity due to clouds or hazes. {\tt The
  radiative-transfer model adopts at the base an internal radiative
  heat flux of $5\rm\,W\,m^{-2}$, similar to that expected for generic
  multi-Gyr-old hot Jupiters (e.g., \citealt{guillot-showman-2002},
  \citealt{fortney-etal-2007}, and others).  Nevertheless, this plays
  little role in the dynamics on the timescale of these simulations,
  as the absorbed stellar flux exceeds this intrinsic flux by a factor
  ranging from hundreds to $>10^4$, depending on the model.}

Treating HD 189733b as a nominal case, we vary the the orbital
semimajor axis and rotation rate over a large range.  We explore
orbital semimajor axes of 0.0313 AU (the actual value for HD 189733b),
0.0789 AU, and 0.1987 AU, corresponding to stellar fluxes incident on
the planet of $4.68\times10^5$, $7.37\times10^4$, and
$1.16\times10^4\rm\,W\,m^{-2}$, respectively\footnote{The stellar
  luminosity adopted in our models is $1.29\times10^{26}\rm\,W$.
  Averaged over the $4\pi$ steradians of the planetary surface, the
  corresponding global-mean effective temperatures assuming zero
  albedo are $1198\rm\,K$, $755\rm\,K$, and $475\rm\,K$ for the
 H, W, and C models, respectively.}.  This
corresponds to a significant variation---greater than a factor of 40
in stellar flux---while still emphasizing close-in planets amenable to
transit observations.

The rotation periods of hot Jupiters are often assumed to be
synchronous with their orbital periods \citep{guillot-etal-1996,
  rasio-etal-1996}.  The spindown timescale from a primordial rotation
rate $\Omega_p$ is
\begin{equation}
\tau\sim Q\left({R_p^3\over GM_p}\right)\Omega_p\left({M_p\over 
M_\star}\right)^2\left({a_{\rm orb}\over R_p}\right)^6
\label{spindown}
\end{equation}
where $Q$, $R_p$ and $M_p$ are the planet's tidal dissipation factor,
radius and mass, $G$ is the gravitational constant, $a_{\rm orb}$ is
the orbital semimajor axis (here considering a circular orbit), and
$M_\star$ is the mass of the star.  Adopting a solar mass for the star,
a Jupiter mass for the planet, and taking the planetary radius as
1.2 Jupiter radii (typical for a hot Jupiter), we obtain
\begin{equation}
\tau\sim 1\times10^6\left({Q\over 10^5}\right)
\left({a_{\rm orb}\over 0.05 {\rm AU}}\right)^6\,{\rm yr}.
\end{equation}
For a Jupiter-like $Q\sim10^5$, this yields $\tau\sim10^6\,$yr for
canonical hot-Jupiter orbital separations of 0.05 AU, but the
timescale increases to $4\times10^9\rm\,yr$---comparable to typical
system ages---for orbital separations of 0.2 AU.  Thus, while this
argument suggests that hot Jupiters should be tidally locked inward of
0.05 AU, sychronization should not be expected outward of 0.2 AU, and
at intermediate distances (perhaps for planets that have experienced
only a few spindown times), the planet may have been significantly
despun but not yet become fully synchronized.  Note that tidal $Q$
values are highly uncertain, and planet radii vary over a wide range
from $\sim$1--2 Jupiter radii, implying that the orbital semimajor
axes over which synchronization is expected are uncertain and may vary
from system to system.  Moreover, it has been suggested, even when
Eq.~(\ref{spindown}) predicts synchronization, that in some cases the
gravitational torque on not only the {\it gravitational} tide but also
on the {\it thermal} tidal response may be important, leading to an
equilibrium configuration with asychronous rotation
\citep{showman-guillot-2002, arras-socrates-2010}.  

Motivated by these considerations, we explore rotation periods varying
by up to a factor of four from the nominal orbital period of HD
189733b, that is, 0.55, 2.2, and 8.8 Earth days\footnote{In this
  paper, 1 day is defined as $86400\rm\,s$.}, corresponding to
rotation rates $\Omega$ of $1.322\times10^{-4}\rm\,s^{-1}$,
$3.3\times10^{-5}\rm\,s^{-1}$, and
$8.264\times10^{-6}\rm\,s^{-1}$. The shortest of these is close to
Jupiter's rotation period of 10 hours.  This is a wider exploration of
rotation rate than considered in previous studies of non-synchronous
rotation \citep{showman-etal-2009, kataria-etal-2013,
  rauscher-kempton-2014, lewis-etal-2014}.  In our non-synchronous
models, the longitude of the substellar point migrates in time $t$ as
$(n-\Omega)t$, where $n$ is $2\pi$ over the orbital period; thus, in
the reference frame of the rotating planet, the entire dayside heating
pattern migrates east or west over time.  We assume circular orbits
with zero obliquity.

In total, these variations constitute a regular grid of models varying
the rotation rate by a factor of 16 and the incident stellar flux by a
factor of over 40.  Figure~\ref{parameter-space} depicts the parameter
space explored.  For each integration, we denote the irradiation level
by H for hot, W for warm, and C for cold (representing models with
orbital semimajor axes of 0.03, 0.08, and $0.2\,$AU respectively) and
rotation rate by \ofast, \omed, and \oslow.  Thus for example
H\oslow is the most highly irradiated, slowly rotating model, while
C\ofast is the least irradiated, most rapidly rotating model.

All models adopt the radius and gravity of HD 189733b ($1.15 R_J$ and
$21.4\rm\,m\,s^{-2}$, respectively),
$c_p=13000\rm\,J\,kg^{-1}\,K^{-1}$, a ratio of gas constant to
specific heat at constant pressure $R/c_p=2/7$, and the ideal-gas
equation of state.  {\tt As is standard in GCMs, gravity and $R/c_p$
  are assumed constant, which is a reasonably good approximation.} The
models are integrated from rest using an initial temperature-pressure
profile from a one-dimensional planetwide-average
radiative-equilibrium calculation.  \citet{liu-showman-2013} showed
the typical hot Jupiter regime does not exhibit significant
sensitivity to initial conditions.

Our nominal grid of models do not include explicit frictional
drag.\footnote{All models include a fourth-order Shapiro filter to maintain
  numerical stability, which smooths grid-scale oscillations and damps
  some kinetic energy at small scales.}  The mechanisms of frictional
dissipation in hot-Jupiter atmospheres are poorly understood and could
include magnetohydrodynamic (ion) drag \citep{perna-etal-2010},
vertical turbulent mixing \citep{li-goodman-2010}, and
breaking small-scale gravity waves \citep[e.g.,][]{lindzen-1981}.  The
atmospheres of planets in the Solar System are generally relatively
inviscid except near the surfaces.  In giant planets such as Jupiter,
magnetohydrodynamic drag at great depth has been suggested as a key
process in braking the interior winds \citep{kirk-stevenson-1987,
  liu-etal-2008, schneider-liu-2009}.  For hot Jupiters as cool as or
cooler than HD 189733b, recent circulation models suggest that
magnetic coupling is unimportant in the observable atmosphere
\citep{rauscher-menou-2013, rogers-showman-2014}, although it may play
a role in the deep atmosphere (pressures greater than tens to hundreds
of bar) where temperatures become hot.  Motivated by this possibility,
we performed a range of sensitivity studies where we included
frictional drag near the bottom of the model, where temperatures are
warmest and magnetohydrodynamic braking is most likely.

We integrated each model until the velocities at low pressure
reached a stable configuration.  In the models without explicit
large-scale drag, the winds in the observable atmosphere
(pressures less than 0.1 bar) become essentially
steady within $\sim$3000 days, and sometimes much less.  
The winds at depth (pressures exceeding 10 bars) are generally 
much weaker than photosphere winds.  Any further increases in wind speed 
beyond our $\sim$3000--6000-day integration periods are likely 
to be modest and confined to pressures well below the mean photosphere,
such that influences on lightcurves and spectra are modest.
When drag is included in the deep atmosphere, it readily allows
the total kinetic energy to equilibrate.  If this drag is confined 
to pressures exceeding 10 bars, it has little effect on the overall 
circulation regime at the photosphere; if the drag extends to 
pressures as low as 1 bar it 
starts to influence the details of the photospheric circulation, although the 
overall qualitative trends identified in this paper are unaffected.  
Although this is reassuring, it is clear that the drag formulation
is one of the greatest uncertainties in current EGP circulation models.

The equations are solved on the cubed-sphere grid
\citep{adcroft-etal-2004}.  The cases at the longest two rotation
periods (2.2 and 8.8 Earth days) adopt a horizontal resolution of C32
(i.e. $32\times32$ finite-volume cells on each cube face),
corresponding to an approximate global resolution of $128\times64$ in
longitude and latitude.  Because of the smaller Rossby deformation
radius in the cases with shortest rotation period (0.55 days), the
dominant lengthscales are smaller, and so we integrated all of these
cases at C64 (i.e., $64\times64$ cells on each cube face),
corresponding to an approximate global resolution of $256\times128$ in
longitude and latitude.  Nevertheless, the behavior of these
high-resolution, rapidly rotating integrations is qualitatively
similar to equivalent cases performed at C32 resolution.  All models
adopt a vertical grid containing $N_r=40$ levels.  The bottom $N_r-1$
levels have interfaces that are evenly spaced in log-pressure between
0.2 mbar and 200 bar; the top level extends from 0 to 0.2 mbar.  These
models generally conserve total angular momentum to $\sim$0.02\%.

\section{Results: Basic circulation regime}
\label{basic-results}

\begin{figure*}
\begin{minipage}[c]{0.3\textwidth}
\includegraphics[scale=0.3, angle=0]{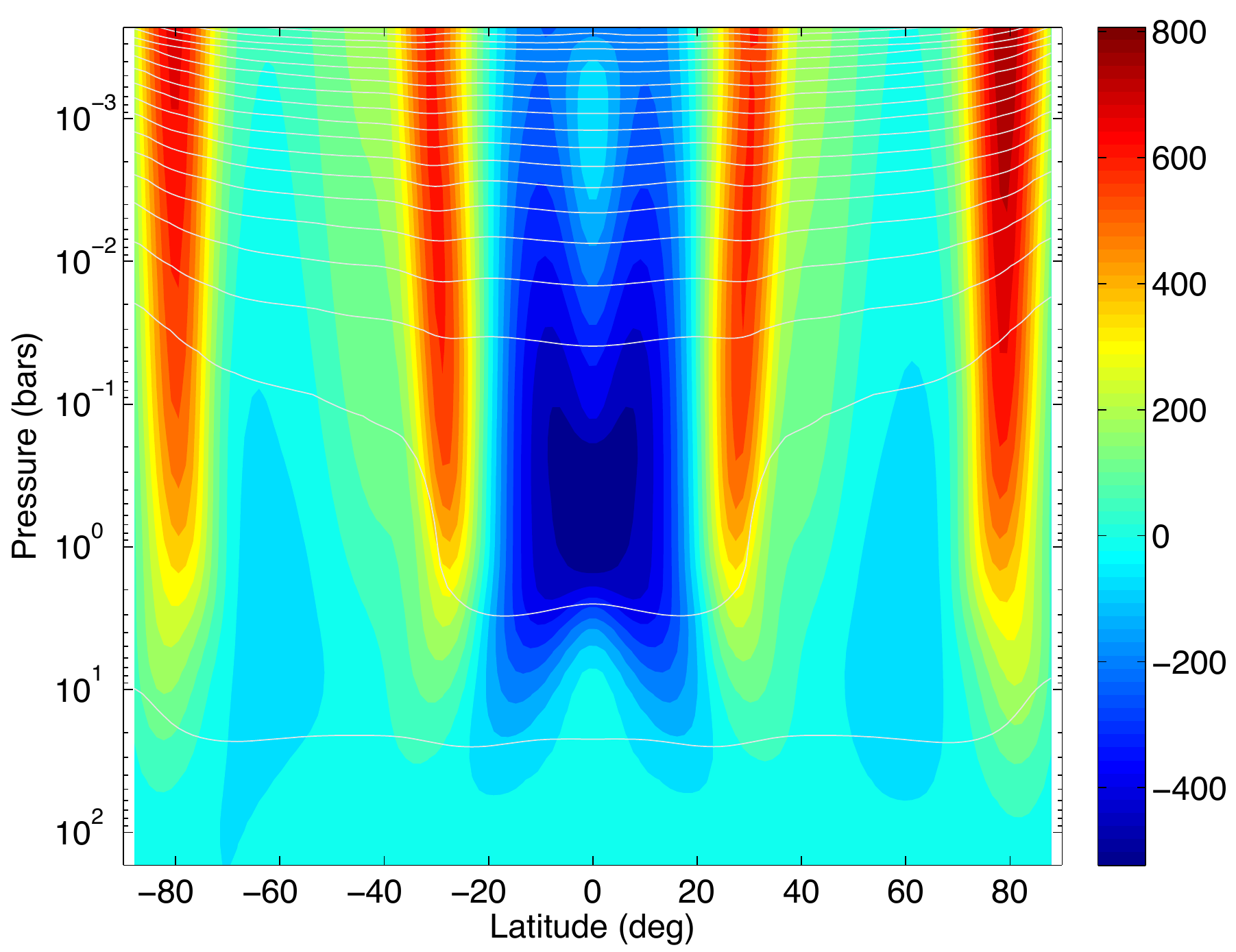}
\put(-90.,120.){\tiny C\ofast}\\
\includegraphics[scale=0.3, angle=0]{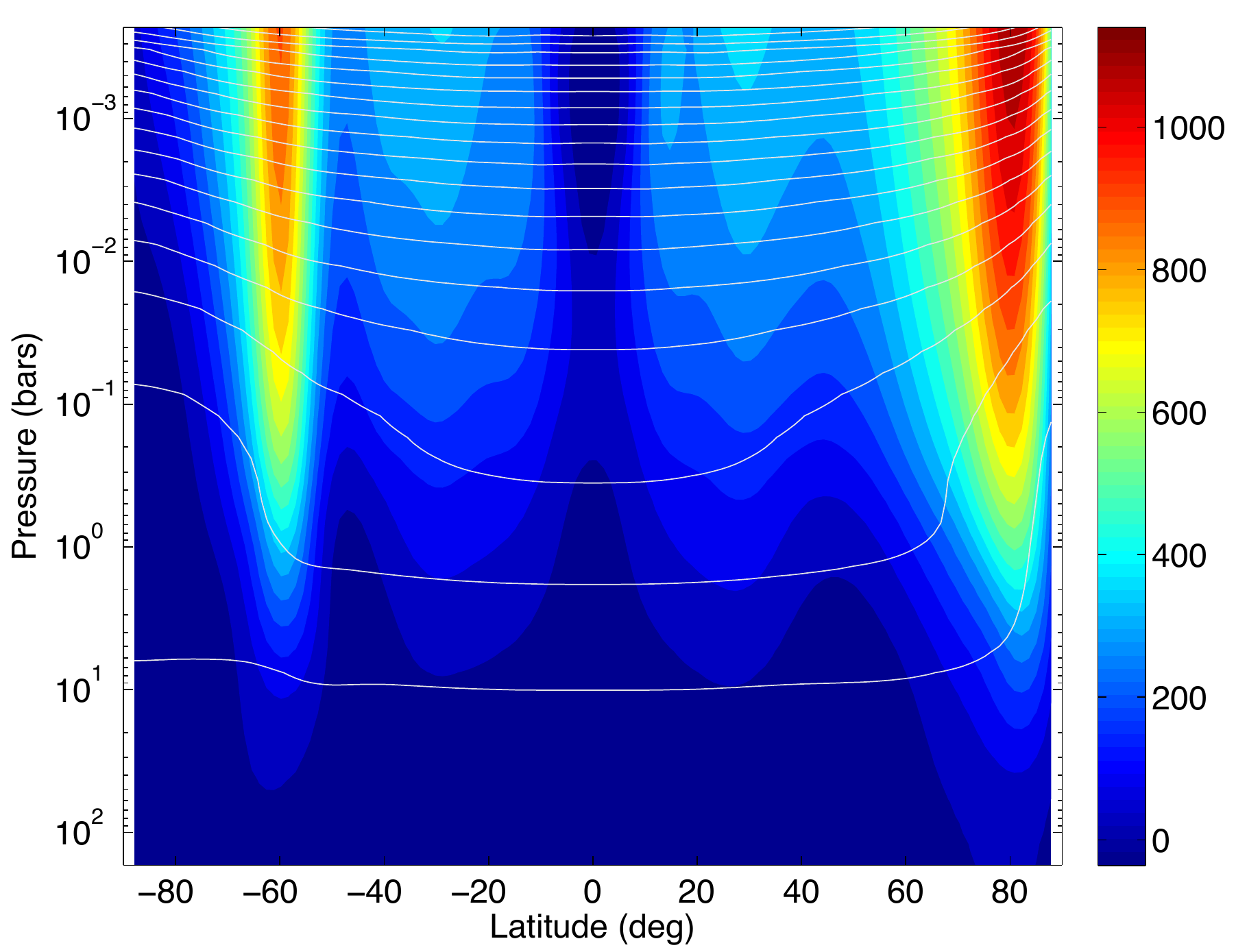}
\put(-90.,120.){\tiny W\ofast}\\
\includegraphics[scale=0.3, angle=0]{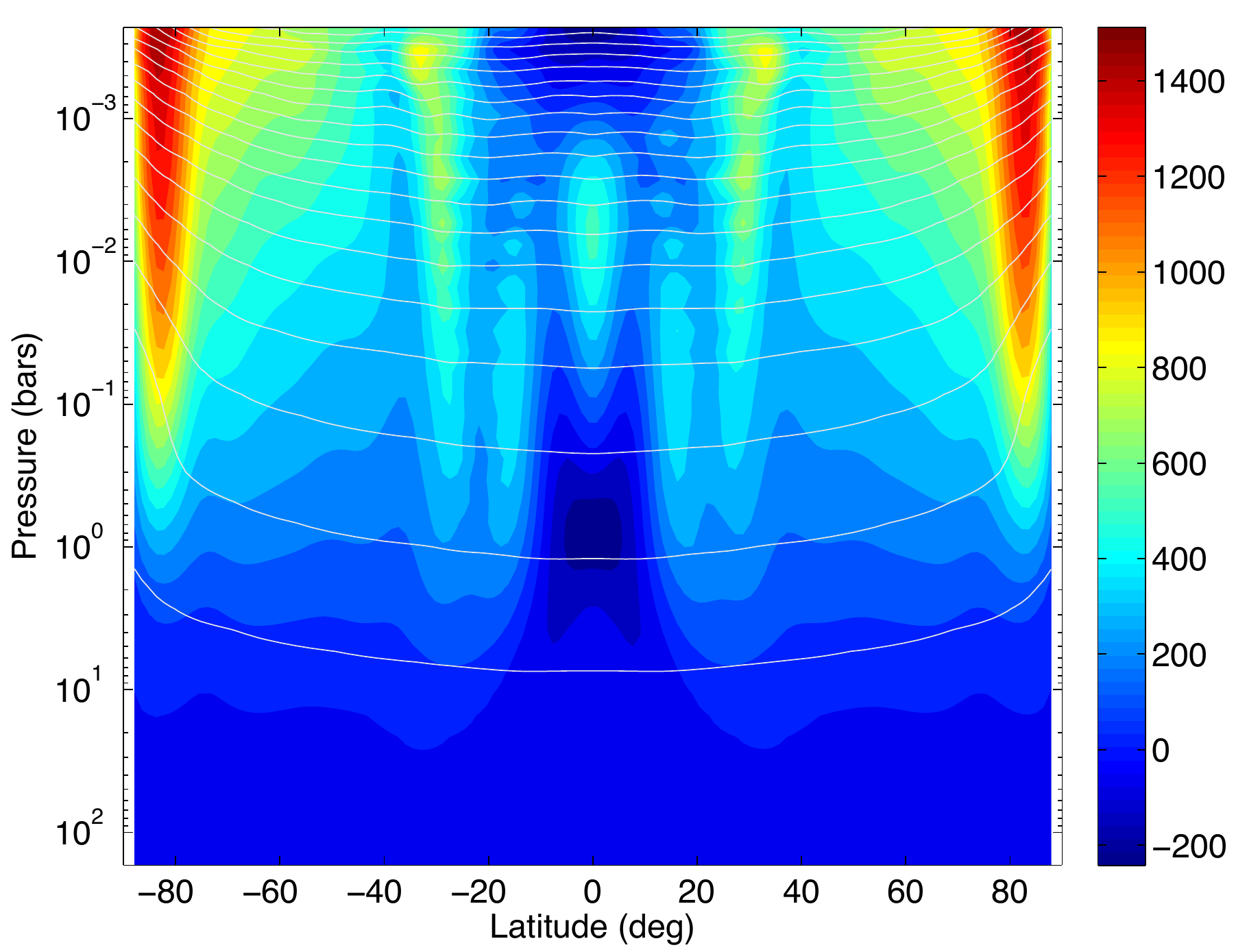}
\put(-90.,120.){\tiny H\ofast}

\end{minipage}
\begin{minipage}[c]{0.3\textwidth}
\includegraphics[scale=0.3, angle=0]{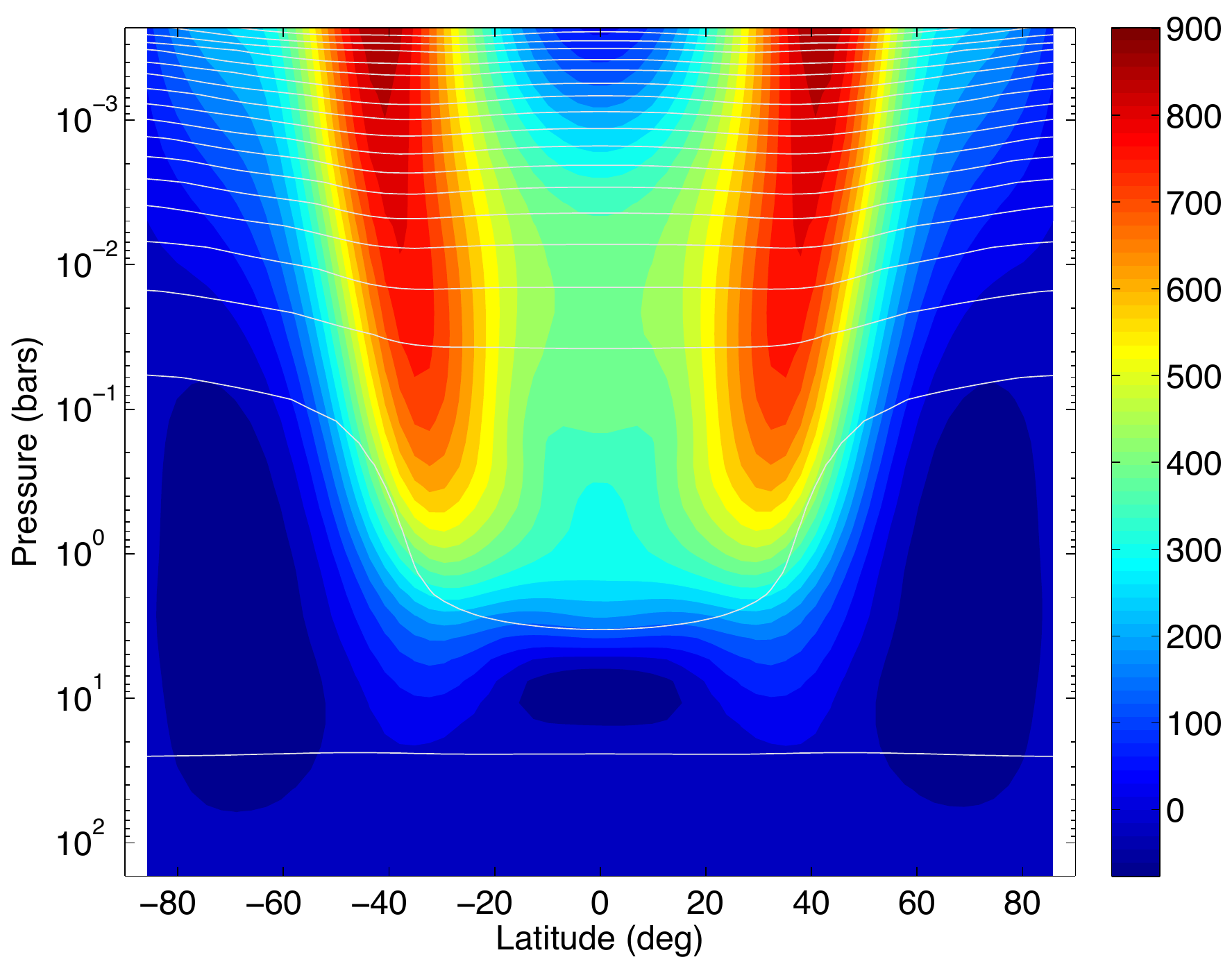}
\put(-90.,120.){\tiny C\omed}\\
\includegraphics[scale=0.3, angle=0]{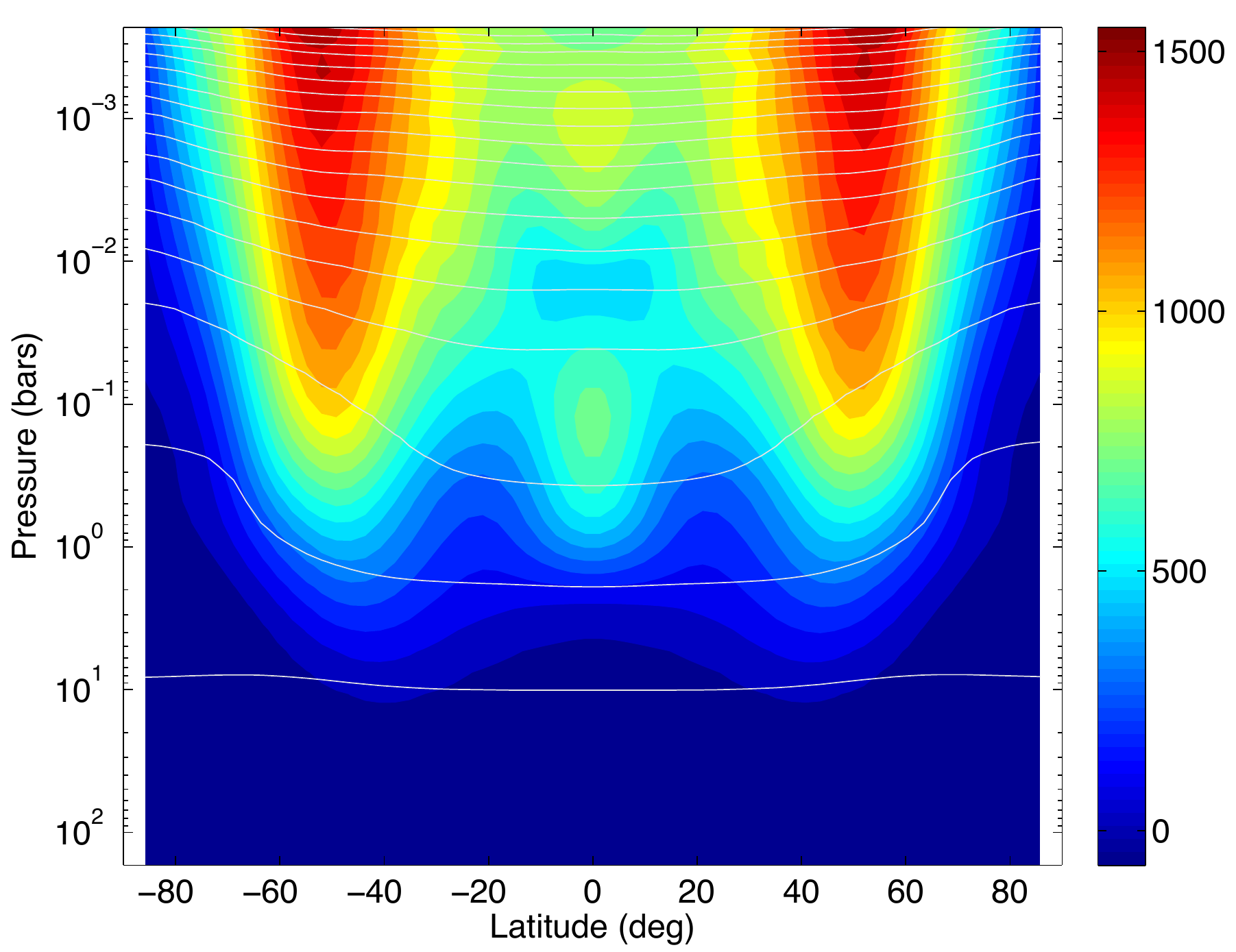}
\put(-90.,120.){\tiny W\omed}\\
\includegraphics[scale=0.3, angle=0]{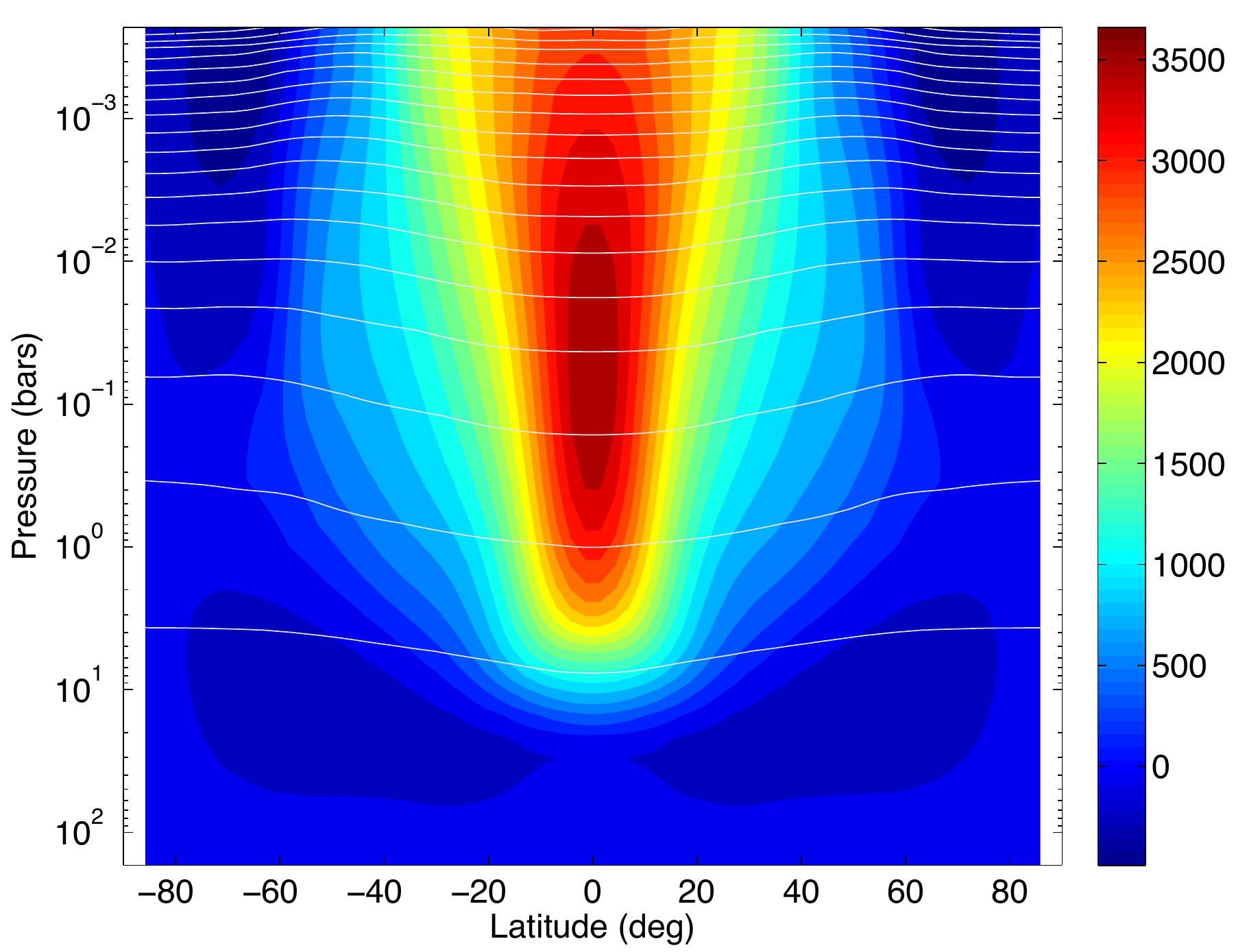}
\put(-90.,120.){\tiny H\omed}
\end{minipage}
\begin{minipage}[c]{0.3\textwidth}
\includegraphics[scale=0.3, angle=0]{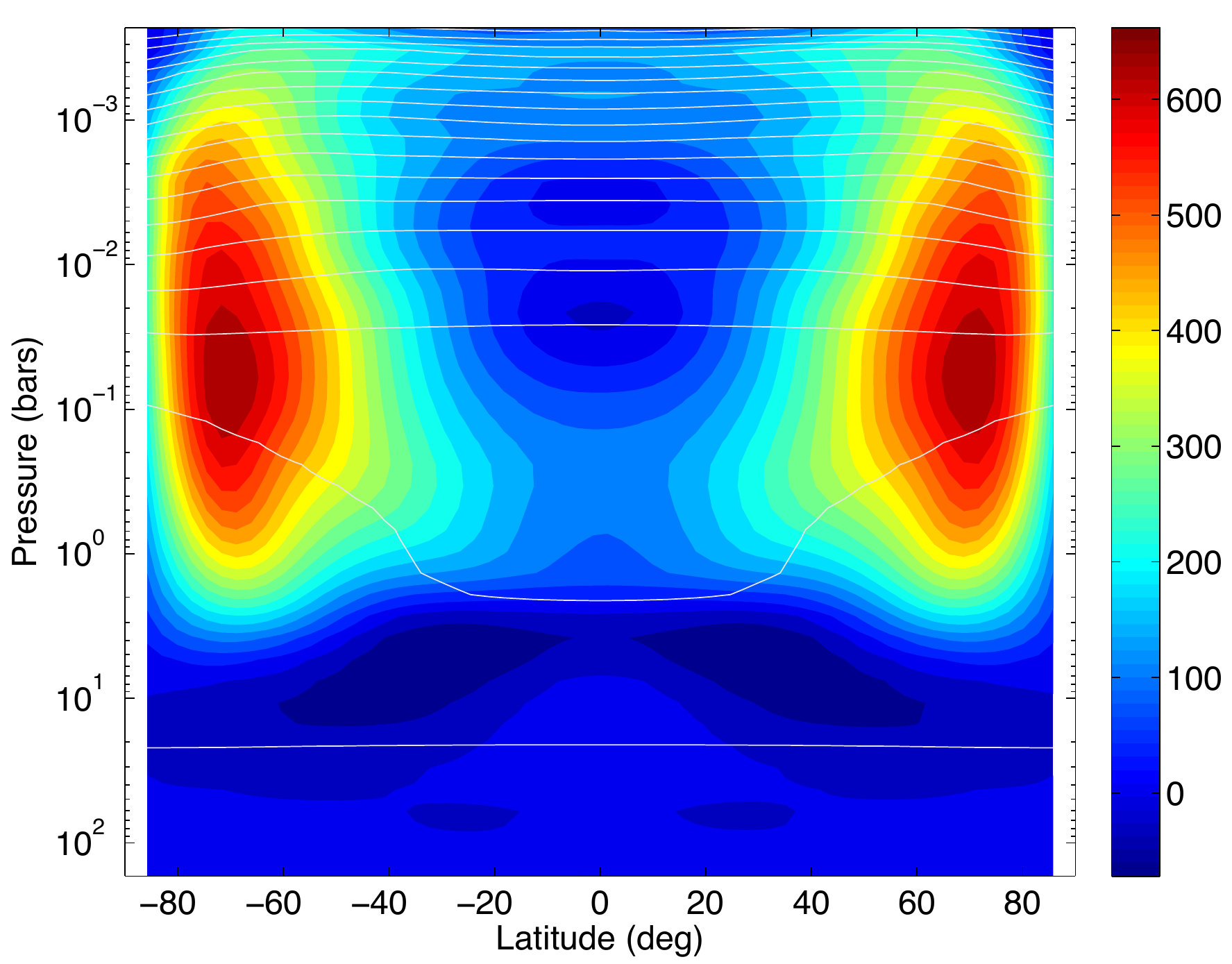}
\put(-90.,120.){\tiny C\oslow}\\
\includegraphics[scale=0.3, angle=0]{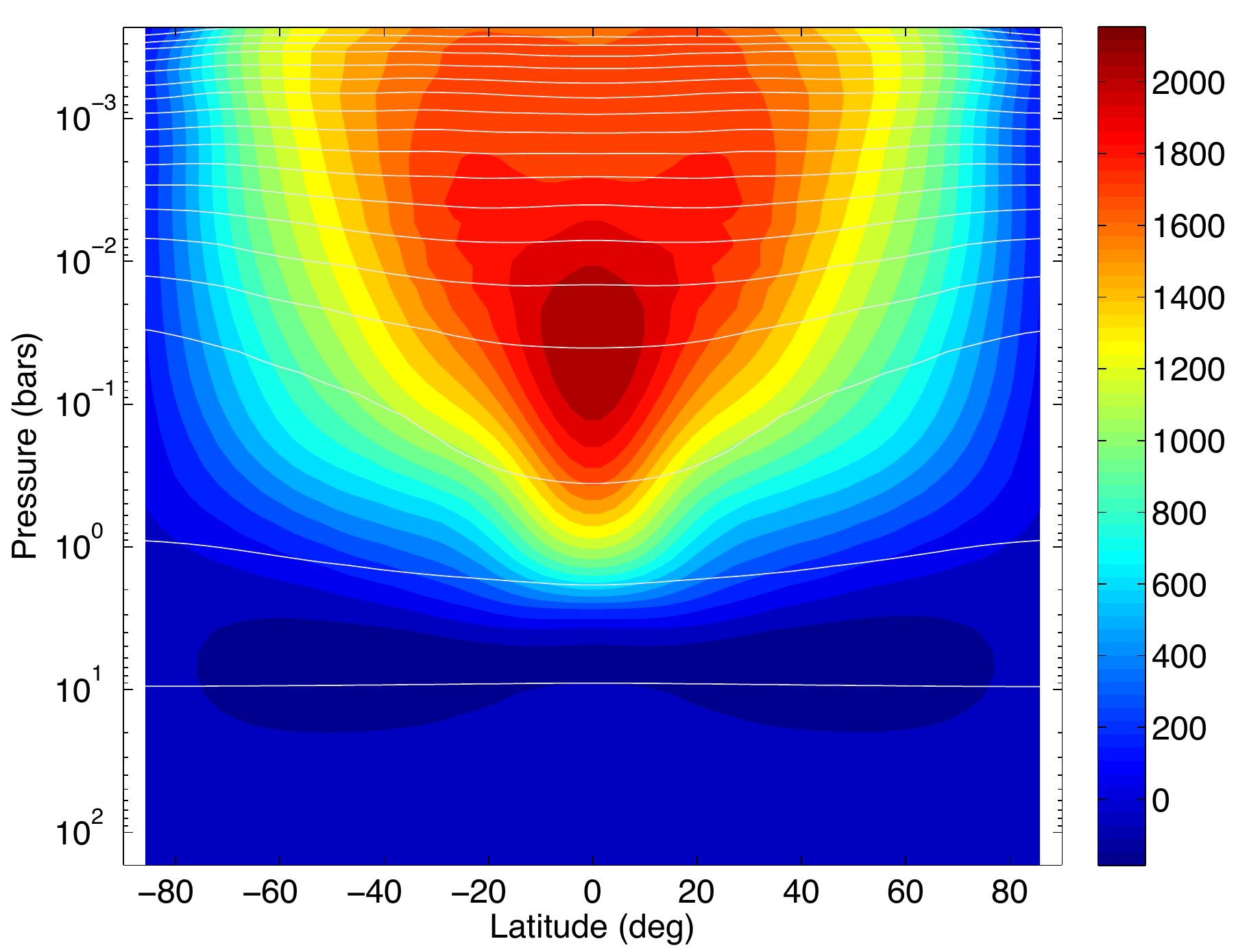}
\put(-90.,120.){\tiny W\oslow}\\
\includegraphics[scale=0.3, angle=0]{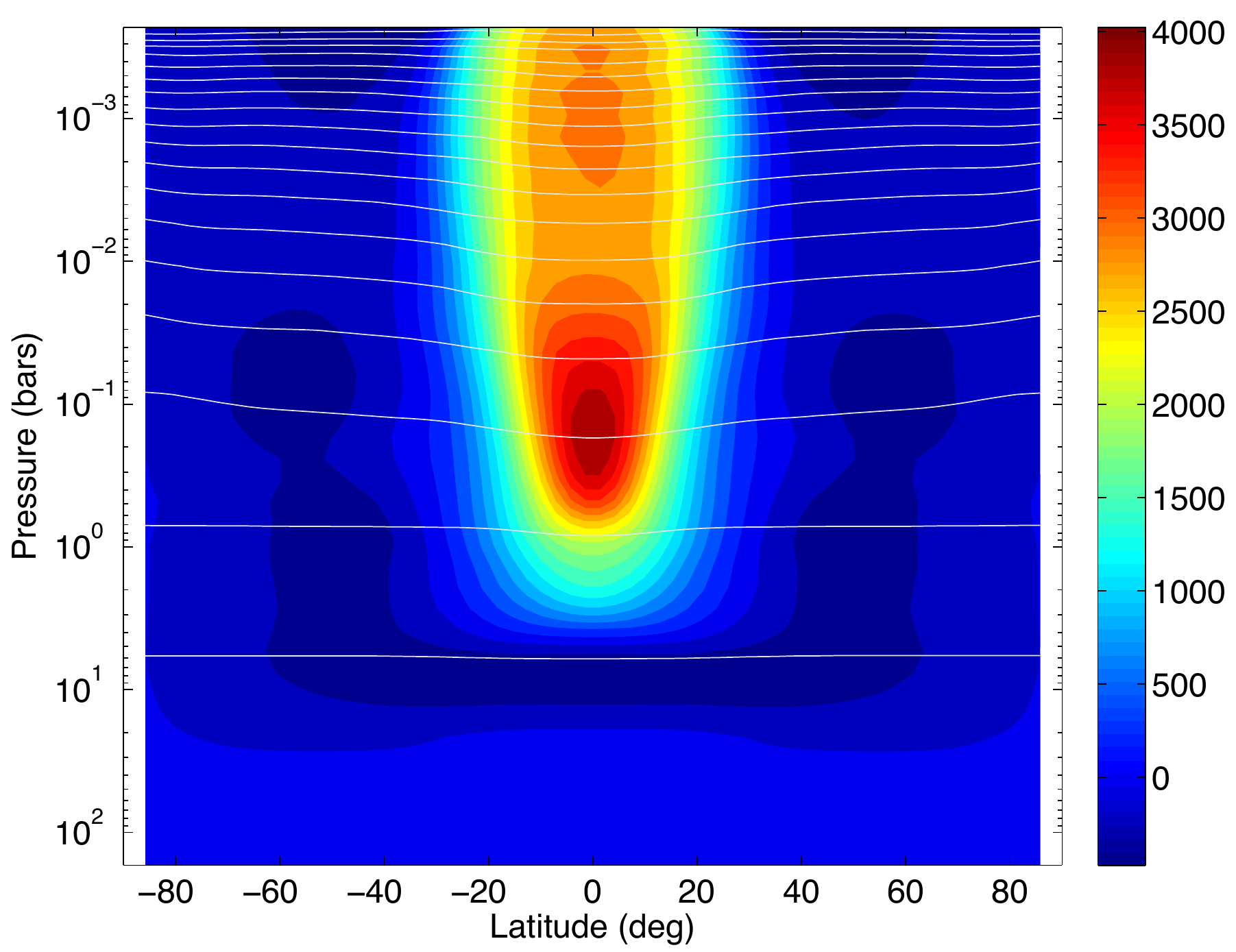}
\put(-90.,120.){\tiny H\oslow}
\end{minipage}
\caption{Zonal-mean circulation for the nine runs in the nominal grid.
  Each panel plots zonal-mean zonal wind (colorscale, $\rm m\,s^{-1}$)
  and zonal-mean potential temperature (white contours) versus
  latitude and pressure.  The left, middle, and right columns adopt
  rotation periods of 0.55, 2.2, and 8.8 days, respectively.  The top,
  middle, and bottom rows adopt orbital semimajor axes of 0.2, 0.08,
  and 0.03 AU, respectively.  Note the regime transition from a flow
  dominated by an equatorial superrotating jet in the bottom right
  (slow rotation, large incident flux) to a flow dominated by
  mid-latitude eastward jets in the middle and top left (fast
  rotation, small incident flux).}
\label{zonal-winds}
\end{figure*}

\begin{figure*}
\begin{minipage}[c]{0.33\textwidth}
\includegraphics[scale=0.465, angle=0]{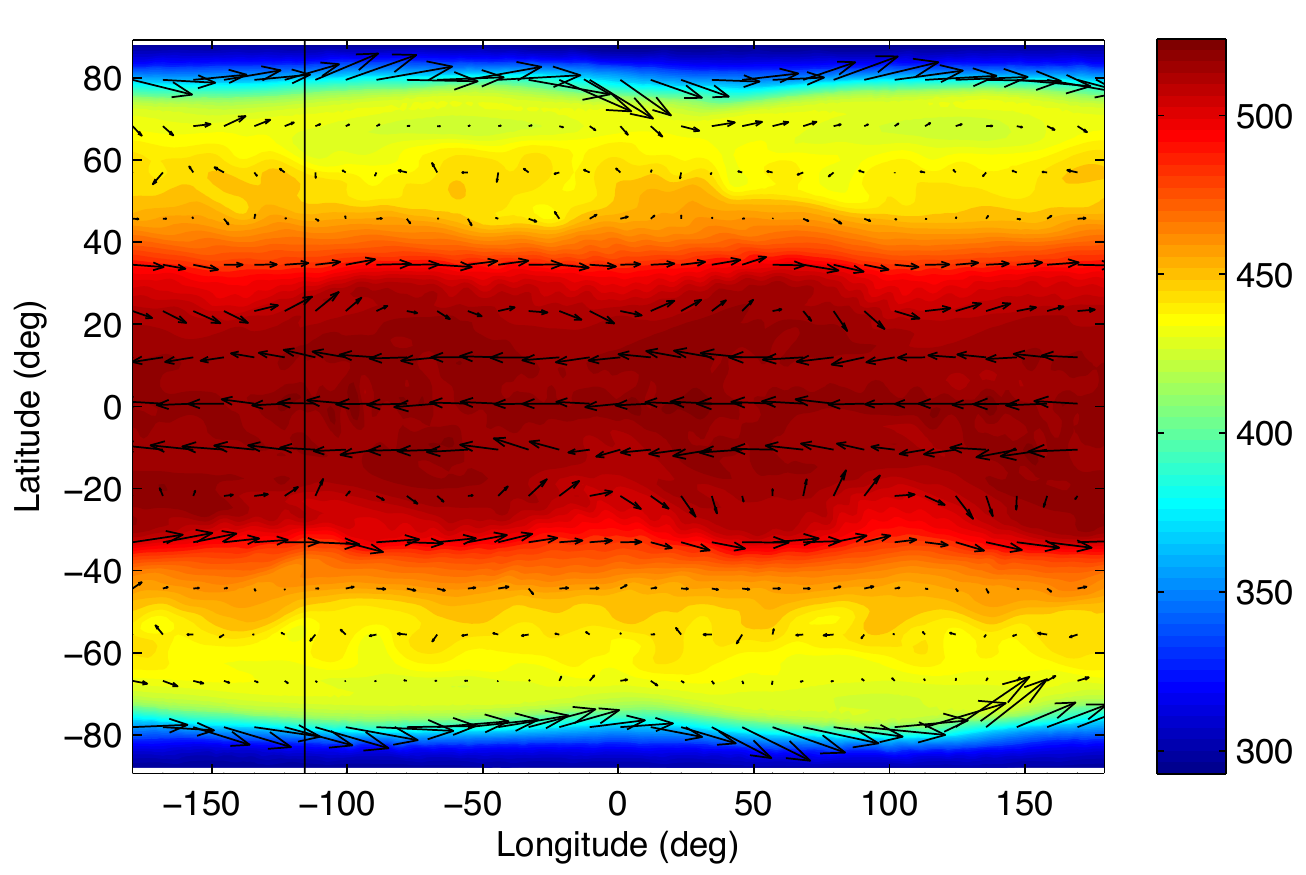}
\put(-105.,118.){\tiny C\ofast}\\
\includegraphics[scale=0.465, angle=0]{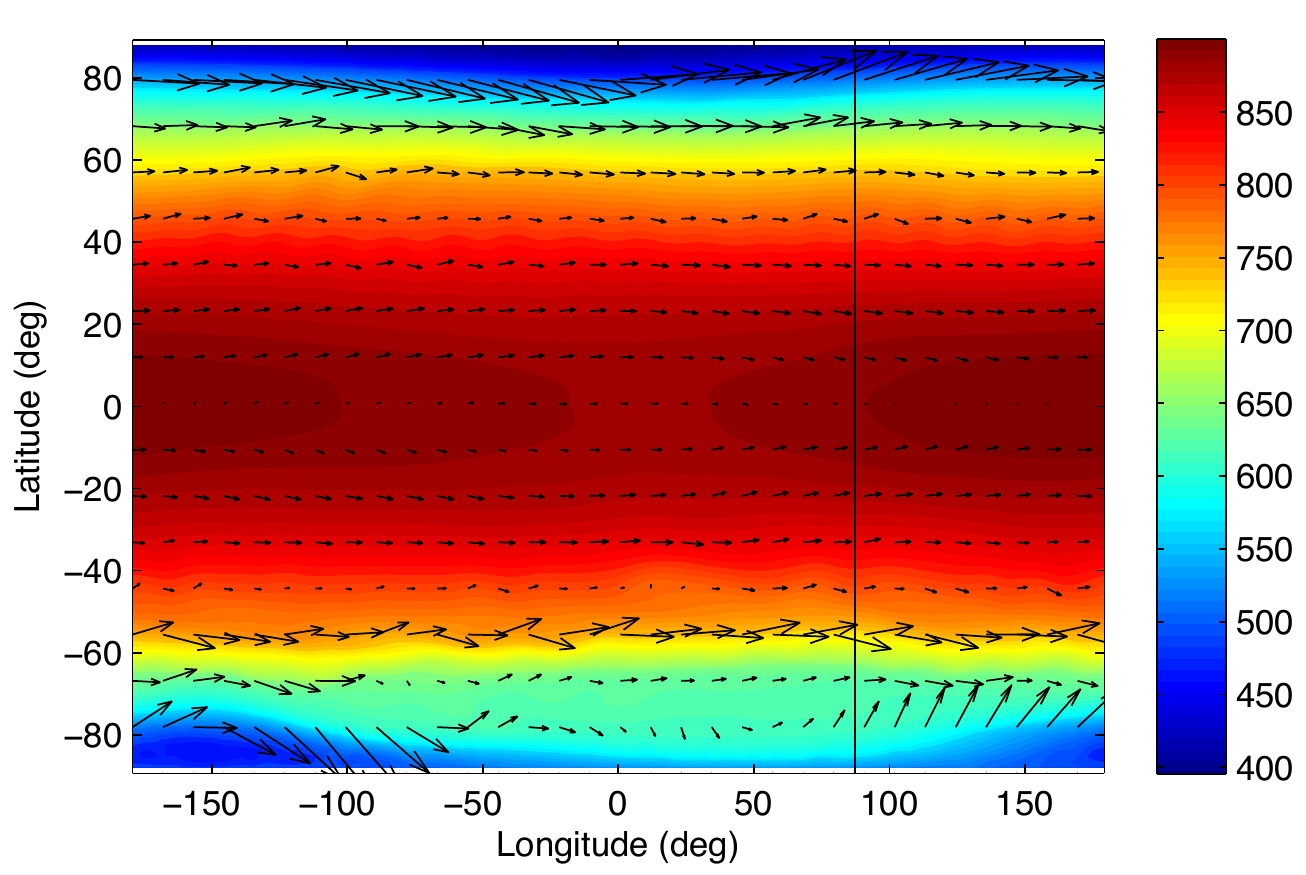}
\put(-105.,118.){\tiny W\ofast}\\
\includegraphics[scale=0.465, angle=0]{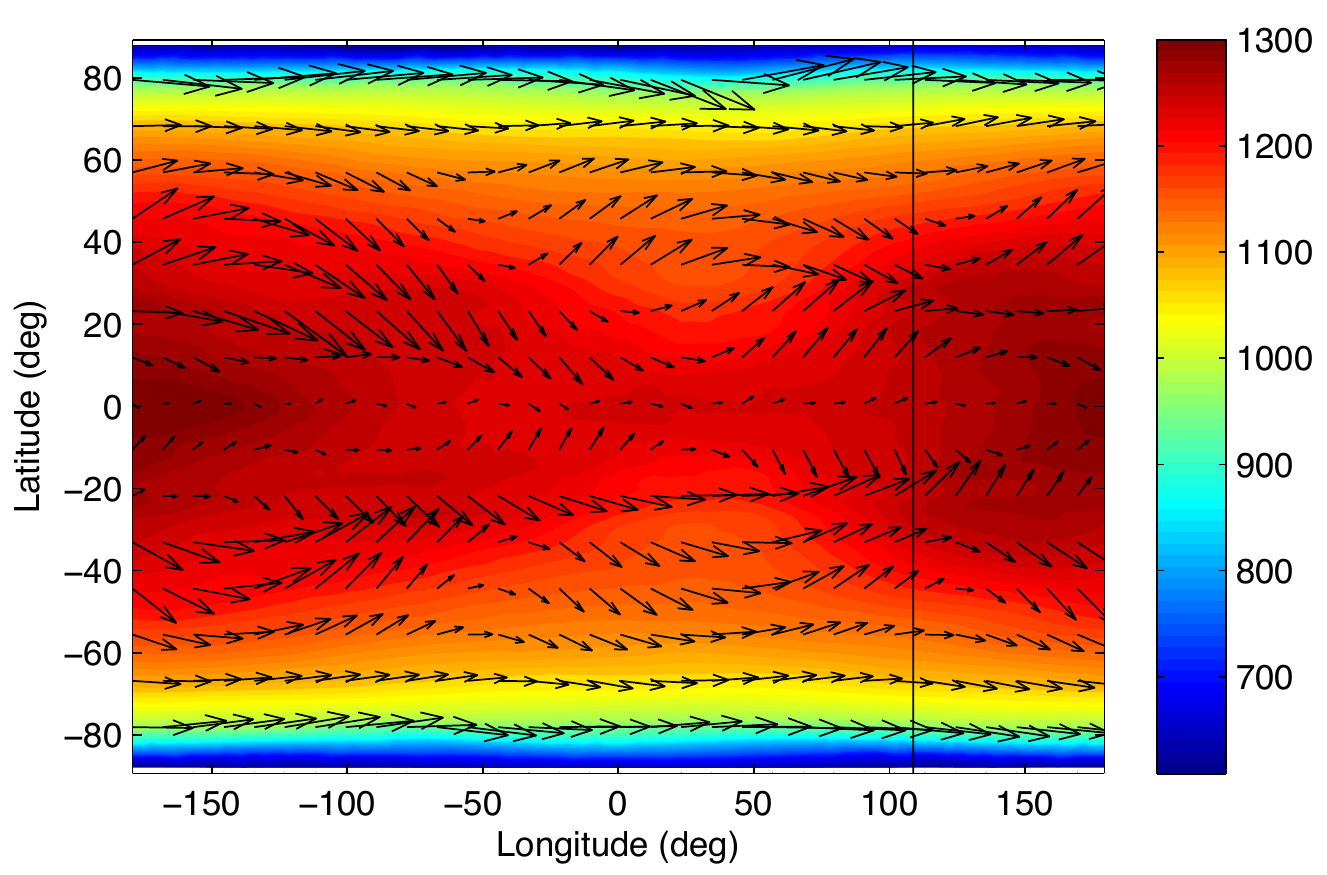}
\put(-105.,118.){\tiny H\ofast}
\end{minipage}
\begin{minipage}[c]{0.33\textwidth}
\includegraphics[scale=0.465, angle=0]{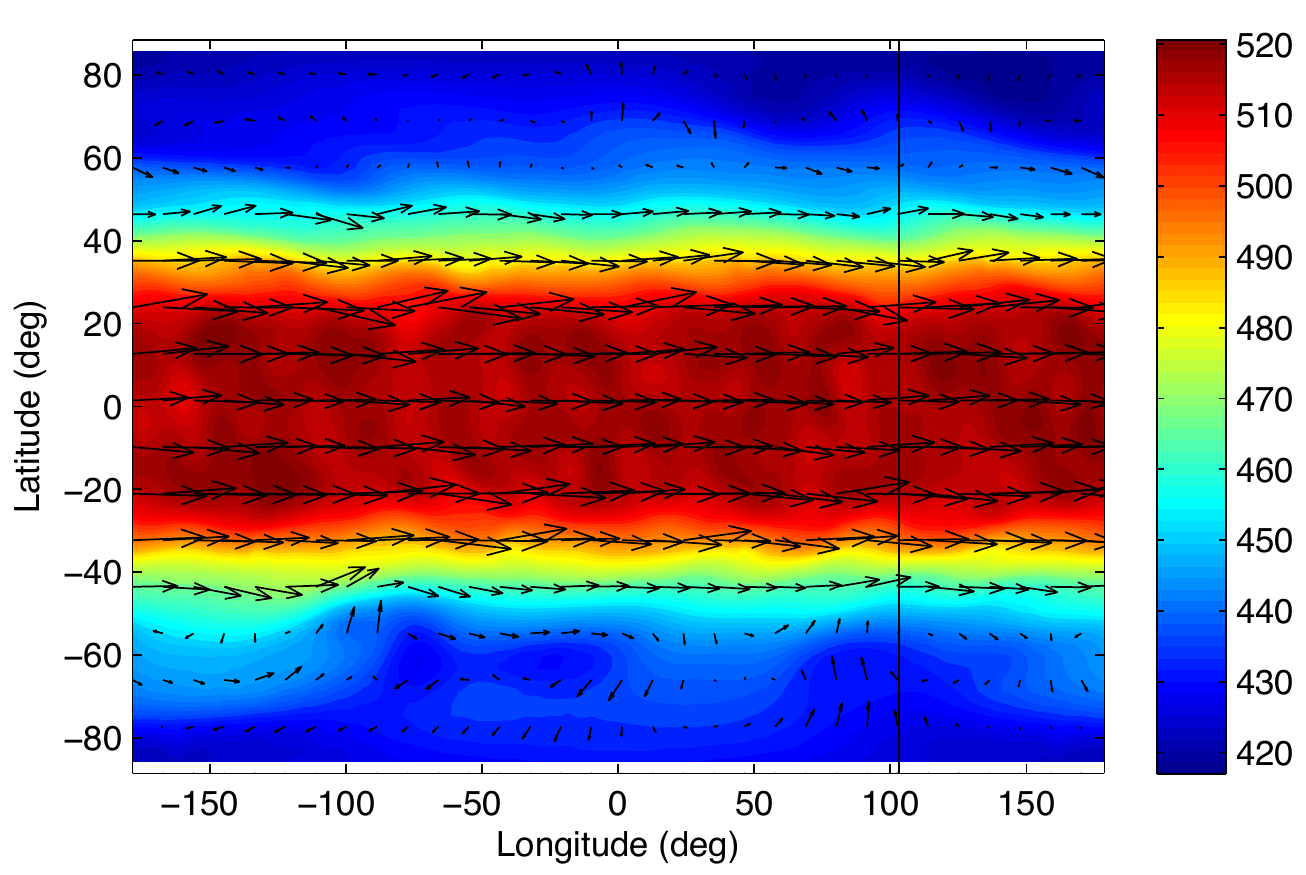}
\put(-105.,118.){\tiny C\omed}\\
\includegraphics[scale=0.465, angle=0]{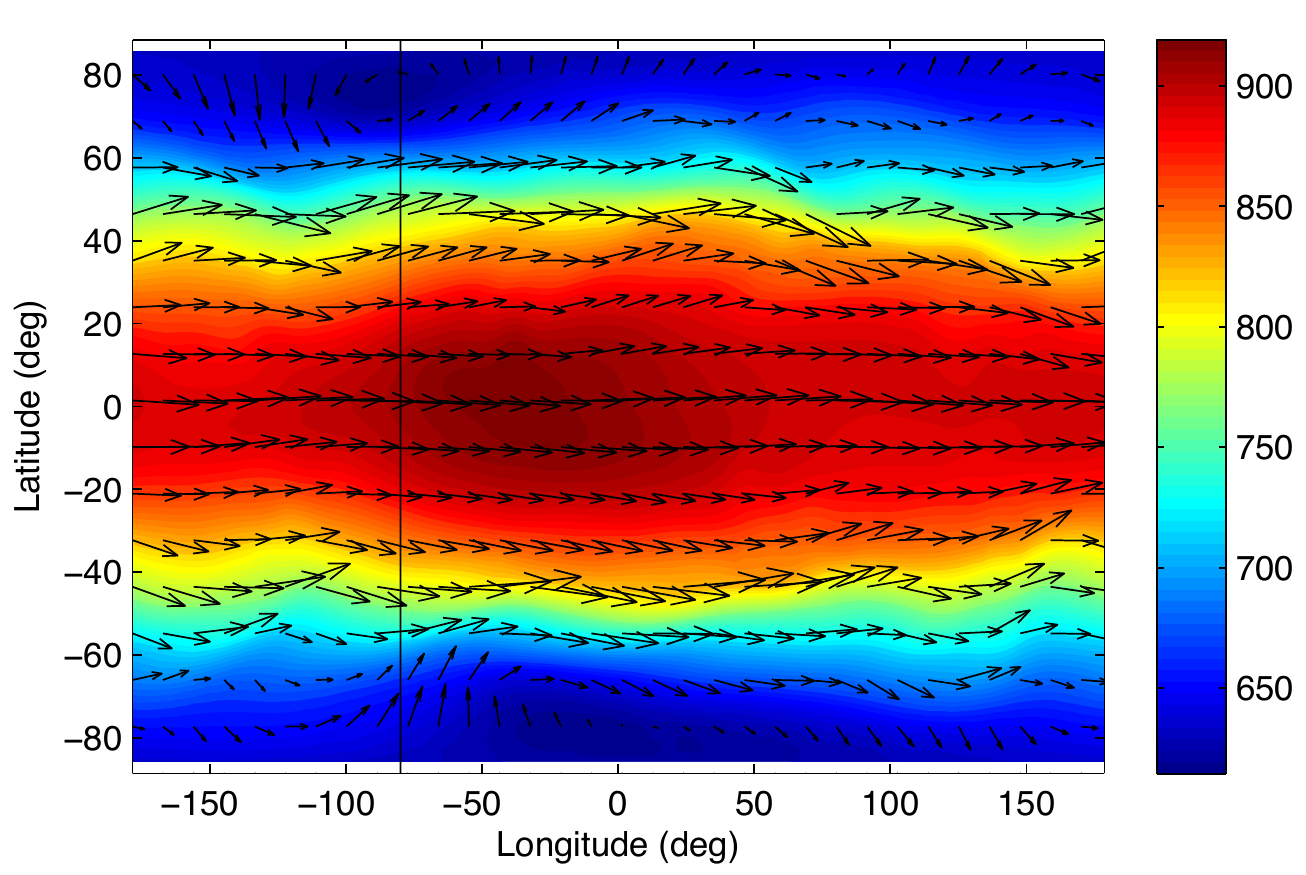}
\put(-105.,118.){\tiny W\omed}\\
\includegraphics[scale=0.465, angle=0]{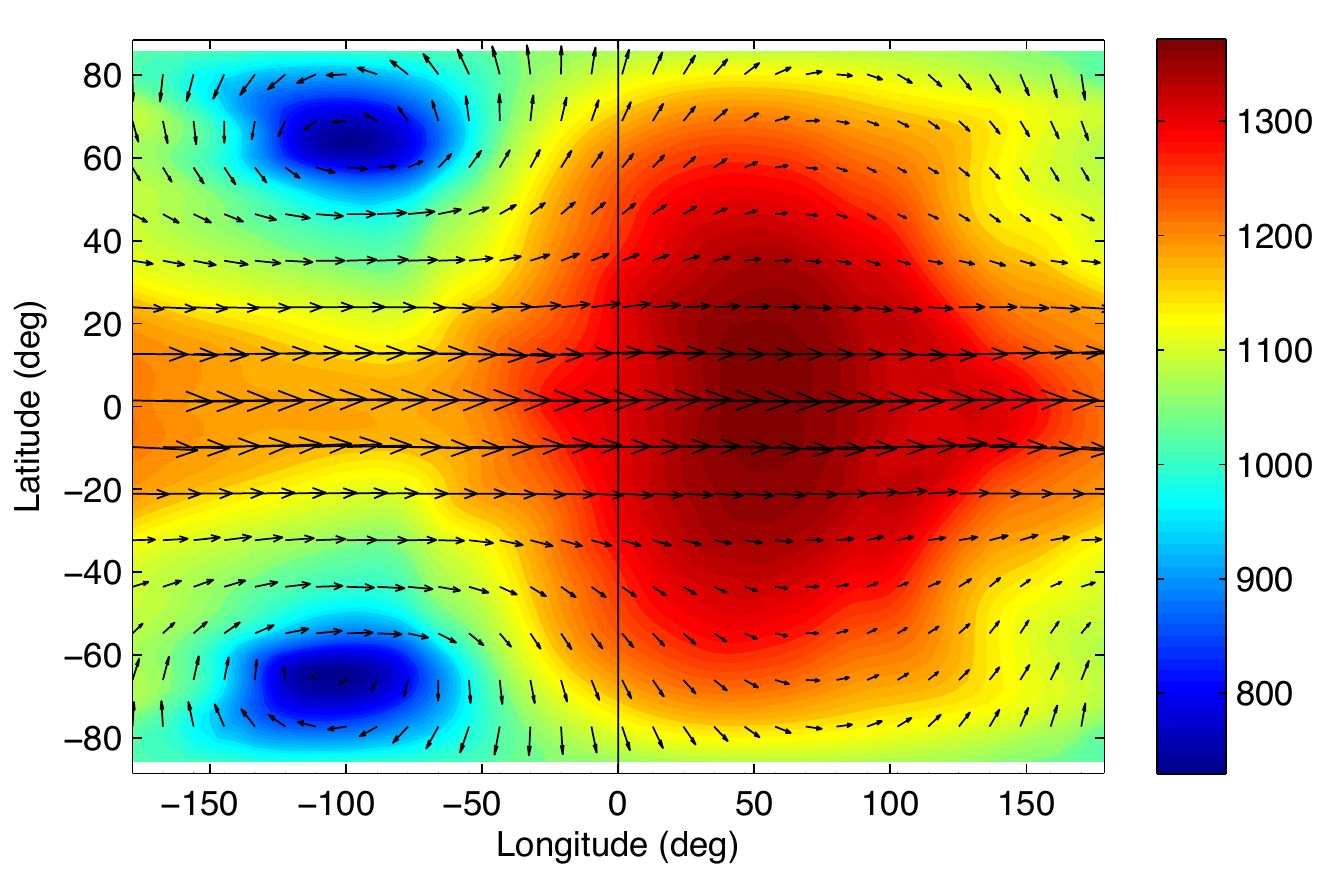}
\put(-105.,118.){\tiny H\omed}
\end{minipage}
\begin{minipage}[c]{0.3\textwidth}
\includegraphics[scale=0.465, angle=0]{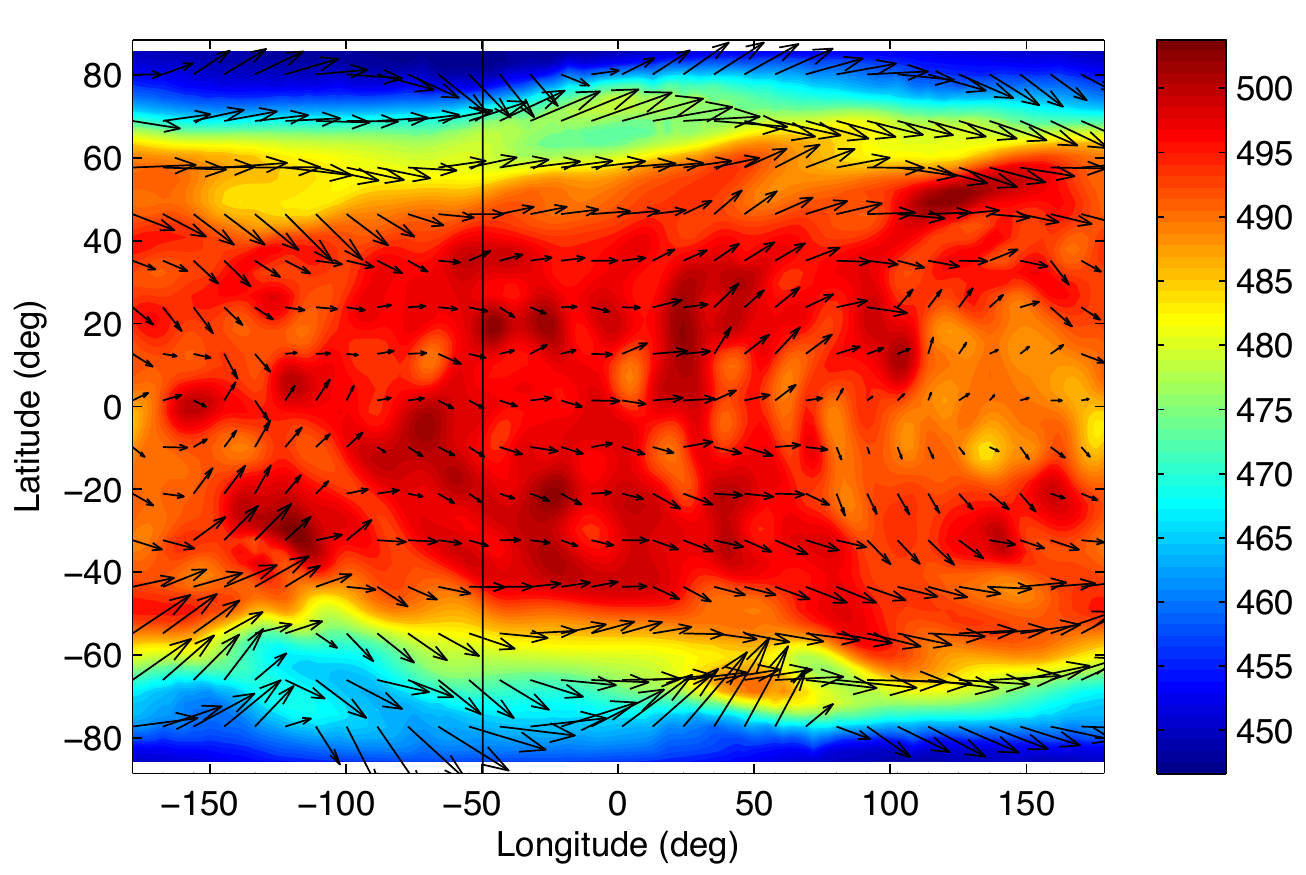}
\put(-105.,118.){\tiny C\oslow}\\
\includegraphics[scale=0.465, angle=0]{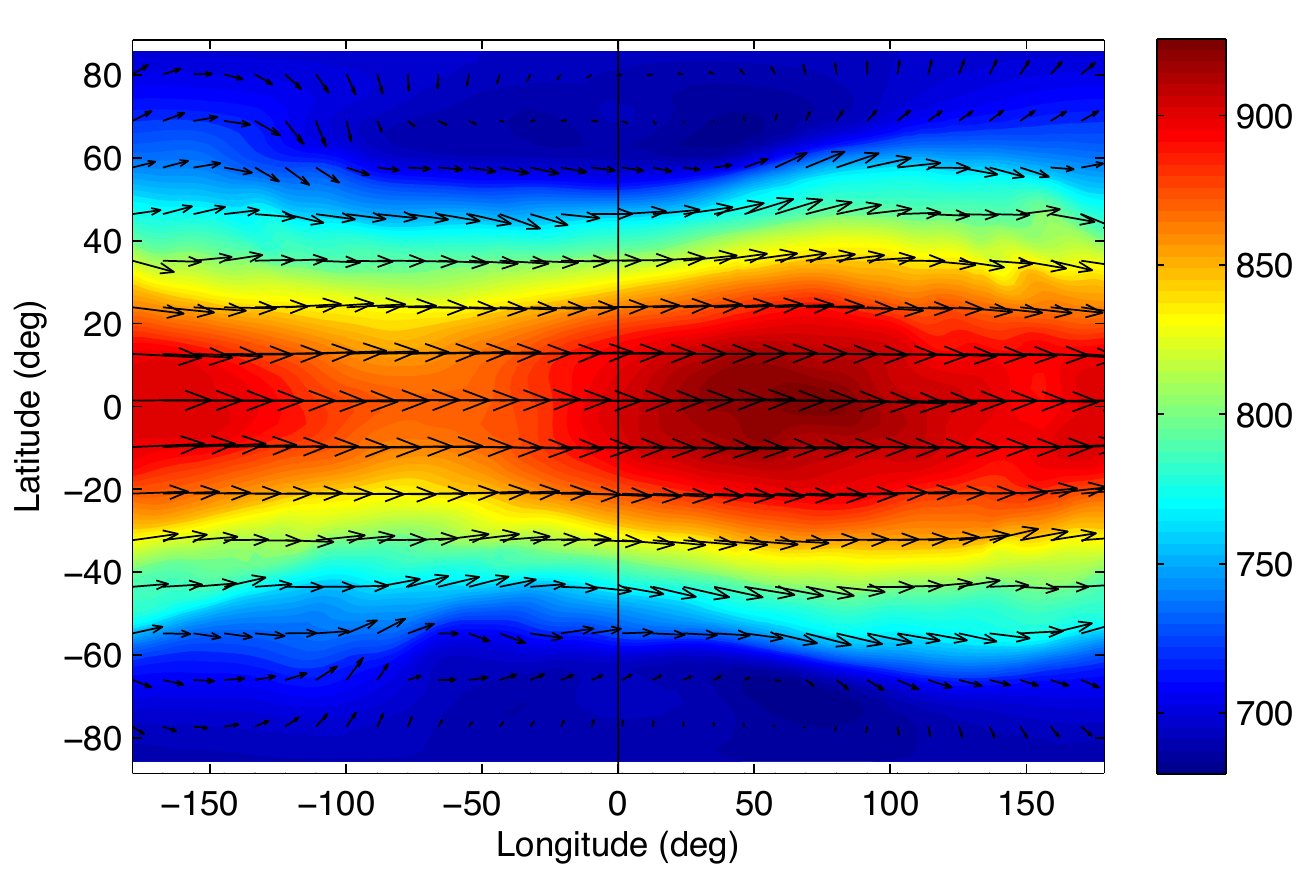}
\put(-105.,118.){\tiny W\oslow}\\
\includegraphics[scale=0.465, angle=0]{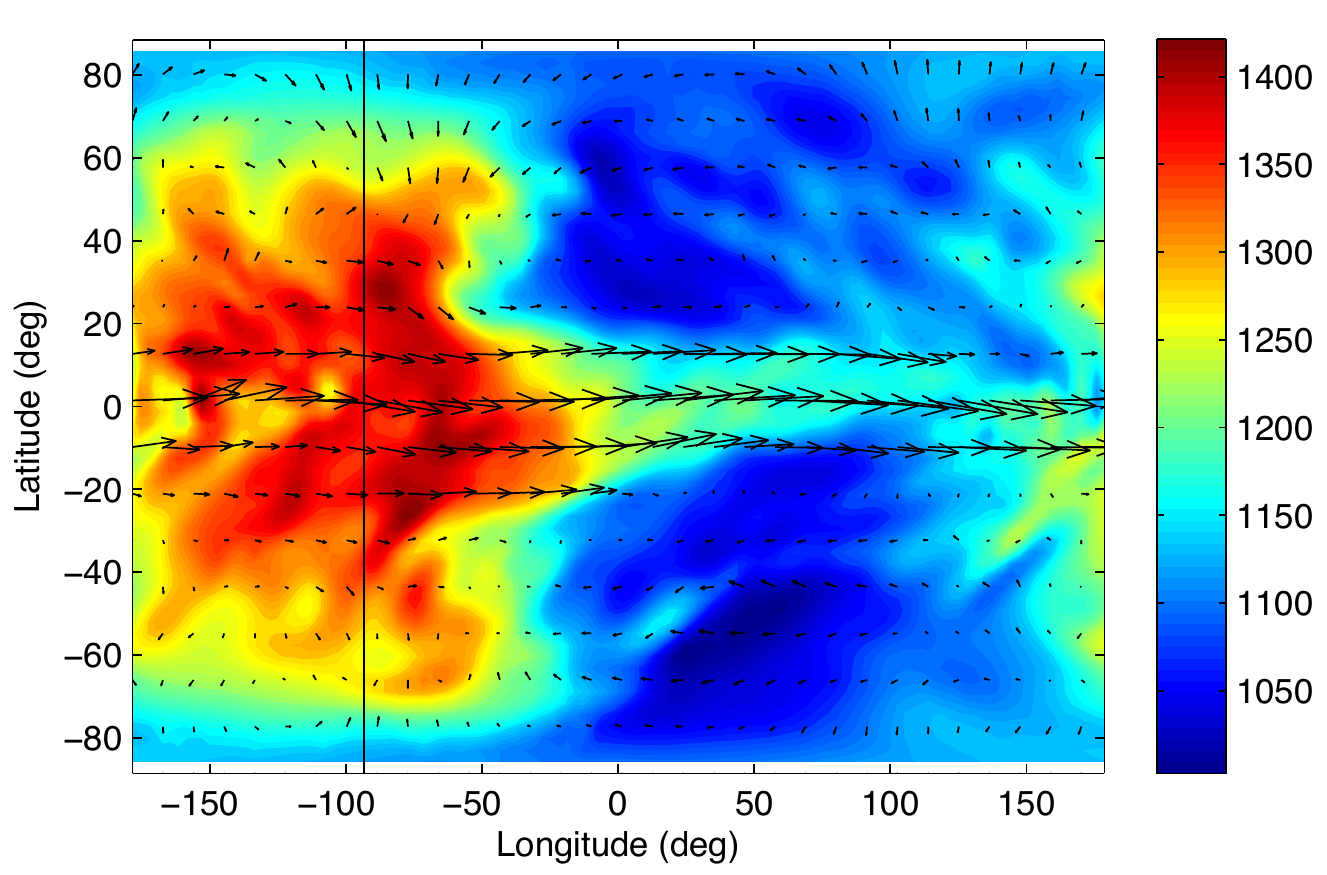}
\put(-105.,118.){\tiny H\oslow}
\end{minipage}
\caption{Temperature (colorscale, K) and winds (arrows) on an isobar
  for the nine runs in the nominal grid.  Each panel plots the
  temperature and winds versus longitude and latitude at 170 mbar.
  The left, middle, and right columns adopt rotation periods of 0.55,
  2.2, and 8.8 days, respectively.  The top, middle, and bottom rows
  adopt orbital semimajor axes of 0.2, 0.08, and 0.03 AU,
  respectively. Vertical black lines in each panel denote the
  substellar longitude at this snapshot.  The hot cases tend to have
  large day-night temperature differences, especially at slow rotation
  rates; cooler cases---particularly at fast rotation rate---have
  minimal temperature differences in longitude but significant
  temperature differences in latitude.}
\label{temp-winds}
\end{figure*}

Our key result is that the circulation undergoes a major
reorganization as the irradiation level and rotation rate are
varied---as predicted by the theoretical arguments in
Section~\ref{theory}.  At high irradiation and slow rotation rates,
the circulation is dominated by a broad equatorial (superrotating) jet
and significant day-night temperature differences at low pressure.
But at low irradiation and/or faster rotation rates, the circulation
shifts to a regime dominated by off-equatorial eastward jets, with
weaker eastward or even westward flow at the equator; day-night
temperature differences are smaller, and the primarily horizontal
temperature differences instead occur between the equator and pole.
This behavior emerges clearly in Figure~\ref{zonal-winds}, which
presents the zonal-mean zonal wind versus latitude and pressure, and
Figure~\ref{temp-winds}, which shows the temperature and wind
structure on the 170-mbar isobar, within the layer that shapes
the IR light curves and spectra.  Here, we discuss basic
structure and trends across our entire ensemble, deferring to
Section~\ref{mechanisms} more detailed diagnostics of the two regimes.

First, consider the highly irradiated, slowly rotating regime of
equatorial superrotation and large day-night temperature contrasts;
this behavior is best developed in the lower-right corner of
Figures~\ref{zonal-winds}--\ref{temp-winds} (models W\oslow, H\omed,
and H\oslow).  The circulation in this portion of the parameter space
resembles that explored extensively in the hot Jupiter literature
\citep[e.g.,][]{showman-guillot-2002, showman-etal-2008a,
  showman-etal-2009, dobbs-dixon-lin-2008, menou-rauscher-2009,
  rauscher-menou-2010, heng-etal-2011}.  Peak zonal-mean zonal wind
speeds reach 2~to $4\rm\,km\,s^{-1}$ in the core of the equatorial
jet, with maximum wind speeds occurring at the equator.  The jet
extends smoothly from the top of the domain ($\sim$0.2 mbar) to a
pressure of $\sim$3--10 bars depending on the model, with the fastest
zonal-mean speeds at a midlevel of 0.1--0.3 bar. In models H\omed and
H\oslow, the equatorial jet is sufficiently narrow for westward
zonal-mean flow to develop at high latitudes; in others (W\oslow), the
equatorial jet extends from pole to pole, such that the zonal-mean
flow within the jet is eastward at essentially all latitudes.  As
shown in Figure~\ref{temp-winds}, day-night temperature differences
are large in the observable atmosphere.  The dayside is characterized
by a broad, hemispheric-scale hot region that is shifted eastward of
the substellar longitude by $\sim$10--$50^{\circ}$ at the highest
irradiation (H\omed and H\oslow) and $\sim$50--$90^{\circ}$ at lower
irradiation (W\oslow).  The amplitude of the offsets depend on pressure.
Longitudinal temperature variations are up to
$\sim$$400\rm\,K$ in the hottest models (H\omed and H\oslow) and
$\sim$100--$200\rm\,K$ in W\oslow.

Next, consider the weakly irradiated and/or rapidly rotating regime
(upper left portion of Figures~\ref{zonal-winds}--\ref{temp-winds};
models C\ofast, C\omed, C\oslow, W\ofast, W\omed, and H\ofast).  In
this regime, superrotation is less dominant, and the circulation
instead becomes dominated by eastward jets at mid-to-high latitudes in
each hemisphere, with weaker eastward or even westward flow at the
equator.  Peak zonal-mean zonal wind speeds reach
$\sim$0.6--$1.4\rm\,km\,s^{-1}$ in the off-equatorial eastward jets.
When the rotation is slow-to-intermediate, only two off-equatorial
eastward jets develop (one in each hemisphere), as seen in models
C\omed, C\oslow, and W\omed.  When the rotation is fast, however, the
dynamical length scales are shorter, and the flow splits into four
mid-latitude eastward jets (two per hemisphere), as in models H\ofast,
W\ofast, and C\ofast.  The phenomenon is best developed in model
C\ofast, the fastest-rotating, lowest-irradiation model of our
ensemble.  In this regime, temperatures are relatively constant in
longitude but vary significantly in latitude (middle and upper left of
Figure~\ref{temp-winds}).  As can be seen from the detailed velocity
patterns in Figure~\ref{temp-winds}, the eastward mid-latitude jets
exhibit quasi-periodic undulations in longitude with zonal wavenumbers
ranging from 2 to $\sim$14, suggesting that the jets are experiencing
dynamical instabilities.  As we will show in Section~\ref{mechanisms},
these instabilities play an important role in maintaining the jets.
The off-equatorial ciculation in these models qualitatively
resembles that of Earth or Jupiter.

The location of the transition, as a function of rotation rate and
stellar flux, agrees well with that predicted in Section~\ref{theory}
(compare Figures~\ref{parameter-space} and \ref{zonal-winds}).  This
provides tentative support for the theoretical arguments presented
there.

The transition between the regimes is continuous and broad.  Models
along the boundary between the two regimes---in particular, C\omed,
W\omed, and H\ofast---exhibit aspects of both regimes.  For example,
C\omed and W\omed exhibit a flow comprising midlatitude eastward jets
(red colors in Figure~\ref{zonal-winds}) embedded in a broad superrotating
flow that includes the equator.  It is likely that, in this regime, 
{\tt diurnal (day-night) forcing and baroclinic instabilities
 both play a strong role in driving the circulation, leading to a 
hybrid between the two scenarios shown in Figure~\ref{schematic}.}

The importance of rotation in the dynamics varies significantly across
our ensemble.  This is characterized by the Rossby number,
$Ro=U/\Omega L$, giving the ratio of advective to Coriolis forces in
the horizontal momentum equation, where $U$ and $L$ are the
characteristic wind speed and horizontal length scale and $\Omega$ is
the planetary rotation rate.  Considering a length scale $L\approx
10^8\rm m$ appropriate to a global-scale flow, our slowest-rotating
models have $Ro = 1.2(U/1000 \rm \,m\,s^{-1})$, implying Rossby
numbers as high as $\sim$4 when irradiation is strongest.  In this
case, Coriolis forces, while important, will be subdominant to
advection in the horizontal force balances.  On the other hand, our
most rapidly rotating models exhibit $Ro = 0.15(U/\rm
\,1000\,m\,s^{-1})$, where we have used $L\approx 5\times10^7\rm \,m$
to account for the shorter flow length scales in those cases
(Figure~\ref{zonal-winds}, left column).  When irradiation is weakest,
wind speeds are typically $\sim$$300\rm\,m\,s^{-1}$
(Figure~\ref{zonal-winds}), implying $Ro\sim 0.05$---smaller than the
value on Earth.  This implies that the large-scale flow is in
approximate geostrophic balance---i.e., a balance between Coriolis and
pressure-gradient forces in the horizontal momentum equation.  Such a
force balance is the dominant force balance away from the equator on
most solar system planets (Earth, Mars, Jupiter, Saturn, Uranus, and
Neptune).  In terms of force balances, these models therefore resemble
these solar system planets more than ``canonical'' hot Jupiters.

\begin{figure}
\includegraphics[scale=0.45, angle=0]{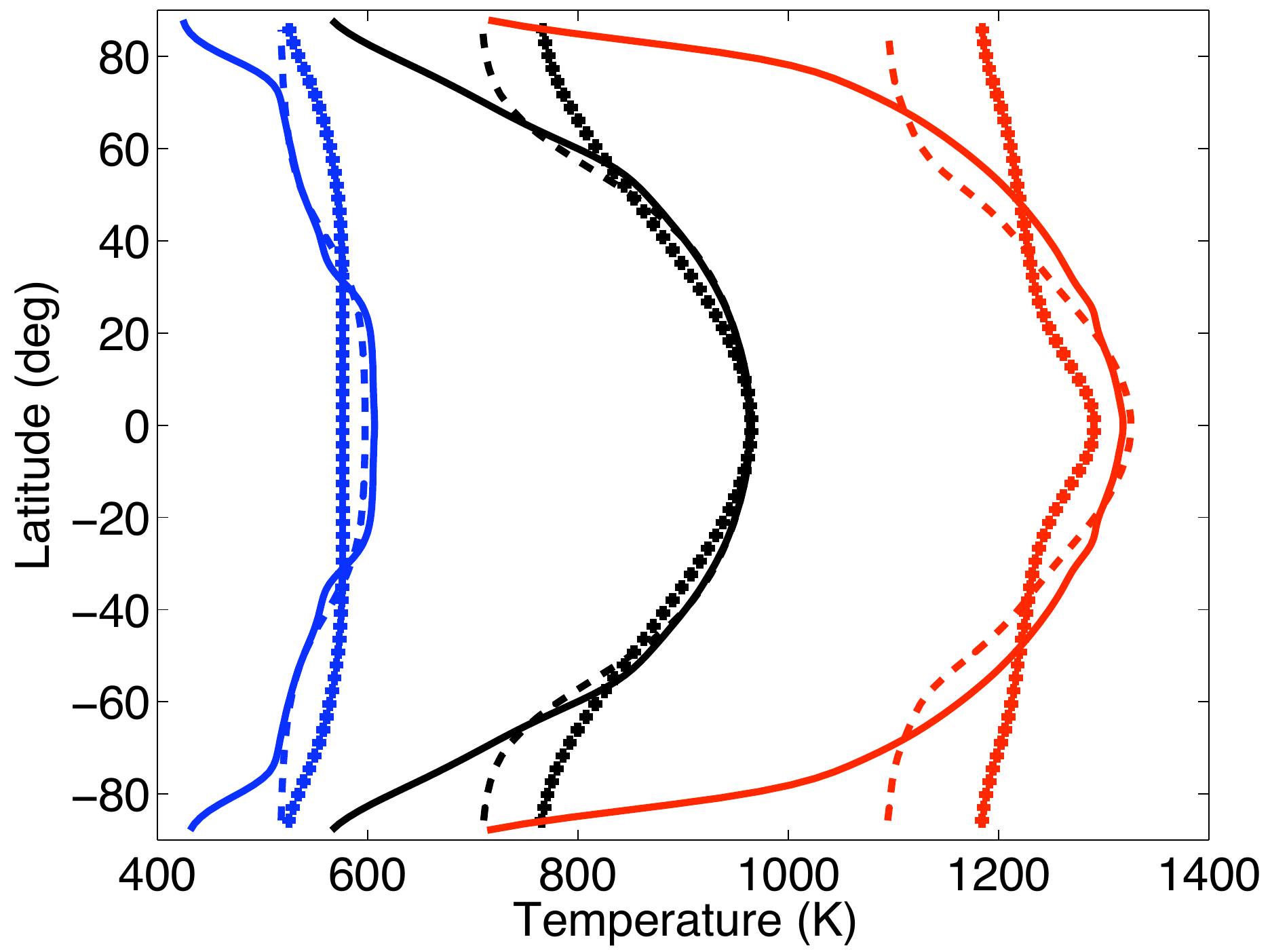}
\caption{Zonal-mean temperature versus latitude, averaged vertically
from 60~mbar to 1~bar.   Red, black, and blue curves represent models
with strong, intermediate, and weak irradiations (i.e.,
H, W, and C respectively).  Solid, dashed, and plus-sign curves
denote models with fast, intermediate, and slow rotation (\ofast, \omed,
and \oslow respectively).  Model W\ofast has been averaged about
the equator.  At a given incident stellar flux, the
equator-to-pole temperature difference increases with rotation rate;
rapidly rotating models have warmer low-latitude regions and colder
poles than slowly rotating models.}
\label{temp-vs-lat}
\end{figure}

The rotation exerts a significant control over the temperature
structure---faster rotation rates weaken the meridional (north-south)
heat flux and lead to greater equator-to-pole temperature differences
in the equilibrated state.  Figure~\ref{temp-vs-lat} shows the
zonal-mean temperature versus latitude, averaged vertically between 60
mbar and 1 bar, for all nine nominal models.
At a given incident stellar flux, rapidly rotating models exhibit
colder poles and warmer mid-to-low-latitude regions than slowly
rotating models.  While all models exhibit net radiative heating at
low latitudes and cooling at the poles, the temperatures shown in
Figure~\ref{temp-vs-lat} imply that the magnitude of this heating
gradient is smaller in the rapidly rotating cases than the slowly
rotating cases.  Essentially, the rapid rotation acts to inhibit
meridional heat transport, forcing the fluid to equilibrate to a state
closer to the (latitudinally varying) radiative equilibrium temperature
profile. 

The tendency of rapidly rotating flows to exhibit larger horizontal
temperature differences (Figure~\ref{temp-vs-lat}) can be understood
with scaling arguments.  The characteristic horizontal
pressure-gradient force has magnitude\footnote{ Hydrostatic balance in
  pressure coordinates with the ideal-gas equation of state is
  $\partial\Phi/\partial\ln p=RT$, which can be expressed to
  order-of-magnitude as $\Delta\Phi_{\rm vert}/\Delta \ln p \sim RT$,
  where $\Phi$ is the gravitational potential on isobars and
  $\Delta\Phi_{\rm vert}$ is the vertical difference in gravitational
  potential that occurs over a vertical range of log-pressures
  $\Delta\ln p$.  Imagine evaluating this expression (across a
  specified range of pressure) at two distinct locations of differing
  temperature separated by horizontal distance $L$.  If the two
  locations have the same gravitational potential at the bottom isobar
  of the layer, then differencing those two expressions implies that
  $\delta\Phi_{\rm horiz}/\Delta\ln p \sim R\delta T_{\rm horiz}$,
  where $\delta\Phi_{\rm horiz}$ is the difference in gravitational
  potential between the two locations at the {\it top} isobar of the
  layer.  The pressure gradient force, to order-of-magnitude, is then
  $\delta\Phi_{\rm horiz}/L \sim R\delta T_{\rm horiz} \Delta\ln p/L$.
  {\tt Note that this is essentially equivalent to taking the
    horizontal gradient of the so-called hypsometric equation; see,
    e.g., \citet[][pp.~69-72]{wallace-hobbs-2006}.}  }
$|\nabla\Phi|\sim R\delta T_{\rm horiz} \Delta\ln p/L$, where $R$ is
the specific gas constant, $\delta T_{\rm horiz}$ is the
characteristic horizontal temperature difference, and $\Delta\ln p$ is
the range of log-pressures over which this horizontal temperature
difference extends vertically (e.g., $\Delta\ln p=1$ if the
temperature differences extend over a layer one scale height thick).
{\tt Here, we consider the case where friction is weak, appropriate
to most atmospheres (away from any solid surface) and to our
simulations.}
In the slowly rotating regime, the pressure-gradient force is balanced
primarily by advection, expressable to order-of-magnitude as $U^2/L$,
where $U$ is a typical wind speed.  On the other hand, in the rapidly
rotating regime, the pressure-gradient force is balanced primarily by
Coriolis forces, $\Omega U$.  Writing the force balance for each of
these cases, it follows that \citep{charney-1963, showman-etal-2013b}
\begin{equation}
{\delta T_{\rm horiz}\over T} \approx 
\begin{cases}
Fr &Ro\gtrsim1,\\
{Fr\over Ro} &Ro\ll1
\end{cases}
\end{equation}
where $Fr\equiv U^2/gH\Delta\ln p$ is a dimensionless number called a
Froude number, which is the {\tt squared} ratio of wind speed {\tt to
  a quantity related to the} gravity wave speed\footnote{\tt The
  horizontal phase speed of long-vertical-wavelength gravity waves is
  approximately $NH$, where $N$ is the Brunt-Vaisala frequency.
  Approximating the atmosphere as vertically isothermal implies that
  $N=g/\sqrt{c_p T}$, from which it follows that
  $NH=\sqrt{gH(R/c_p)}$, i.e., the gravity wave speed squared is $gH$
  times a dimensionless factor not too different from unity.}; here,
$H=RT/g$ is the atmospheric scale height and $g$ is gravity.  The key
point is that $\delta T_{\rm horiz}/T$ is significantly greater---by a
factor $Ro^{-1}$---in the rapidly rotating regime than in the slowly
rotating regime.  Inserting numbers appropriate to our most strongly
irradiated simulations ($U\approx 1000\rm\,m\,s^{-1}$,
$g=23\rm\,m\,s^{-2}$, $H\approx 200\rm\,km$, $\Delta\ln p=2$) yields
$\delta T_h \sim 100\rm \,K$ when rotation is slow but $\delta T_h
\sim 600\rm\,K$ when rotation is fast.  These values agree {\tt
reasonably well} with our simulation results (compare plus-sign and solid
red curves in Figure~\ref{temp-vs-lat}).

The structure of isentropes---surfaces of constant entropy---differs
considerably across our ensemble and provides important clues about
the dynamical stability of the atmosphere.  To show this,
Figure~\ref{zonal-winds} plots in contours the zonal-mean potential
temperature for our simulations.\footnote{Potential temperature is
  defined as $\theta=T(p_0/p)^{R/c_p}$, where $T$ is temperature and
  $p_0$ is a reference pressure.  It is a measure of entropy.  Since
  surfaces of constant entropy are equivalent to those of constant
  $\theta$, we use the term ``isentropes'' to refer to these
  iso-surfaces.}  Because the atmosphere is stably stratified
throughout the domain, potential temperature increases upward.  In our
highly irradiated, slowly rotating models, zonal-mean isentropes are
relatively flat (lower right corner of Figure~\ref{zonal-winds}),
indicating that the zonal-mean temperature varies only modestly on
constant-pressure surfaces (consistent with Figure~\ref{temp-vs-lat}).
In our poorly irradiated and/or rapidly rotating models, however, the
isentropes slopes becomes large in local regions.  The slopes become
particularly tilted at pressures of $\sim$0.1--1 bar and latitudes of
$\sim$20--$40^{\circ}$ in our poorly irradiated models (C\oslow,
C\omed, C\ofast), but the large slopes are confined closer to the
poles in some other models (e.g., W\ofast and H\ofast).  In some
cases, individual isentropes vary in pressure by $\sim$3 scale heights
over horizontal distances of only $\sim$$20^{\circ}$ latitude.  In
stratified atmospheres, sloping isentropes indicate a source of
atmospheric potential energy that can be liberated by atmospheric
motions.  Strongly sloping isentropes---with vertical
variations exceeding a scale height over horizontal distances
comparable to a planetary radius---often indicate that the atmosphere
is dynamically unstable, in particular to baroclinic instabilities.
We return to this issue in Section~\ref{mechanisms}.

\section{Mechanisms}
\label{mechanisms}

 Here, we present further diagnostics that clarify the dynamical mechanisms
occurring in each regime.

\subsection{Slow rotation, high irradiation: equatorial superrotation}
\label{superrotate}

The slowly rotating, strongly irradiated regime of large day-night
temperature differences and fast equatorial superrotation is
exemplified by models H\omed, H\oslow, and W\oslow.  As described in
Section~\ref{theory}, \citet{showman-polvani-2011} showed that the
day-night forcing generates standing, planetary-scale waves that
transport angular momentum from mid-to-high latitudes to the equator,
driving the equatorial superrotation.

\begin{figure*}
\begin{minipage}[c]{0.5\textwidth}
\includegraphics[scale=0.49, angle=0]{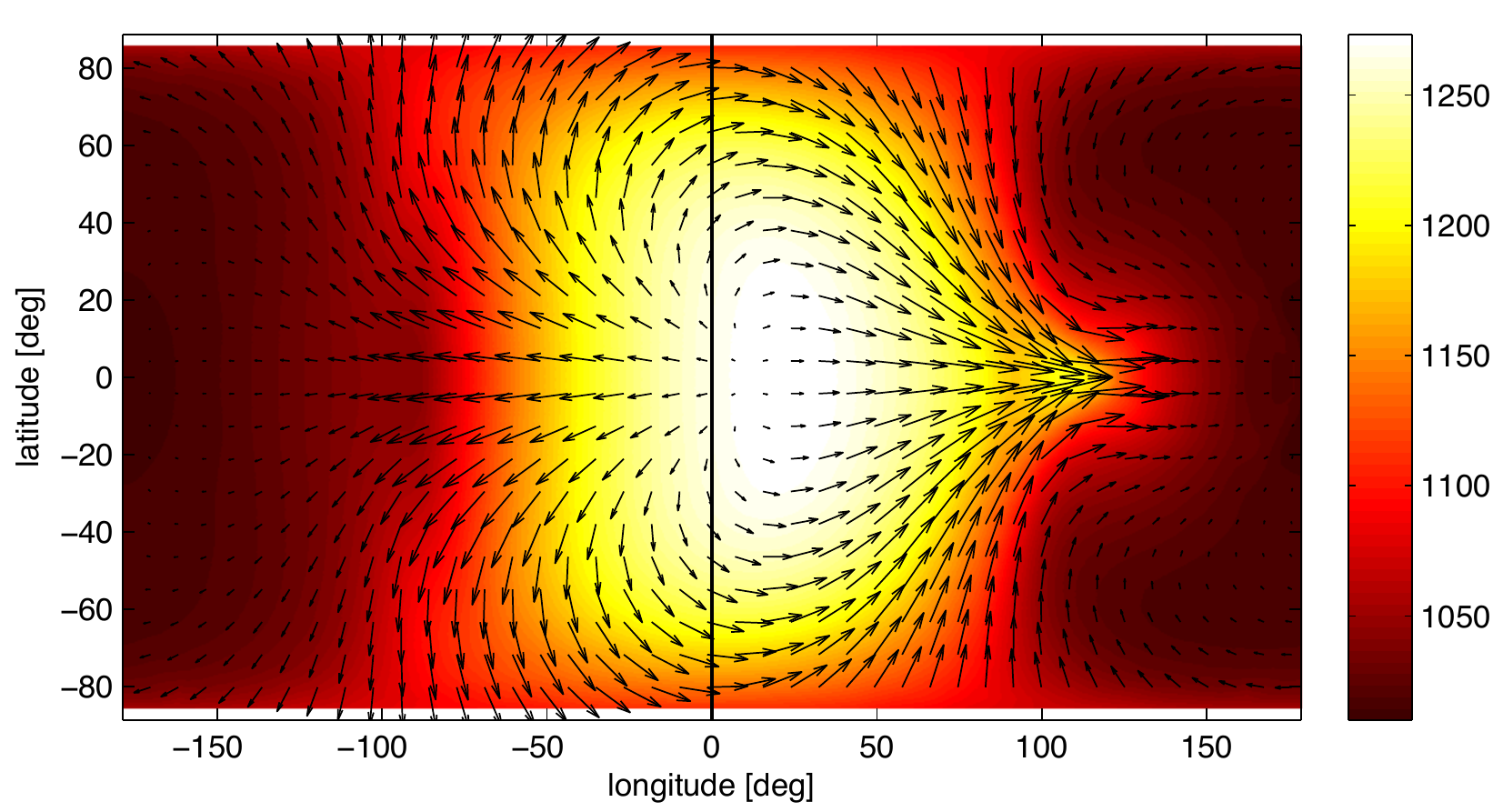}
\put(-236.,110.){\normalsize (a)}
\put(-135.,124.){\tiny H\omed}

\includegraphics[scale=0.49, angle=0]{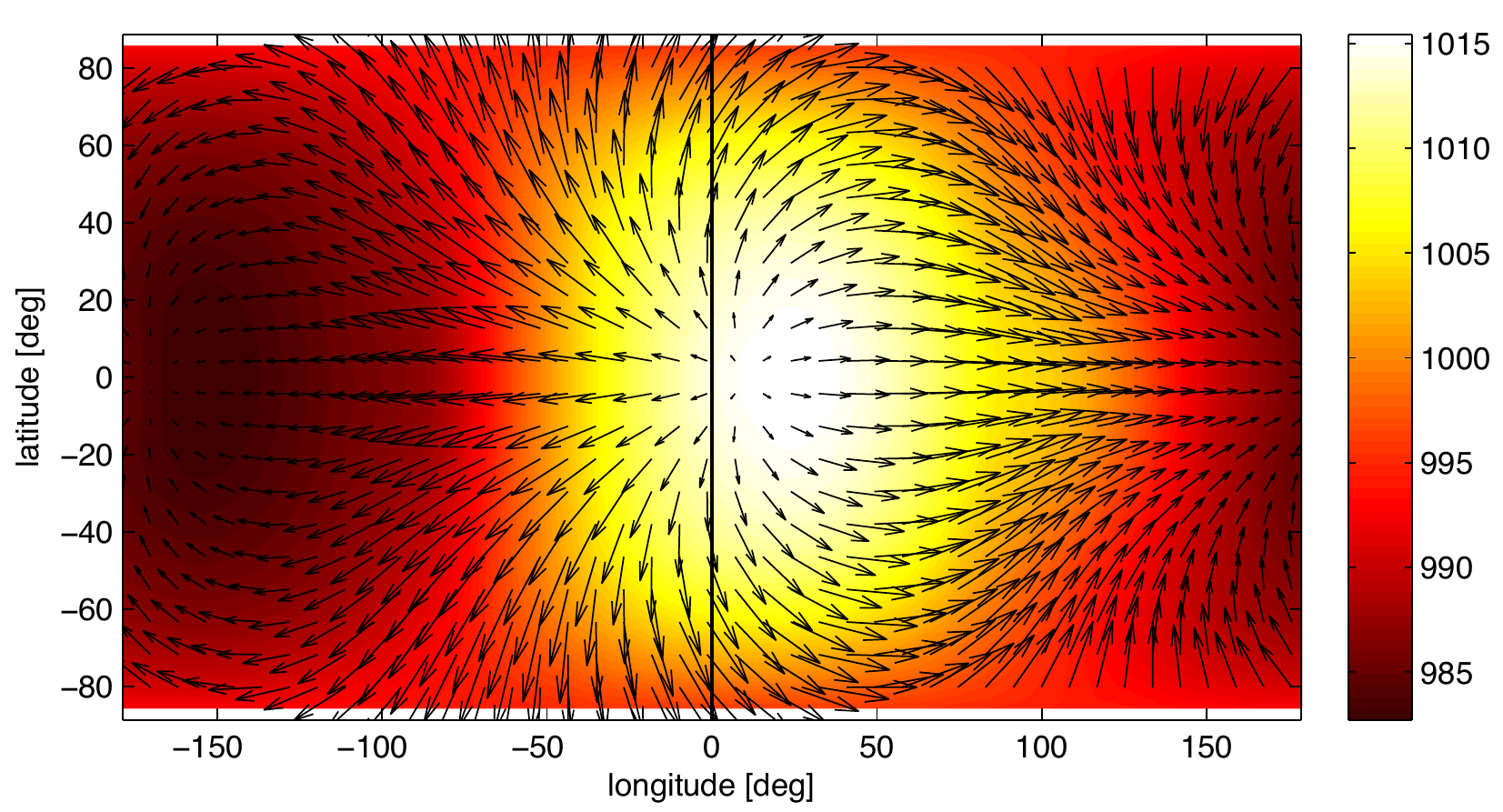}
\put(-236.,110.){\normalsize (c)}
\put(-135.,125.){\tiny W\oslow}

\includegraphics[scale=0.49, angle=0]{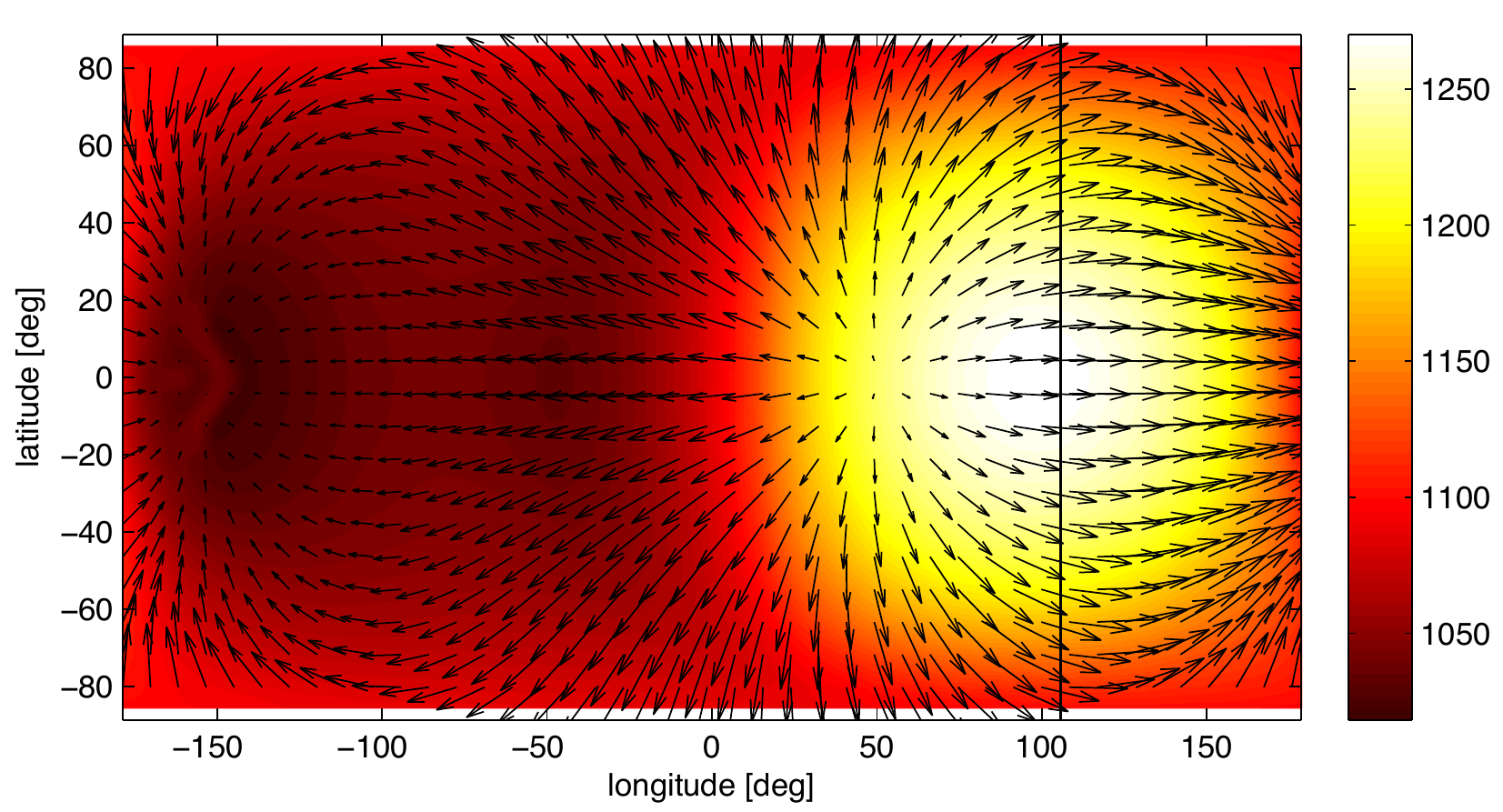}
\put(-236.,110.){\normalsize (e)}
\put(-135.,125.){\tiny H\oslow}
\end{minipage}
\begin{minipage}[c]{0.5\textwidth}
\includegraphics[scale=0.47, angle=0]{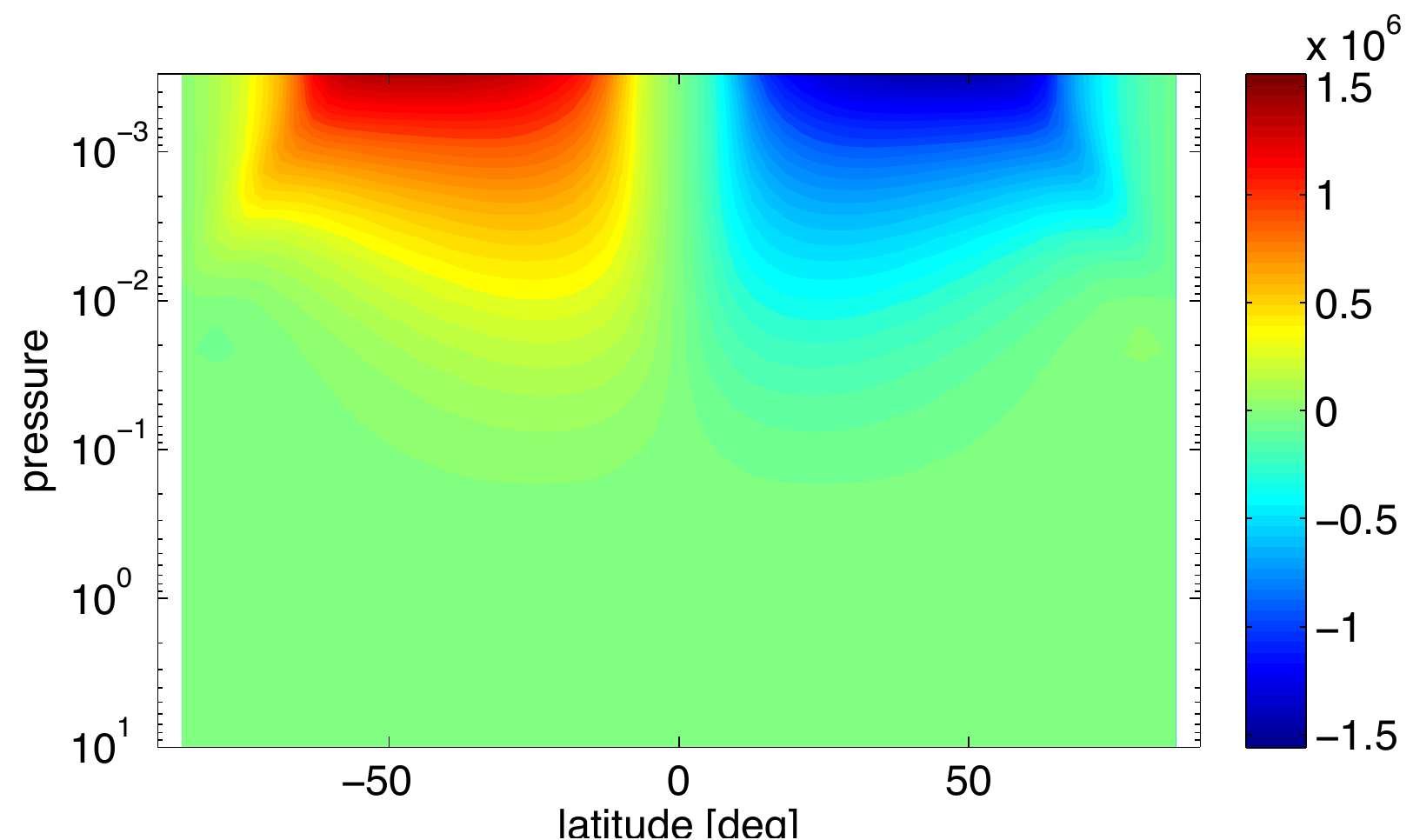}
\put(-223.,110.){\normalsize (b)}
\put(-125.,124.){\tiny H\omed}

\includegraphics[scale=0.47, angle=0]{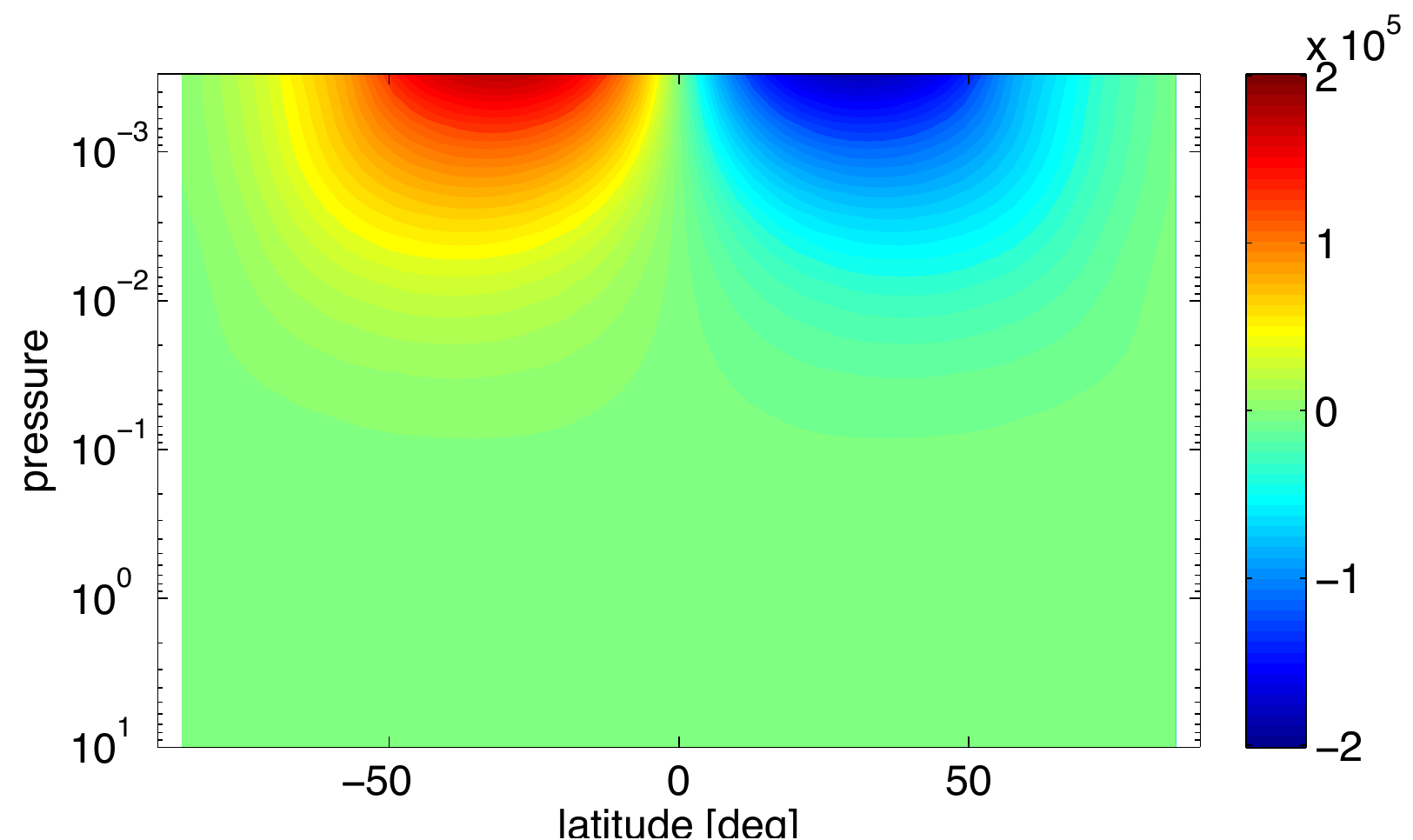}
\put(-223.,110.){\normalsize (d)}
\put(-125.,124.){\tiny W\oslow}

\includegraphics[scale=0.47, angle=0]{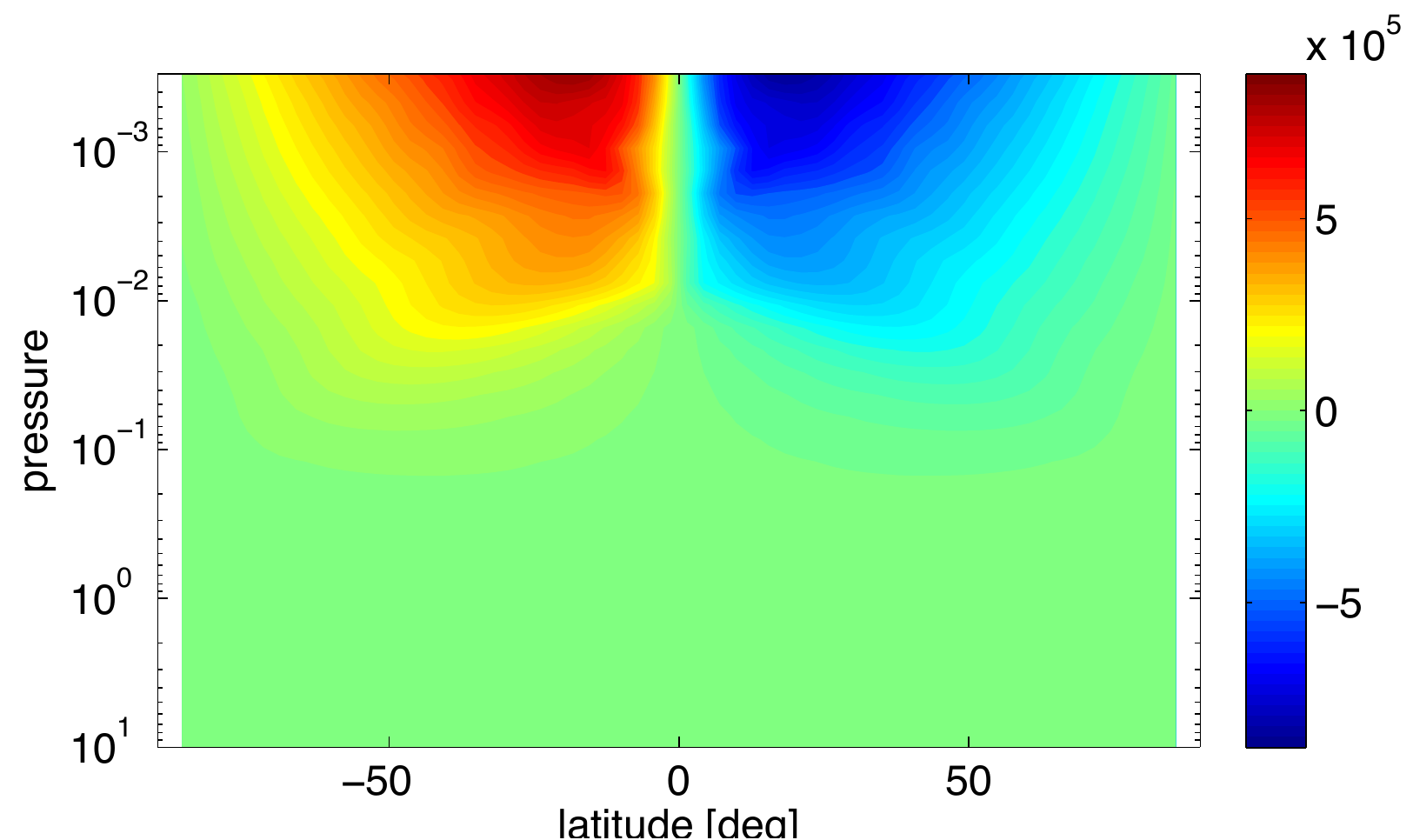}
\put(-222.,110.){\normalsize (f)}
\put(-125.,124.){\tiny H\oslow}

\end{minipage}

\caption{{\tt Structure of our three highly irradiated, slowly
    rotating models during the spin-up phase, after the day-night
    heating contrast has induced a global wave response, but before
    strong zonal jets have developed.  Models are H\omed (top),
    W\oslow (middle), and H\oslow (bottom) at times of 0.9, 1.4, and
    0.9 Earth day, respectively. Left column shows temperature
  (colorscale, K) and winds (arrows) versus longitude and latitude on
  the 85-mbar level; right column shows zonal-mean meridional momentum
  transport, $\overline{u'v'}$, versus latitude and pressure.}  All
  three models show the development of a standing-wave,
  ``Matsuno-Gill'' pattern that causes a transport of angular momentum
  to the equator and subsequently generates equatorial superrotation.
  The solid lines in (a), (c), and (e) denote the substellar
  longitude.  The top two rows are synchronously rotating models,
  whereas the bottom row exhibits asynchronous rotation.  }
\label{gill-pattern}
\end{figure*}

Figure~\ref{gill-pattern} demonstrates the existence of such a
standing wave pattern in the transient spin-up phases of our highly
irradiated, slowly rotating models.  These are snapshots shown at
early times, after the day-night forcing has had time to trigger a
global wave response but before the equatorial jet has spun up to high
speed.  In synchronously rotating models H\omed and W\oslow, strong
east-west divergence occurs along the equator from a point east of the
substellar point (Figure~\ref{gill-pattern}a and c, respectively).  At
northern latitudes, Coriolis forces lead to clockwise curvature of the
flow on the dayside and counterclockwise curvature on the nightside
(with reversed directions at southern latitudes).  Globally, these
flows exhibit a striking similarity to the analytic standing-wave
solutions of \citet[][compare our Figures~\ref{gill-pattern}a and c to
  the top middle and top left panels of their Figure
  3]{showman-polvani-2011}.  The zonal divergence (with an eastward
displacement) along the equator is precisely the behavior expected for
a steady, forced, damped equatorial Kelvin wave, whereas the
high-latitude behavior is analogous to the steady, forced, damped
equatorial Rossby wave.  Linear, analytic solutions show that the two
wave components tend to be more distinct when the radiative or drag
timescales are longer, and less distinct when they are shorter; the
behavior in our simulations is toward the latter limit.  In contrast,
H\oslow (Figure~\ref{gill-pattern}e) is an asynchronous model where
the substellar longitude migrates to the east over time.  No analytic
solutions of the asynchronous case have been published in the
hot-Jupiter literature, but the similarity of the wind patterns to the
synchronous case (particularly W\oslow) is evident.  The main
difference is that, although the thermal pattern tracks the heating
(and is thus centered near the substellar longitude), there is a
time-lag in the wind response, such that the flow divergence point
lies {\it west} (rather than east) of the substellar point.  Still,
taken as a whole, Figure~\ref{gill-pattern} provides strong evidence
that the mechanism of \citet{showman-polvani-2011} is occurring in
these simulations.

The planetary-scale waves shown in Figure~\ref{gill-pattern} lead to a
pattern of eddy velocities that transport angular momentum from the
mid-latitudes to the equator, which allows the development of
equatorial superrotation.  This can be seen visually from the
velocities in Figure~\ref{gill-pattern} {(\tt left column)}, which
exhibit a preferential northwest-southeast orientation in the northern
hemisphere and southwest-northeast orientation in the southern
hemisphere.  As a result, one expects the eddy velocity correlation
$\overline{u'v'}$ to be negative in the northern hemisphere and
positive in the southern hemisphere, where $u$ and $v$ are the zonal
and meridional winds, the primes denote deviations from the zonal
average, and the overbar denotes a zonal average.  {\tt This is
  explicitly demonstrated in Figure~\ref{gill-pattern} (right column),
  which shows $\overline{u'v'}$ versus latitude and pressure for the
  three simulations in the left column.  Note that $\overline{u'v'}$
  represents the meridional transport of zonal (relative) momentum per
  unit mass by eddies (negative implying southward momentum transport
  and positive implying northward momentum transport).  A strong
  equatorward momentum flux occurs at pressures less than 0.1~bar, with
  spatial patterns that are quite similar in 
  all three cases.   The momentum fluxes are largest in
  H\omed (Figure~\ref{gill-pattern}b) and somewhat smaller in H\oslow
  (Figure~\ref{gill-pattern}f), consistent with the fact that the
  eddy-velocity phase tilts are better correlated in the former than
  the latter.  (That is, the eddy velocity tilts are more strongly
  organized in the poleward/westward to equatorward/eastward direction
  in H\omed than in H\oslow; the weaker correlation in H\oslow may
  result from the effect of slower rotation and/or non-synchronous
  rotation on the structure of the wave modes.)  Model W\oslow
  exhibits the weakest momentum fluxes, presumably because of the
  weaker stellar forcing in that case.} 

More formally, the zonal-mean zonal momentum equation of the primitive
equations using pressure as a vertical coordinate can be written
\begin{eqnarray}
\nonumber
{\partial\overline{u}\over\partial t}=\overline{v}\left[ f - {1\over 
a\cos\phi}{\partial(\overline{u}\cos\phi)\over\partial\phi}\right] - 
\overline{\omega}{\partial\overline{u}\over\partial p} + \overline{X}\\
- {1\over a\cos^2\phi}{\partial (\overline{u'v'}\cos^2\phi) \over\partial\phi}
- {\partial(\overline{u'\omega'})\over\partial p}
\label{eulerian-zonal-momentum}
\end{eqnarray}
where $a$ is the planetary radius, $f=2\Omega \sin\phi$ is the
Coriolis parameter, $\phi$ is latitude, $\omega = dp/dt$ is the
vertical velocity in pressure coordinates (i.e., the rate of change of
pressure with time following an air parcel), with $d/dt$ being the
total (material) derivative in three dimensions, and $X$ represents
any frictional terms.  The terms on the righthand side represent
meridional momentum advection by the mean flow, vertical momentum
advection by the mean flow, frictional drag, horizontal convergence of
eddy momentum, and vertical convergence of eddy momentum.

\begin{figure*}
\includegraphics[scale=0.65, angle=0]{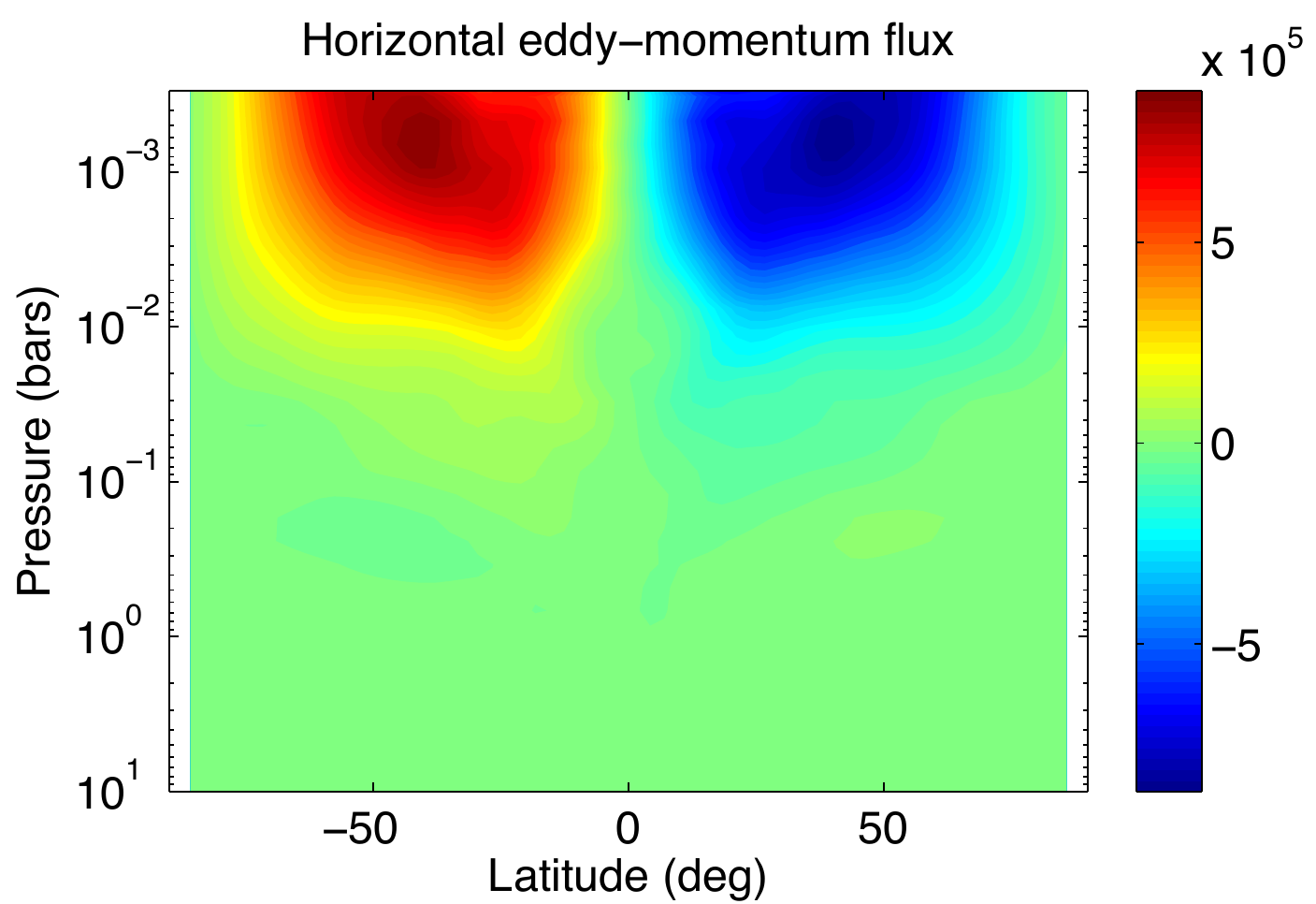}
\put(-252.,175.){\normalsize (a)}
\includegraphics[scale=0.65, angle=0]{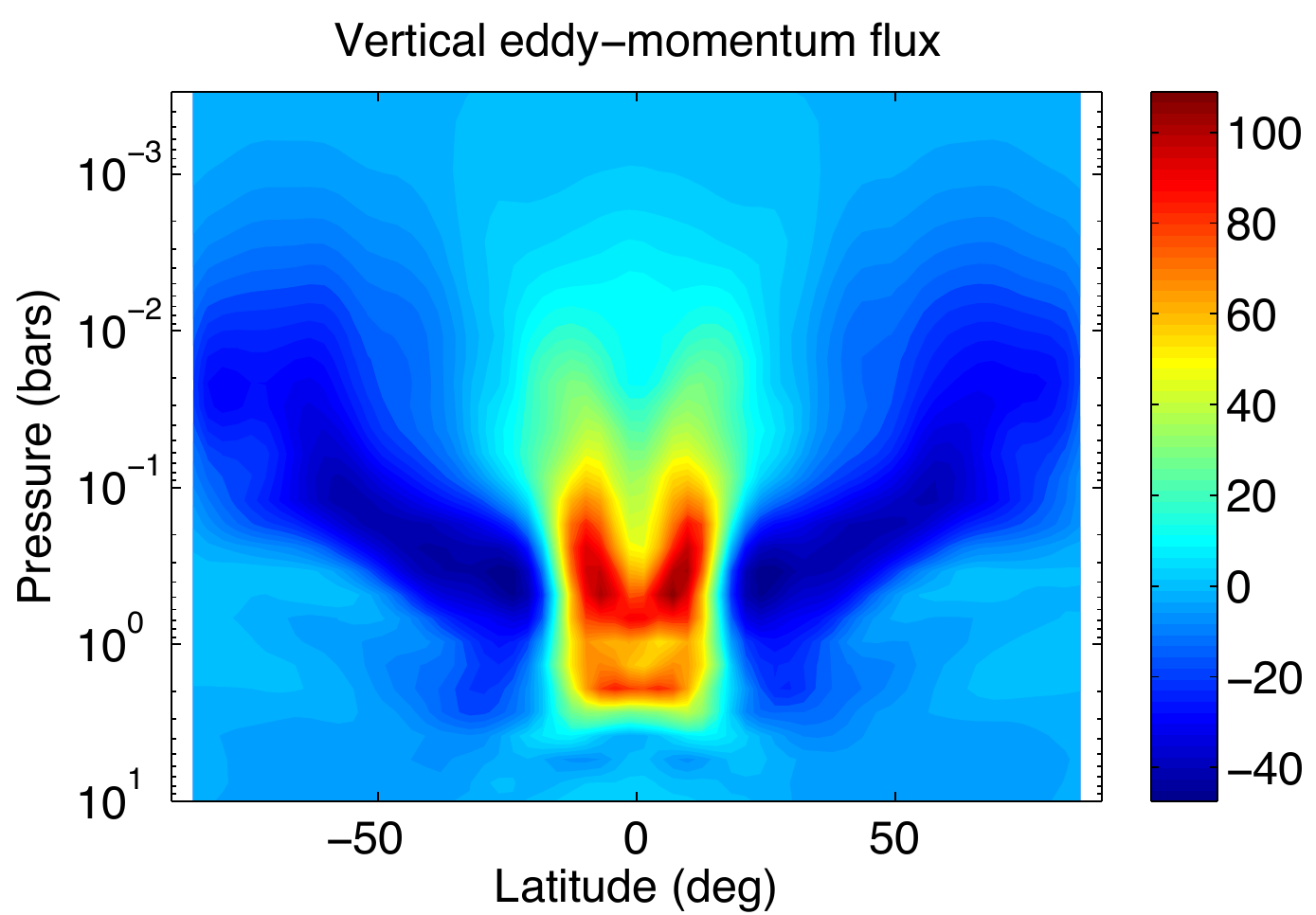}
\put(-250.,175.){\normalsize (b)}\\
\includegraphics[scale=0.65, angle=0]{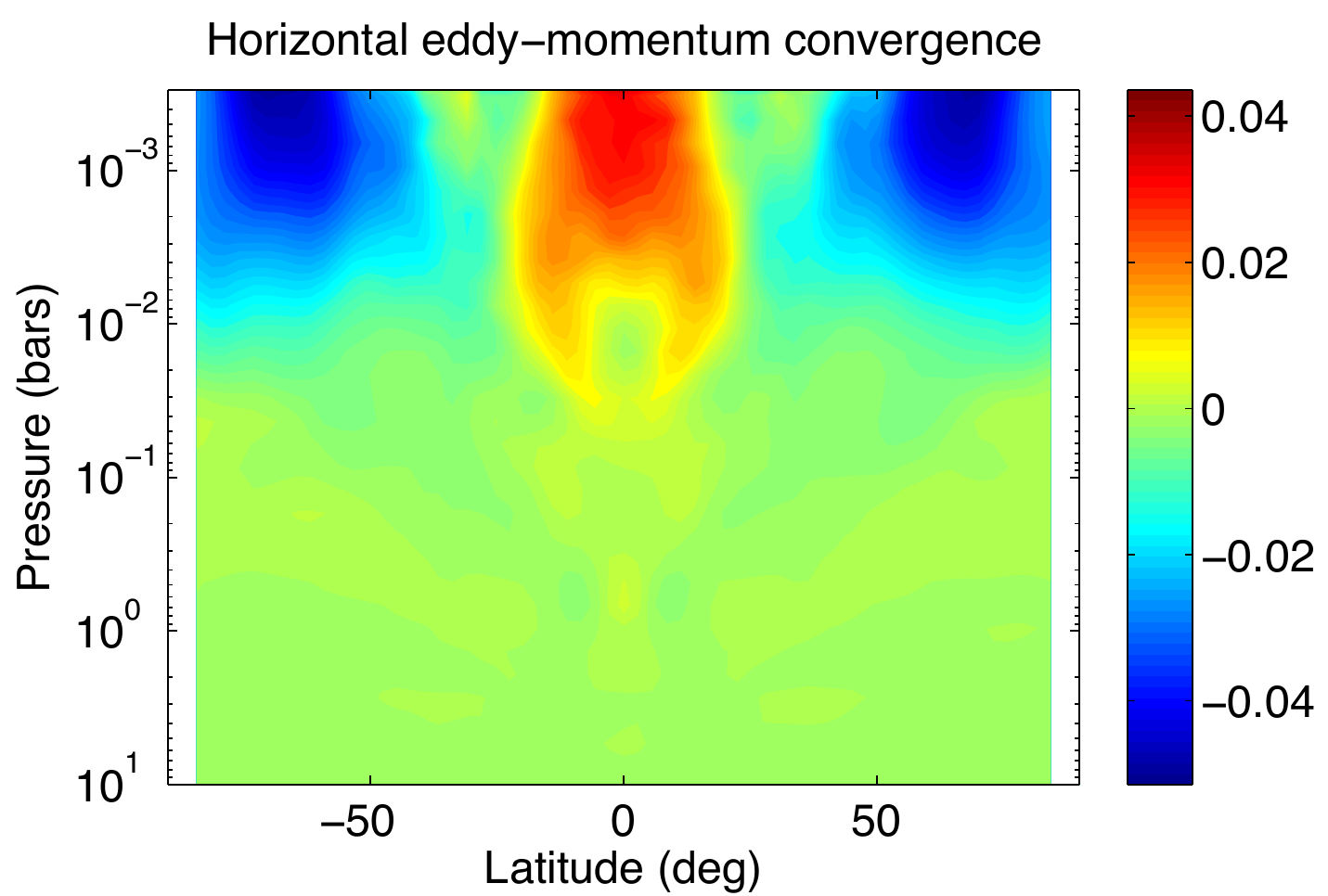}
\put(-255.,175.){\normalsize (c)}
\includegraphics[scale=0.65, angle=0]{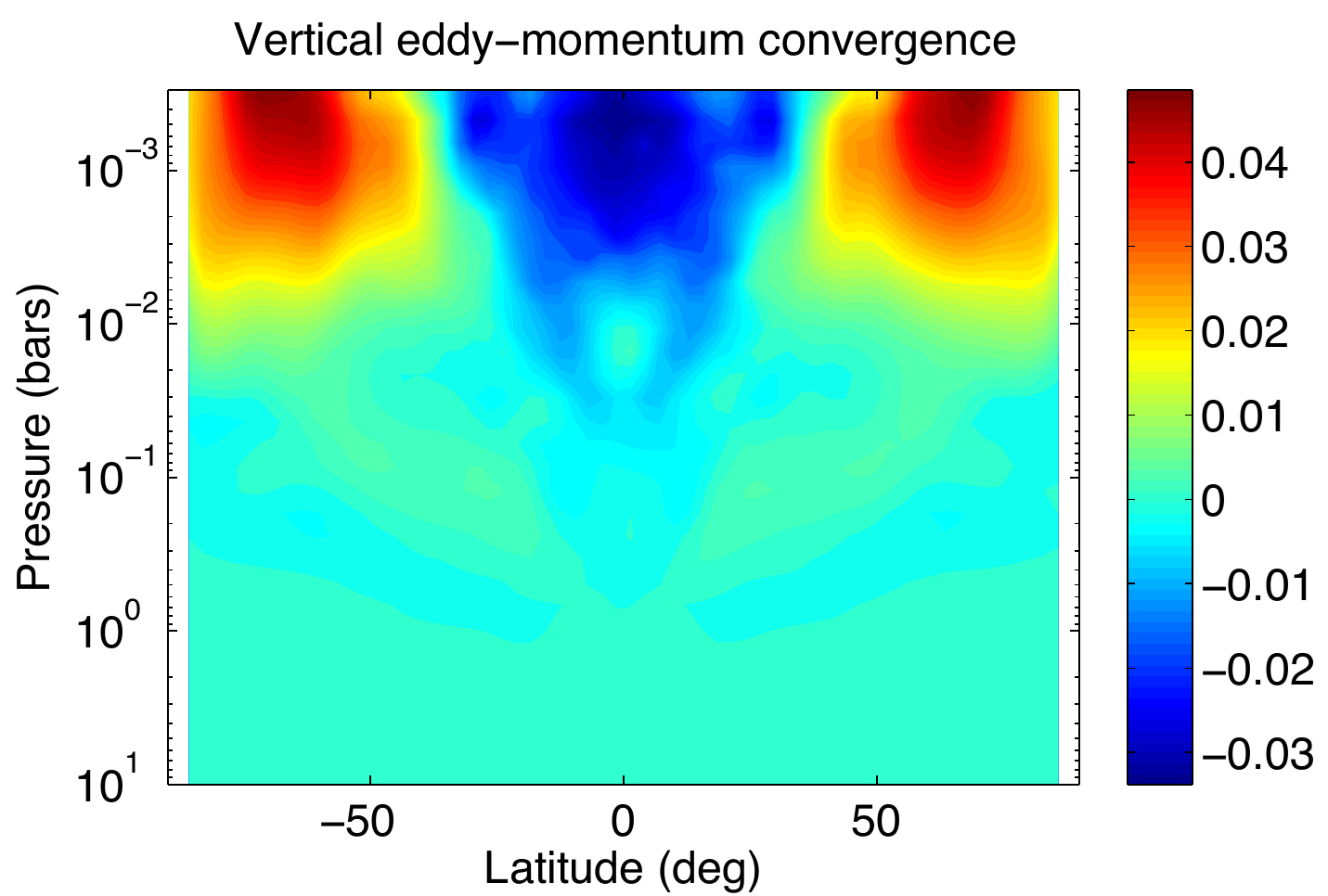}
\put(-255.,175.){\normalsize (d)}
\caption{Zonal eddy-momentum fluxes and convergences for our slowly
  rotating, highly irradiated model, H\oslow, averaged in time once
  the simulation has reached equilibrium at pressures less than a few
  bars.  Panels (a) and (b) show the horizontal and vertical eddy
  fluxes, $\overline{u'v'}$ and $\overline{u'\omega'}$ (units of $\rm
  m^2\,s^{-2}$ and $\rm m\,Pa\,s^{-2}$ respectively). Panels (c) and
  (d) show the eddy-momentum convergences from
  Equation~(\ref{eulerian-zonal-momentum}),
  $-(a\cos\phi)^{-1}\partial(\overline{u'v'}\cos^2\phi)/\partial\phi$
  and $-\partial(\overline{u'\omega'})/\partial p$ (units of $\rm
  m\,s^{-2}$).  {\tt Large-scale} waves transport eddy momentum to
  equator and downward.  Plots show the upper part of the domain from
  0.3 mbar to 10 bars.  Note that positive $\omega$ indicates {\it
    downward} motion, such that, in panel (b), a downward momentum
  flux is red, while an upward momentum flux is blue.}
\label{mom-fluxes-superrotate}
\end{figure*}

To illustrate how the thermally forced, planetary-scale waves maintain
the equatorial jet, Figure~\ref{mom-fluxes-superrotate} shows the
resulting eddy fluxes and accelerations for our slowly rotating,
highly irradiated model (H\oslow).  These are long-time averages after
the simulation has reached equilibrium at pressures less than a few
bars. Panels (a) and (b) show the horizontal and vertical eddy fluxes,
$\overline{u'v'}$ and $\overline{u'\omega'}$, whereas panels (c) and
(d) show the eddy-momentum convergences from
Equation~(\ref{eulerian-zonal-momentum}), namely
$-(a\cos\phi)^{-1}\partial(\overline{u'v'}\cos^2\phi)/\partial\phi$
and $-\partial(\overline{u'\omega'})/\partial p$.  Planetary-scale
waves transport eddy momentum to the equator
(Figure~\ref{mom-fluxes-superrotate}a), especially at pressure less
than $\sim$1 bar.  At pressures less than $\sim$0.1~bar, this leads to
a torque that causes an eastward acceleration at the equator and
westward acceleration at high latitudes
(Figure~\ref{mom-fluxes-superrotate}c), as expected from the arguments
surrounding Figure~\ref{gill-pattern} and from the theory of
\citet{showman-polvani-2011}. {\tt Indeed, the equilibrated,
  time-averaged meridional momentum fluxes shown in
  Figure~\ref{mom-fluxes-superrotate}a strongly resemble the structure
  and magnitudes of those at early times shown in
  Figure~\ref{gill-pattern}f.}  Considering now the vertical fluxes,
Figure~\ref{mom-fluxes-superrotate}b shows that, at the equator, eddy
momentum is transported {\it downward} from the upper regions
($\sim$1~mbar--1~bar) to pressures exceeding a few bars.  This
downward momentum transport induces a {\it westward} eddy acceleration
at the equator, shown in panel (d), which largely cancels the eastward
acceleration resulting from meridional momentum convergence.  The
residual eddy acceleration (sum of panels (c) and (d)) is very weakly
eastward at the equator at most pressures, thereby maintaining the
equatorial superrotation.  Meanwhile, at high latitudes, the vertical
eddy flux is upward (panel b), leading to an eastward eddy
acceleration that largely cancels the westward eddy acceleration
caused by the meridional convergence. {\tt In steady state, the net
  eddy acceleration---that is, the sum of the lower two panels in
  Figure~\ref{mom-fluxes-superrotate}---is balanced by a combination of Coriolis,
  horizontal mean advection, and vertical mean advection terms.  At
  the equator, the Coriolis and horizontal mean advection terms are
  negligible, and the weak net eddy acceleration is balanced primarily
  by vertical advection (i.e., $-\overline{\omega}
  \partial\overline{u}/\partial p$).}

The rotation rate and stellar heating pattern both exert significant
control over the equatorial jet width.  A comparison of models H\omed
and W\oslow---which are both synchronously rotating---allows a
comparison of the effects of rotation alone.  Interestingly, W\oslow,
which has a rotation rate four times slower than H\omed, exhibits an
equatorial jet twice as wide, extending nearly from pole to pole
(Figure~\ref{zonal-winds}).
This agrees with the theory of \citet{showman-polvani-2011},
which predicts for synchronously rotating planets that the meridional
half-width of the equatorial jet is comparable to the equatorial
Rossby deformation radius, $(NH/\beta)^{1/2}$, where $N$ is the
Brunt-Vaisala frequency, $H$ is scale height, and $\beta = df/dy$ is
the gradient of the Coriolis parameter with northward distance $y$.
In contrast, a comparison of models W\oslow$\,$ and H\oslow$\,$  allows a
comparison of stellar-heating pattern at constant rotation rate.
Although the stellar insolation is fixed in longitude in W\oslow, it
migrates eastward over time in H\oslow $\,$ due to the asynchronous
rotation.  Despite the identical rotation rates, the equatorial
jet is much narrower in H\oslow than W\oslow.  This suggests that
the asynchronous thermal forcing alters the nature of the wave
modes that are generated, leading to differing equatorial jet
widths and speeds.

\subsection{Fast rotation, weak irradiation: off-equatorial eastward jets}
\label{offeq}

In comparison to the closest-in hot Jupiters, EGPs that are rapidly
rotating and relatively far from their stars will exhibit relatively
weak diurnal (day-night) forcing, and will exhibit a circulation that
develops primarily in response to the need to transport heat from low
to high latitudes.  Here we describe the dynamics of this regime
in further detail, focusing for concreteness on model C\ofast.

\begin{figure*}
\includegraphics[scale=0.55, angle=0]{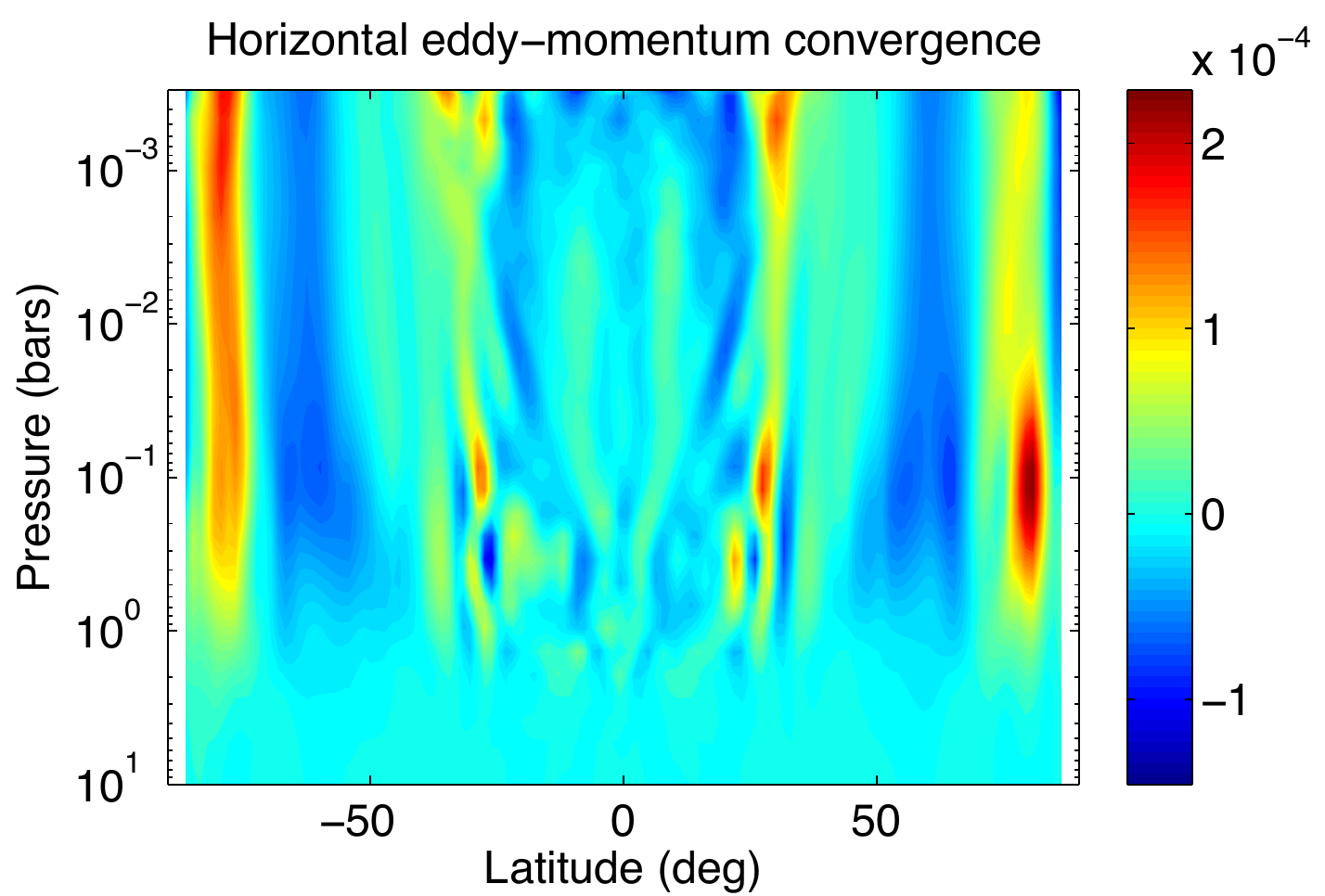}
\put(-215.,140.){\normalsize (a)}
\includegraphics[scale=0.55, angle=0]{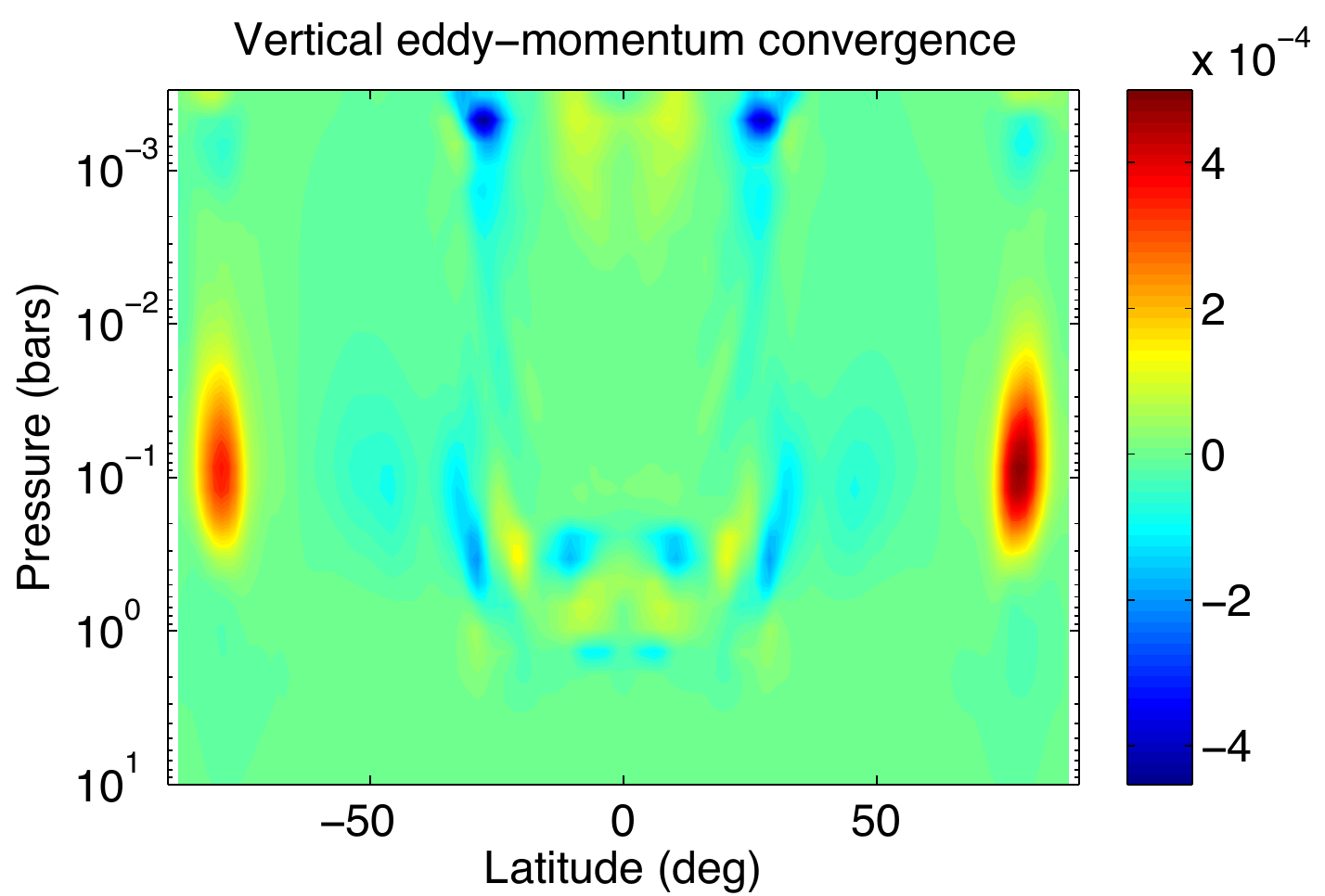}
\put(-215.,140.){\normalsize (b)}\\
\includegraphics[scale=0.55, angle=0]{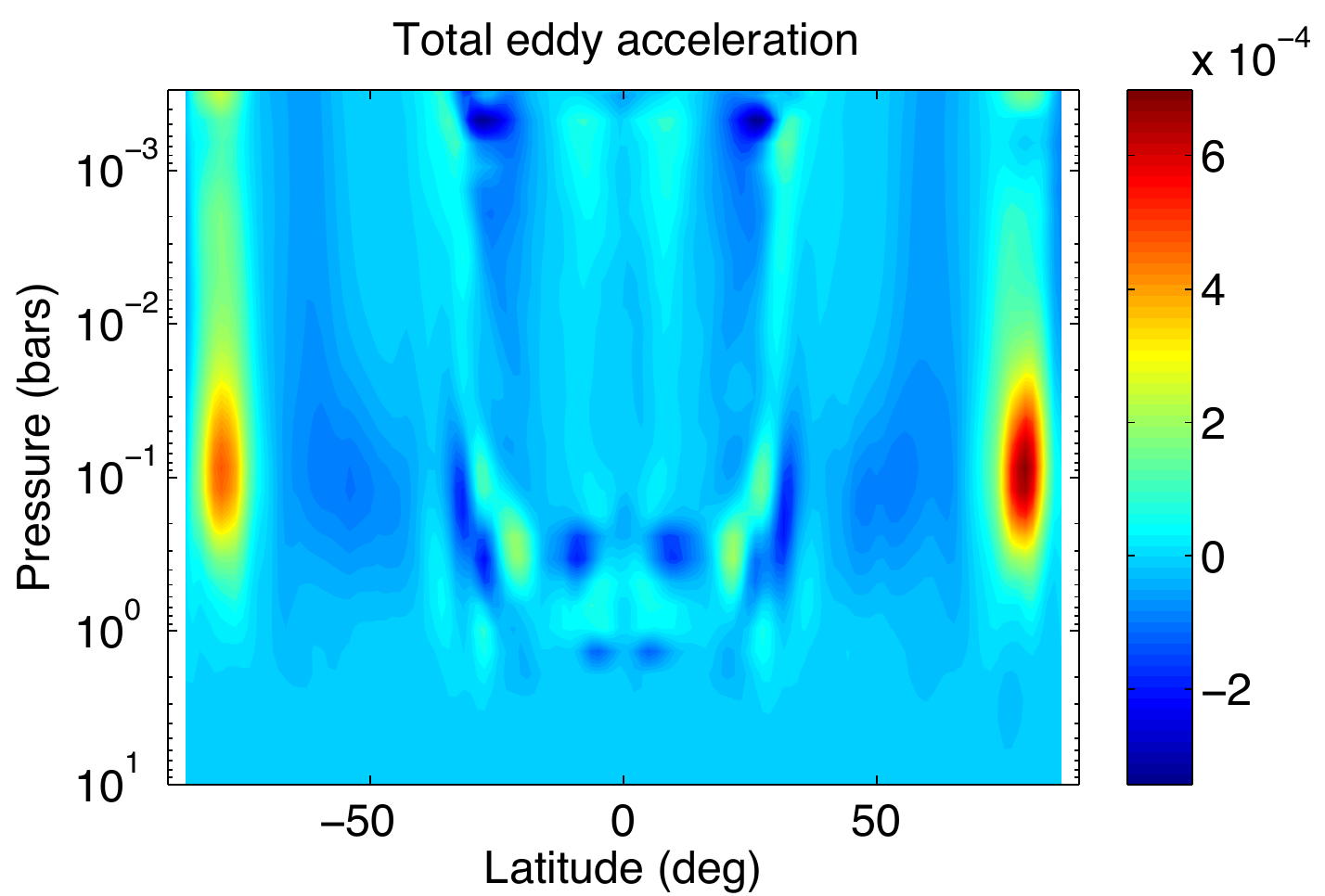}
\put(-215.,140.){\normalsize (c)}
\includegraphics[scale=0.55, angle=0]{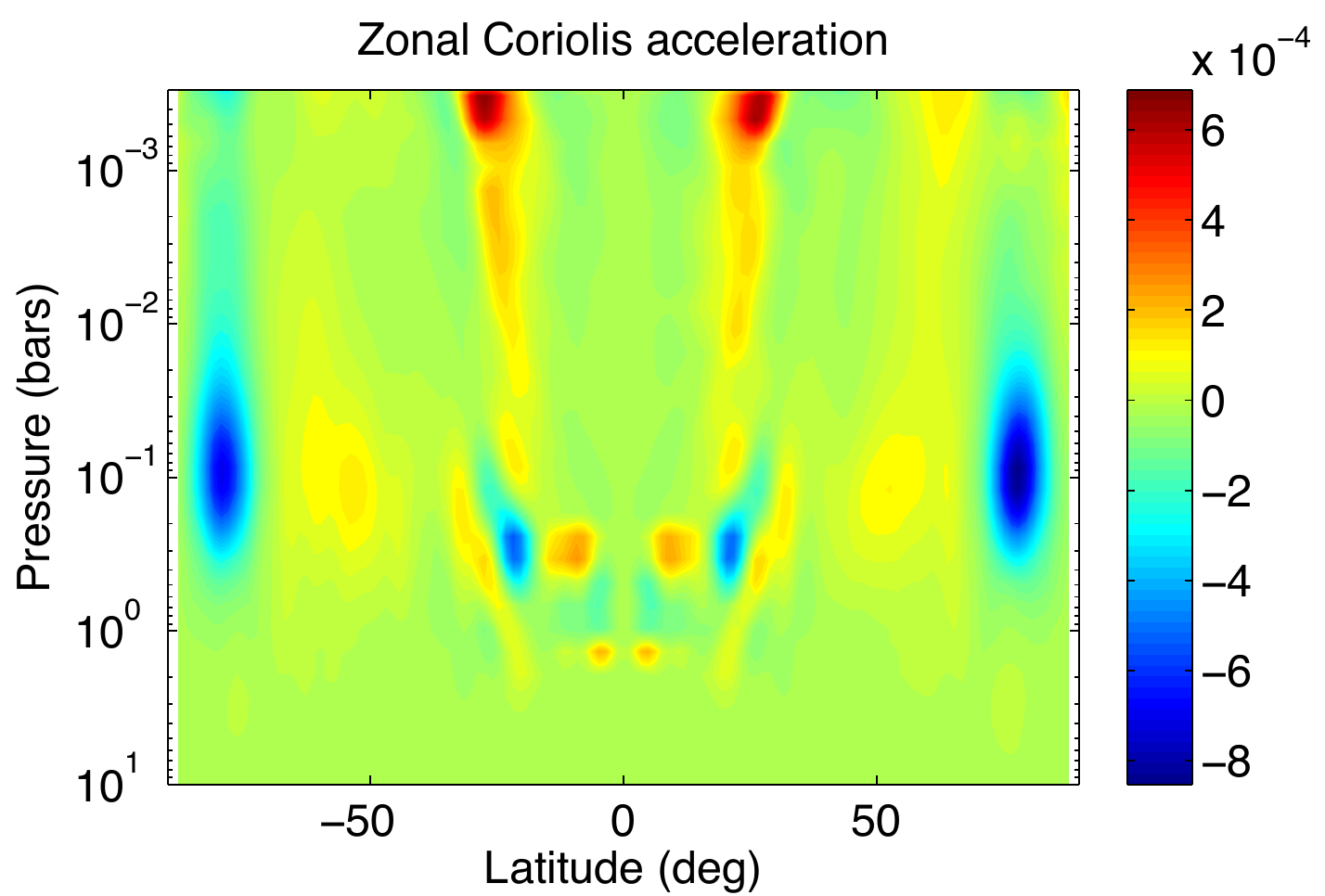}
\put(-215.,140.){\normalsize (d)}\\
\includegraphics[scale=0.55, angle=0]{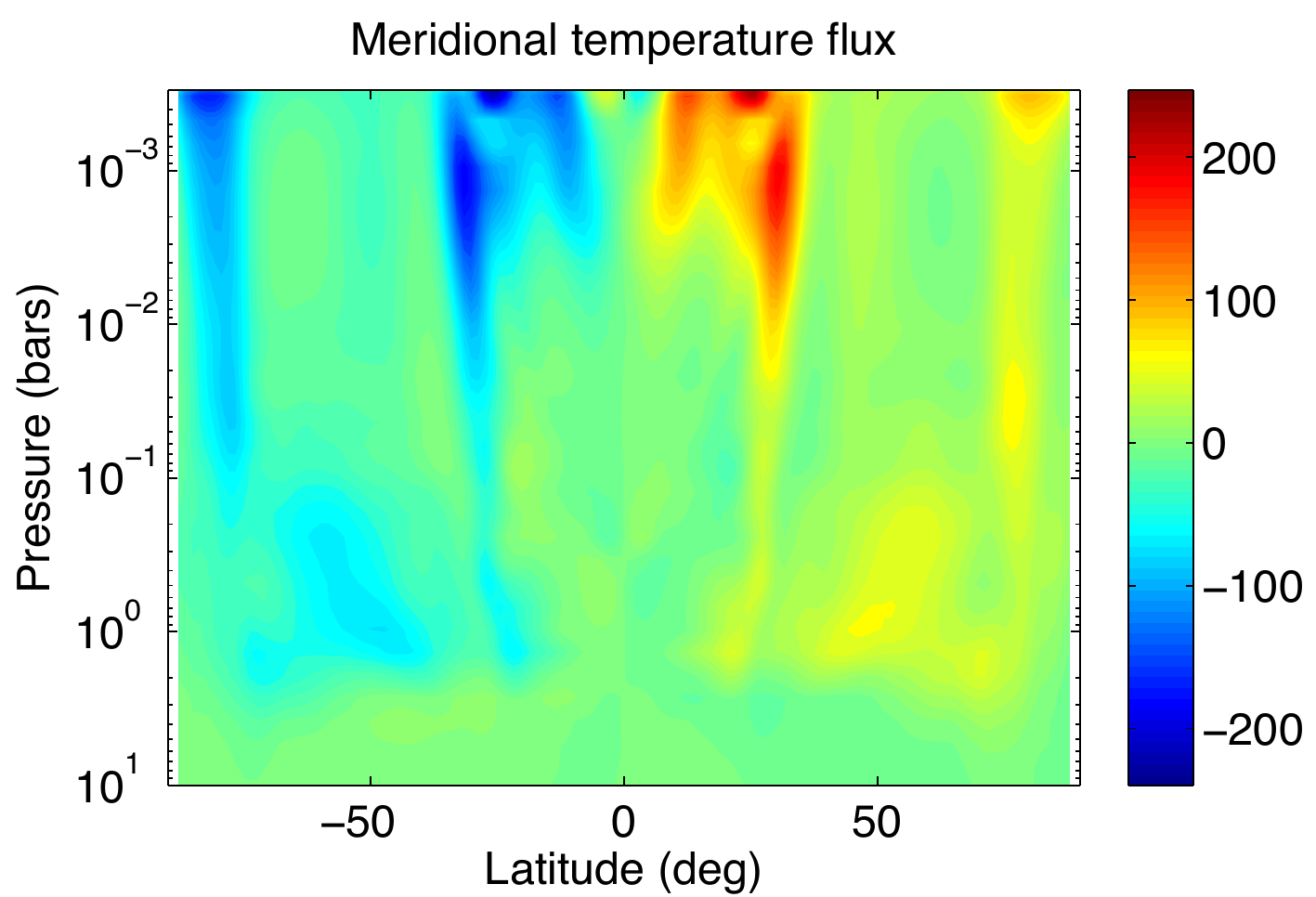}
\put(-215.,140.){\normalsize (e)}
\includegraphics[scale=0.55, angle=0]{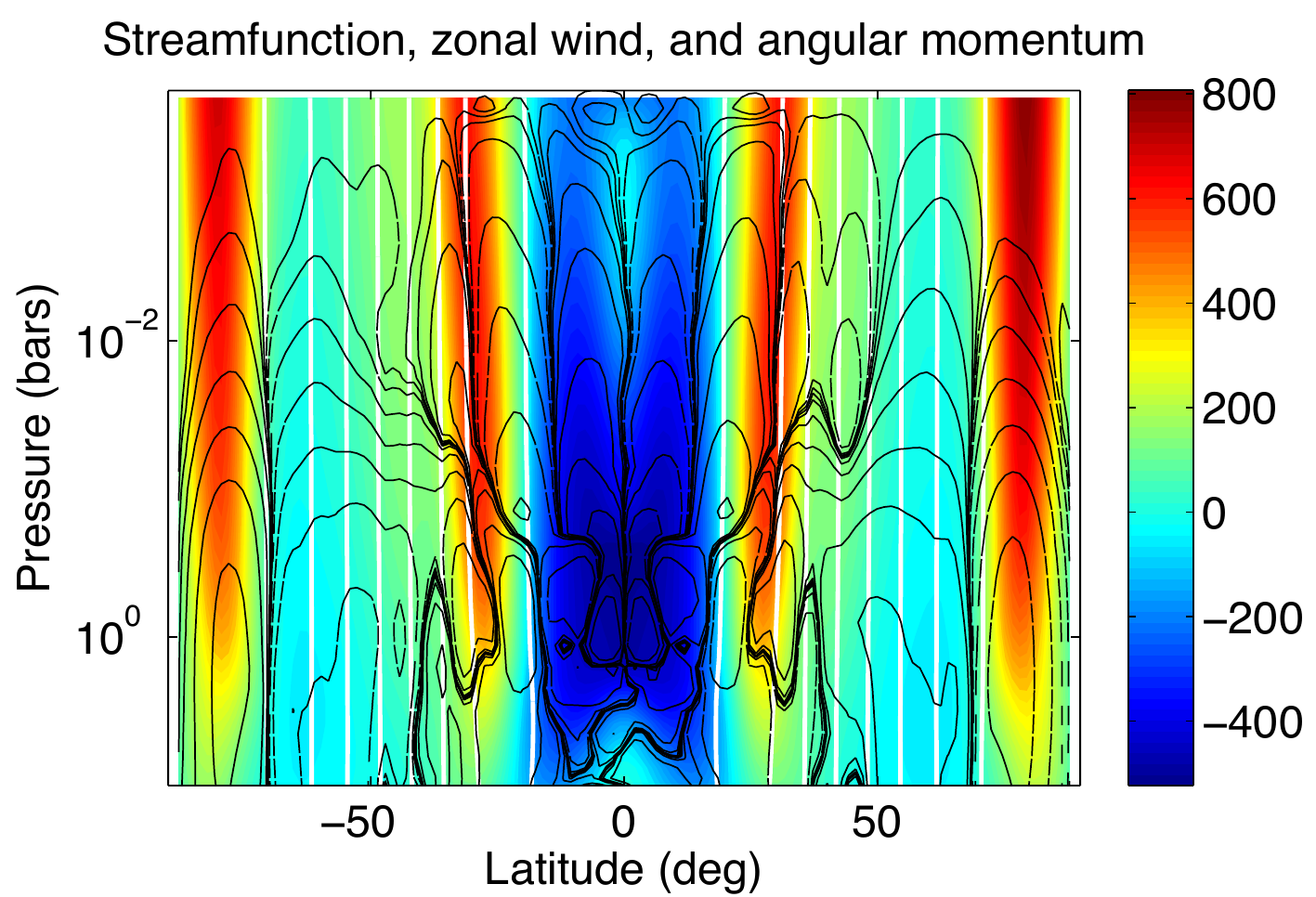}
\put(-220.,140.){\normalsize (f)}\\
\caption{\small Zonal-mean circulation {\tt diagnostics} for model
  C\ofast, our coolest, most rapidly rotating model, averaged in time
  once the simulation has reached equilibrium at pressures less than a
  few bars.  Panels (a) and (b) show the horizontal and vertical
  eddy-momentum convergence from
  Equation~(\ref{eulerian-zonal-momentum}),
  $-(a\cos\phi)^{-1}\partial(\overline{u'v'}\cos^2\phi)/\partial\phi$
  and $-\partial(\overline{u'\omega'})/\partial p$.  Panel (c) shows
  their sum. (d) shows the Coriolis acceleration, $f\overline{v}$.
  The top four panels all have units $\rm m\,s^{-2}$.  (e) shows the
  meridional eddy temperature flux, $\overline{v'T'}$ (units $\rm
  m\,K\,s^{-1}$), illustrating that large meridional eddy heat
  transport occurs across the eastward jets.  (f) shows the zonal-mean
  streamfunction (black contours) overlain on the zonal-mean zonal
  wind (colorscale, $\rm m\,s^{-1}$).  Solid is clockwise (with ten
  log-spaced contours between $1.3\times10^8$ and $1.3\times10^{13}\rm
  kg\,s^{-1}$) and dashed is counterclockwise (with ten log-spaced
  contours between $-1.3\times10^8$ and $-1.3\times10^{13}\rm
  kg\,s^{-1}$).  White contours depict angular momentum per unit mass
  with respect to the planet's rotation axis (with ten linearly spaced
  contours between $3.36\times10^{10}$ and $8.99\times10^{10}\rm
  m^2\,s^{-1}$).  }
\label{midlat2D}
\end{figure*}

In the rapidly rotating regime, the meridional heat transport is
accomplished by baroclinic eddies that develop in the mid-to-high
latitudes, which---as predicted by the theory in
Section~\ref{theory}---transport momentum into their latitude of
generation, thereby producing and maintaining the off-equatorial zonal
jets that dominate this regime.  Figure~\ref{midlat2D} illustrates
this behavior for model C\ofast.  Panels (a) and (b) show the
time-mean, zonal-mean horizontal and vertical eddy momentum
convergences, that is, the terms $-(a\cos^2\phi)^{-1}
\partial(\overline{u'v'}\cos^2\phi)/\partial\phi$ and
$-\partial(\overline{u'\omega'})/\partial p$, respectively from
Equation~(\ref{eulerian-zonal-momentum}).  Figure~\ref{midlat2D}c
shows their sum.  There is a strong correlation of the
accelerations---particularly the horizontal eddy-momentum
convergence---with the jet latitudes (compare Figure~\ref{midlat2D}a and
b to the zonal winds in
Figure~\ref{zonal-winds} or \ref{midlat2D}f).  Specifically, the
eddies produce strongest eastward accelerations primarily within the
eastward jets, thereby maintaining them.

A mean-meridional circulation develops in response to the thermal and
eddy forcing.  This circulation organizes into a series of cells that
extend coherently from the top of the domain ($\sim$0.2 mbar) to
pressures exceeding 1 bar.  The streamfunction $\psi$, defined by
$\overline{v}=g (2\pi a \cos\phi)^{-1} \partial \overline{\psi}/
\partial p$ and $\overline{\omega} = -g (2\pi a^2\cos\phi)^{-1}
\partial{\overline \psi}/\partial \phi$, where $a$ is planetary
radius, is contoured in Figure~\ref{midlat2D}f; solid and dashed
contours denote clockwise and counterclockwise motion, respectively.
The cells are meridionally narrow, with approximately five cells per
hemisphere.  In each hemisphere, adjacent cells alternate in sign,
representing thermally direct and indirect circulations analogous to
the terrestrial Hadley and Ferrel cells, respectively.

\begin{figure*}
\includegraphics[scale=1.05, angle=0]{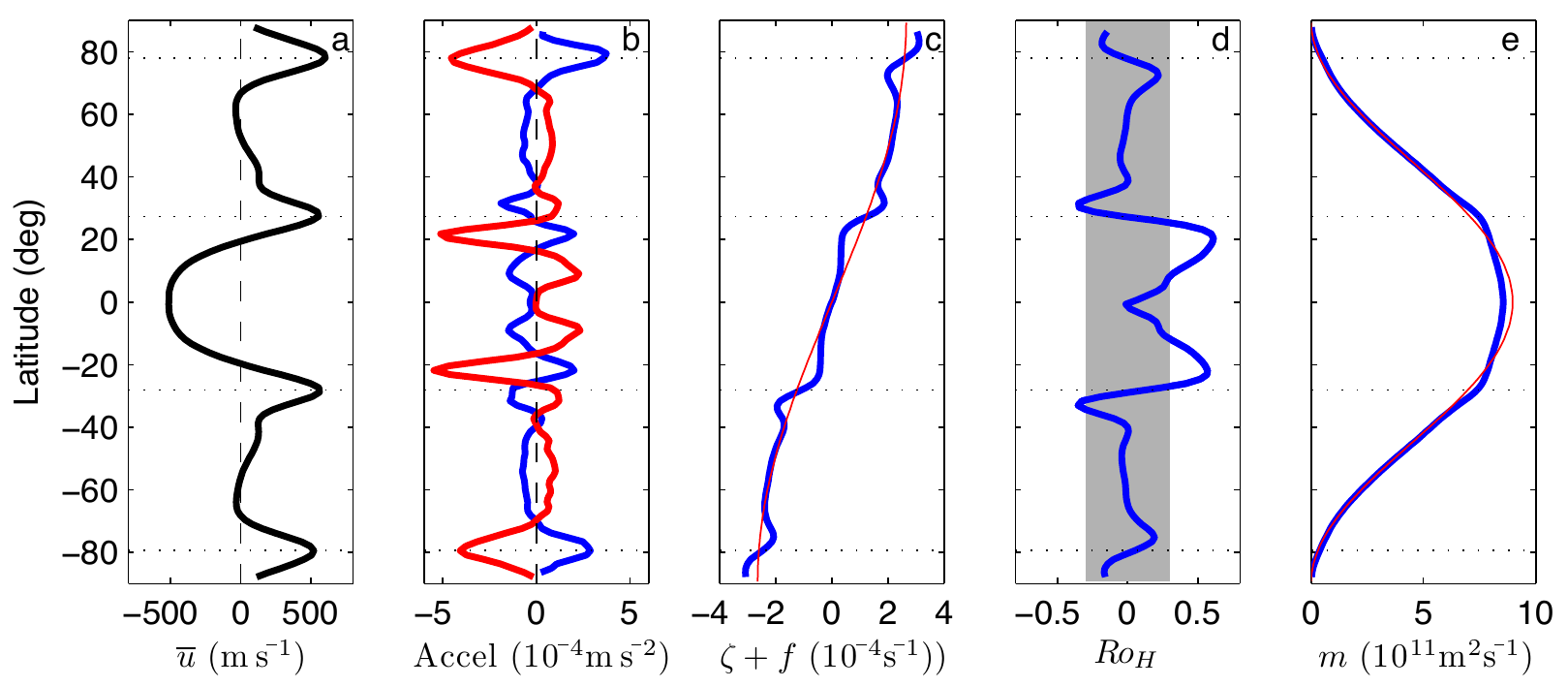}
\caption{Zonal-mean, time-mean state of model C\ofast at 240 mbar.  (a)
Zonal-mean zonal wind. (b) Total zonal eddy acceleration (sum of last two
terms in Equation~\ref{eulerian-zonal-momentum}) (blue) and 
Coriolis acceleration $f\overline{v}$ (red).  The strong anticorrelation
between these quantities indicates a good balance between them, especially
poleward of $30^\circ$ latitude.  (c) Zonal-mean absolute vorticity, 
$\overline{\zeta}+f$ (blue) and the Coriolis parameter $f$ itself (red).
(d) The Rossby number $Ro_H = -\overline{\zeta}/f$.  Grey strip corresponds
to Rossby numbers with magnitudes less than 0.3, indicating that advection
is small relative to the Coriolis acceleration. (e) Zonal-mean angular momentum
per unit mass with respect to the planetary rotation axis, $m$ (blue)
and the angular momentum of solid-body rotation, $m_{\rm solid}=\Omega a^2
\cos^2\phi$ (red).  }
\label{midlat1D}
\end{figure*}

Our models show that, in steady state, the zonal eddy acceleration is
approximately balanced (away from the equator) by the zonal Coriolis
acceleration associated with this mean-meridional circulation.  This
is demonstrated in Figure~\ref{midlat2D}d, which shows the time-mean
$f\overline{v}$ for model C\ofast.  It can be seen that, in most
regions, the Coriolis acceleration exhibits a strong anticorrelation
with the total eddy acceleration shown in Figure~\ref{midlat2D}c,
indicating an approximate balance between the two.  This balance is
also evident in Figure~\ref{midlat1D}b, which shows the total zonal
eddy acceleration (blue) and Coriolis acceleration (red) versus
latitude at 200 mbar in model C\ofast.

To understand this behavior, we can, in a statistical steady state,
rewrite the zonal momentum balance (\ref{eulerian-zonal-momentum}) as
\begin{equation}
(f+\overline{\zeta})\overline{v}=f(1-Ro_H)\overline{v}=
\overline{\omega}{\partial\overline{u}
    \over\partial p} + S
\label{mombal}
\end{equation}
where $S$ represents the sum of the two zonal eddy accelerations (the
last two terms in Equation~\ref{eulerian-zonal-momentum}), $\zeta =
{\bf k}\cdot\nabla\times{\bf v}$ is the relative vorticity, ${\bf k}$
is the vertical (upward) unit vector, and ${\bf v}$ is the horizontal
velocity.  In the middle expression, we have defined a Rossby number
for the mean-meridional flow, $Ro_H = -\overline{\zeta}/f$
\citep[cf][]{held-2000, walker-schneider-2006,
  schneider-bordoni-2008}; note that, to order of magnitude, the
relative vorticity scales as $\zeta \sim U/L$, where $U$ is a
characteristic zonal wind speed and $L$ is a horizontal length scale;
thus, $\zeta/f$ essentially equals $Ro\sim U/fL$.  Thus, in the
rapidly rotating regime, the lefthand side of Equation~(\ref{mombal})
simplifies to $f\overline{v}$.  Likewise, the mean vertical advection
term $\overline{\omega}\, \partial{\overline{u}}/\partial p$ can be
expressed to order-of-magnitude as
$\overline{\omega}\,\overline{u}/\Delta p$, where $\Delta p$ is the
pressure thickness of the circulation; the zonal-mean continuity
equation implies that $\overline{\omega}/\Delta p \sim \overline{v}/L$, and
thus this term is also $Ro$ smaller than $f\overline{v}$.  Thus, when
the Rossby number is small, Equation~(\ref{mombal}) leads to the
balance\footnote{This balance is well known in the extratropical Earth
  regime; see, e.g., \citet[][p.~319]{holton-2004},
  \citet{hartmann-2007}, and \citet{karoly-etal-1998}.  Observations
  and models of Jupiter and Saturn also suggest this force balance in
  and above the cloud deck \citep{delgenio-etal-2007,
    delgenio-barbara-2012, showman-2007, lian-showman-2008}.}
\begin{equation}
f\overline{v} \approx S.
\label{geostrophicbal}
\end{equation}
This balance
implies that the strength of the mean-meridional circulation in the
extratropics is controlled by the eddy-momentum convergence; large
eddy momentum convergence leads to a strong meridional circulation and
vice versa.  Figure~\ref{midlat1D}b shows that, poleward of $\sim$$30^{\circ}$
latitude, the balance (\ref{geostrophicbal}) is quite good.  Over this
latitude range, the deviations of the zonal-mean absolute vorticity
$\overline{\zeta}+f$ from $f$ itself are generally small
(Figure~\ref{midlat1D}c); in agreement, the corresponding
extratropical Rossby number $Ro_H$ is almost everywhere less than
$\sim$0.2 (Figure~\ref{midlat1D}d).

The dynamics of this rapidly rotating regime can be
understood by considering the angular momentum per unit mass about the
planetary rotation axis, $m=(\Omega a\cos\phi + u)a\cos\phi$.  Under
small Rossby number, the second term is small compared to the first,
which means that dynamics provides only minor perturbations to the
planetary (solid-body) contribution.  In this case, contours of
constant angular momentum are parallel to the rotation axis
(essentially vertical in the context of a thin, hydrostatic
atmosphere).  In this regime, any mean-meridional circulation must
cross angular-momentum contours.  This can only happen in the presence
of eddy-momentum convergences that alter the zonal-mean angular
momentum of the air as it moves meridionally.  Figure~\ref{midlat2D}f
shows contours of $m$ in white, and it can be seen that poleward of
$30^{\circ}$, the meridional circulation indeed strongly crosses the
angular-momentum contours, and that the $m$-contours are almost
vertical, as expected.

At low latitudes, in contrast, the dynamics deviate significantly from
the rapidly rotating regime described above.  Equatorward of
$30^\circ$, {\tt the angular momentum becomes more slowly varying}
with latitude (Figure~\ref{midlat2D}f and \ref{midlat1D}e).  This
behavior indicates that eddy-momentum convergences do not strongly
alter the angular momentum of air as it flows meridionally {\tt
(at least in comparison to the situation at higher latitudes)}, so that
the meridional circulation is closer to the limit where the angular
momentum is conserved following the flow \citep[cf][]{held-hou-1980}.
Consistent with this picture, the Rossby number $Ro_H$ reaches values
as high as 0.6 near $20^\circ$ latitude (Figure~\ref{midlat1D}c),
indicating that---unlike the situation poleward of
$30^\circ$---meridional momentum advection plays an important role in
the zonal momentum budget.  This low-latitude regime is analogous to
that of Earth's Hadley circulation, particular the summer cell (e.g.,
\citealt{held-hou-1980}, \citealt{walker-schneider-2006},
\citealt{schneider-bordoni-2008}; see review in
\citealt{showman-etal-2013b}) and suggests that thermal driving may be
an equal or more important factor in controlling the amplitude of this
cell than the amplitude of the eddy-momentum convergences.  Indeed, it
can be seen that, like Earth's Hadley circulation, the cell extending
from $\sim$15--$30^\circ$ in each hemisphere is thermally direct, with
ascent on the equatorward flank and descent on the poleward flank
(Figure~\ref{midlat2D}f).

The relative location of the zonal jets, eddy-momentum convergences,
and meridional circulation cells suggests the following picture for
the circulation.  At high latitudes, the zonal winds are eddy driven.
The eastward jet at $\sim$$80^\circ$ latitude occurs at precisely the
same latitude as the prograde eddy-momentum convergences, which extend
coherently---as does the jet---from the top of the model to pressures
of $\sim$1 bar (compare Figure~\ref{midlat1D}a and b, or
Figure~\ref{midlat2D}a and f).  Equation~(\ref{geostrophicbal}) then
implies that the jet is co-located with a Ferrel cell, in which air
flows equatoward across the jet at pressures $\lesssim 1\rm\,bar$
(Figure~\ref{midlat2D}f).  Likewise, the local minimum in zonal wind
from $\sim$50--$70^\circ$ latitude corresponds well to a broad region
of westward eddy acceleration and poleward mean-meridional
circulation.  This is precisely the picture that has been suggested
for the off-equatorial zonal jets on Jupiter and Saturn near cloud
level \citep{showman-2007, delgenio-etal-2007, delgenio-barbara-2012,
  lian-showman-2008, schneider-liu-2009}, and would plausibly result
from the generation of Rossby waves near the core of the eastward jet
and their equatorward propagation into the zonal-flow minimum, where
they would break and/or become absorbed at their critical levels
\citep[e.g.][]{dritschel-mcintyre-2008}.

On the other hand, the eastward jet at $\sim$$30^\circ$ latitude is
more complex.  Over much of its vertical extent (from $\sim$0.2--100
mbar), the eddy acceleration is eastward on the jet's poleward flank
but westward on its equatorward flank.  Conversely, the Coriolis
acceleration (associated with the mean meridional circulation) is
westward on the jet's poleward flank and eastward on its equatorward
flank.  In the streamfunction (Figure~\ref{midlat2D}f), it is clear
that, at $p\lesssim 0.1\,$bar, the jet's poleward flank corresponds to
a Ferrel cell whereas its equatorward flank corresponds to the Hadley
cell, with a transition latitude close to the jet axis.  This suggests
that the jet is a hybrid, corresponding to an eddy-driven jet on its
poleward flank and a subtropical jet---i.e., a jet at the poleward
edge of a Hadley circulation that is driven by the Coriolis
acceleration in the poleward flowing air---on its equatorward flank.
Interestingly, the midlatitude local maximum of zonal-mean zonal wind
in Earth's troposphere corresponds to just such a hybrid.

The transition from Hadley to Ferrel cell that occurs near the axis of
the $30^\circ$-latitude jet suggests that the latitude of this
jet---and therefore the width of the the constant-angular-momentum
region---is controlled by the location where the jet first becomes
baroclinically unstable.  Just such a criterion has been proposed as a
controlling factor in the width of the Hadley circulation on Earth
\citep[e.g.][]{held-2000, lu-etal-2007, frierson-etal-2007}.  In the
context of a simple two-layer model of a background flow that
conserves angular momentum in its upper branch, the lowest latitude of
instability is \citep{held-2000}
\begin{equation}
\phi_H \approx \left({g D\Delta\theta_v\over \Omega^2 a^2 \theta_0}\right)^{1/4}
\label{hadley-width}
\end{equation}
where $D$ is the layer thickness and $\Delta\theta_v/\theta_0$ is the
fractional variation in potential temperature occurring vertically
across the circulation.  Inserting the radius and gravity of HD
189733b, $\Omega_{\rm fast}=1.3\times10^{-4}\rm\,s^{-1}$, $D\approx
300\rm\,km$ appropriate to the multi-scale-height deep circulation in
C\ofast, and $\Delta\theta_v/\theta_0\approx 1$ yields $\phi_H \approx
30^\circ$.  The agreement with the jet latitude in the simulation is
encouraging, though we caution that the two-layer model is crude and
its applicability to a continuously stratified, compressible
atmosphere extending over many scale heights is perhaps open to
question.  Nevertheless, the hypothesis is worth testing further in
future work.

Our most rapidly rotating, weakly illuminated model, C\ofast,
illuminates the dynamical continuum from hot EGPs to Jupiter and
Saturn themselves.  Numerous one-layer turbulence studies have
suggested that, in rapidly rotating, turbulent, geostrophic flows, the
existence of zonal jets is controlled by Rhines scaling, with a
meridional jet spacing $L\sim \pi (U/\beta)^{1/2}$ and approximately
$N_{\rm jet}\sim a(\beta/U)^{1/2}$ zonal jets from pole to pole, where $U$ is
the jet speed \citep[for a review, see][] {vasavada-showman-2005}.
Inserting $U=600\rm\,m\,s^{-1}$ and
$\beta=1.6\times10^{-12}\rm\,m^{-1}\,s^{-1}$ (appropriate to the
off-equatorial jets in C\ofast) yields $L\approx 60,000\rm\,km$ and
$N_{\rm jet}\approx 4$, in agreement with the existence of four eastward jets in
this simulation.  Inserting $U\approx 30\rm\,s^{-1}$ appropriate for
Jupiter's midlatitudes yields $N_{\rm jet}\approx 16$, similar to the
number of eastward jets on Jupiter.  This and other similarities suggests
that the dynamical regime of C\ofast resembles Jupiter in important ways.
Presumably, models like C\ofast but with even weaker stellar irradiation
would exhibit weaker wind speeds and therefore more jets, approaching
Jupiter even more closely.   

\subsection{Discussion}

The above analysis suggests that the equatorial jet direction is
controlled by the relative amplitudes of the equatorial versus
extratropical wave driving associated with day-night and equator-pole
heating gradients, respectively.  The day-night (diurnal) forcing
drives equatorial waves that attempt to cause equatorial superrotation
\citep[cf][]{showman-polvani-2011}.  In contrast, any baroclinic
instabilities induced by the equator-to-pole forcing (i.e., the
meridional gradient in zonal-mean heating) induce Rossby waves that
propagate meridionally; these waves can become absorbed at critical
layers on the equatorward flanks of the off-equatorial subtropical
jets, causing a {\it westward} acceleration in the equatorial region
\citep{randel-held-1991}.  The net equatorial jet direction depends on
which effect dominates.  In our models W\oslow, H\omed, and H\oslow,
the former effect is far stronger, and the equatorial jet is fast and
eastward.  On the other hand, in C\ofast $\,$ and W\ofast, the latter is
stronger, leading to (weak) westward flow at the equator.  The
amplitudes of both wave sources change gradually with rotation rate
and incident stellar flux, leading to the transitional behavior
described in Section~\ref{basic-results} for models C\omed, W\omed,
and H\ofast.

This control of equatorial jet direction by equatorial versus
mid-latitude wave driving is dynamically similar to that observed in
idealized GCMs of Earth and solar-system giant planets that
independently vary a zonally symmetric equator-pole forcing and an
imposed tropical forcing associated with zonally varying tropical
heating anomalies or tropical convection \citep{suarez-duffy-1992,
  saravanan-1993, kraucunas-hartmann-2005, liu-schneider-2010,
  liu-schneider-2011}.  In these models, equatorial superrotation occurs
when the tropical wave forcing dominates over the mid-latitude wave
forcing, while equatorial subrotation (i.e., westward flow) occurs
in the reverse case.

These ideas explain the absence of strong superrotation on C\ofast
$\,$ despite its existence on Jupiter and Saturn.  Models of Jupiter
and Saturn have consistently shown that strong convection is a crucial
ingredient in causing their equatorial superrotation
\citep[e.g.][]{heimpel-etal-2005, schneider-liu-2009, kaspi-etal-2009,
  lian-showman-2010}.  Jupiter is in a regime where the convection is
dynamically important because it is driven by a heat flux comparable
to the absorbed solar flux.  However, our EGP models have absorbed
stellar fluxes ranging from hundreds (C series) to $\sim$$10^4$ (H
series) times greater than the expected internal convected fluxes.
This suggests that internal convection is dynamically unimportant for
the photosphere-level atmospheric circulation of warm and hot EGPs.
Indeed, our models lack a representation of such convection, implying
that when the diurnal forcing is weak, all that remains is the
equator-pole forcing, leading to westward equatorial flow as long as
rotation is not too slow.

\section{Observable implications}
\label{observables}

Here, we present IR spectra and light curves calculated for our nominal grid of
models following the methods described in \citet{fortney-etal-2006b},
\citet{showman-etal-2008a}, and \citet{showman-etal-2009}.
Essentially, given the $T(p)$ profile at each vertical column of the
model's 3D grid, we calculate the local emergent flux radiating toward
Earth from that patch of surface.  We then sum all such contributions
from every surface patch visible from Earth at a given point in the
planet's orbit to obtain an integrated, planet-averaged,
wavelength-dependent flux as received at Earth versus orbital phase.
Note that this approach naturally accounts for any limb darkening or
brightening (associated with longer path lengths through the
atmosphere in regions near the planetary limb as viewed from Earth).
To compare with previous observed and synthetic light curves
\citep[e.g.][]{showman-etal-2009, lewis-etal-2010, kataria-etal-2013,
  kataria-etal-2014}, we calculate light curves in the Spitzer and
WFC3 bandpasses, integrating the emergent flux appropriately over each
wavelength-dependent instrument bandpass.  To ensure self-consistency, the
spectra and lightcurves are calculated using the same radiative
transfer model and opacities used in the SPARC model itself.
Nevertheless, we use a greater resolution of 196 spectral bins so that
spectral features may be better represented.

\begin{figure*}
\begin{minipage}[c]{0.3\textwidth}
\includegraphics[scale=0.36, angle=0]{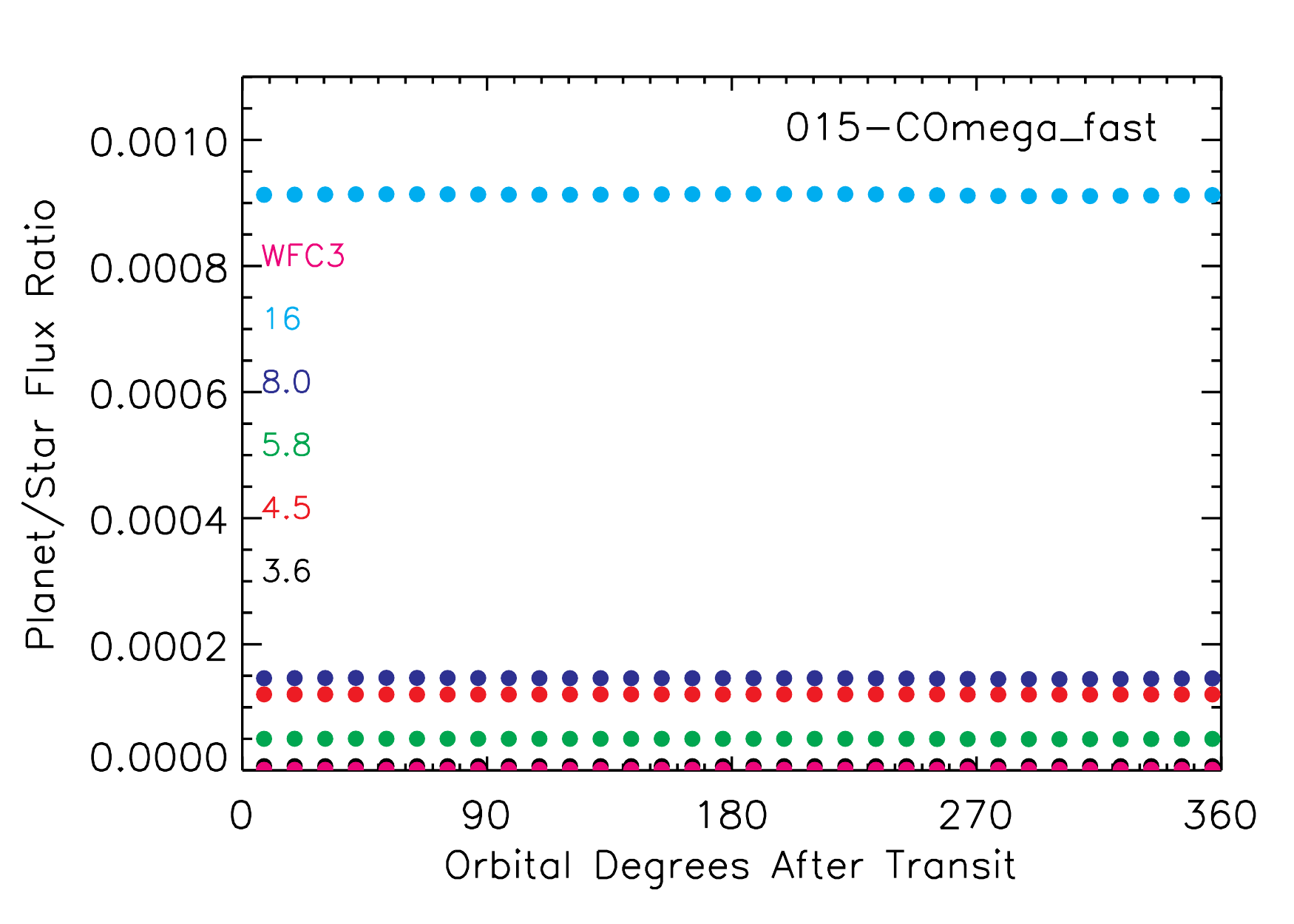}
\includegraphics[scale=0.36, angle=0]{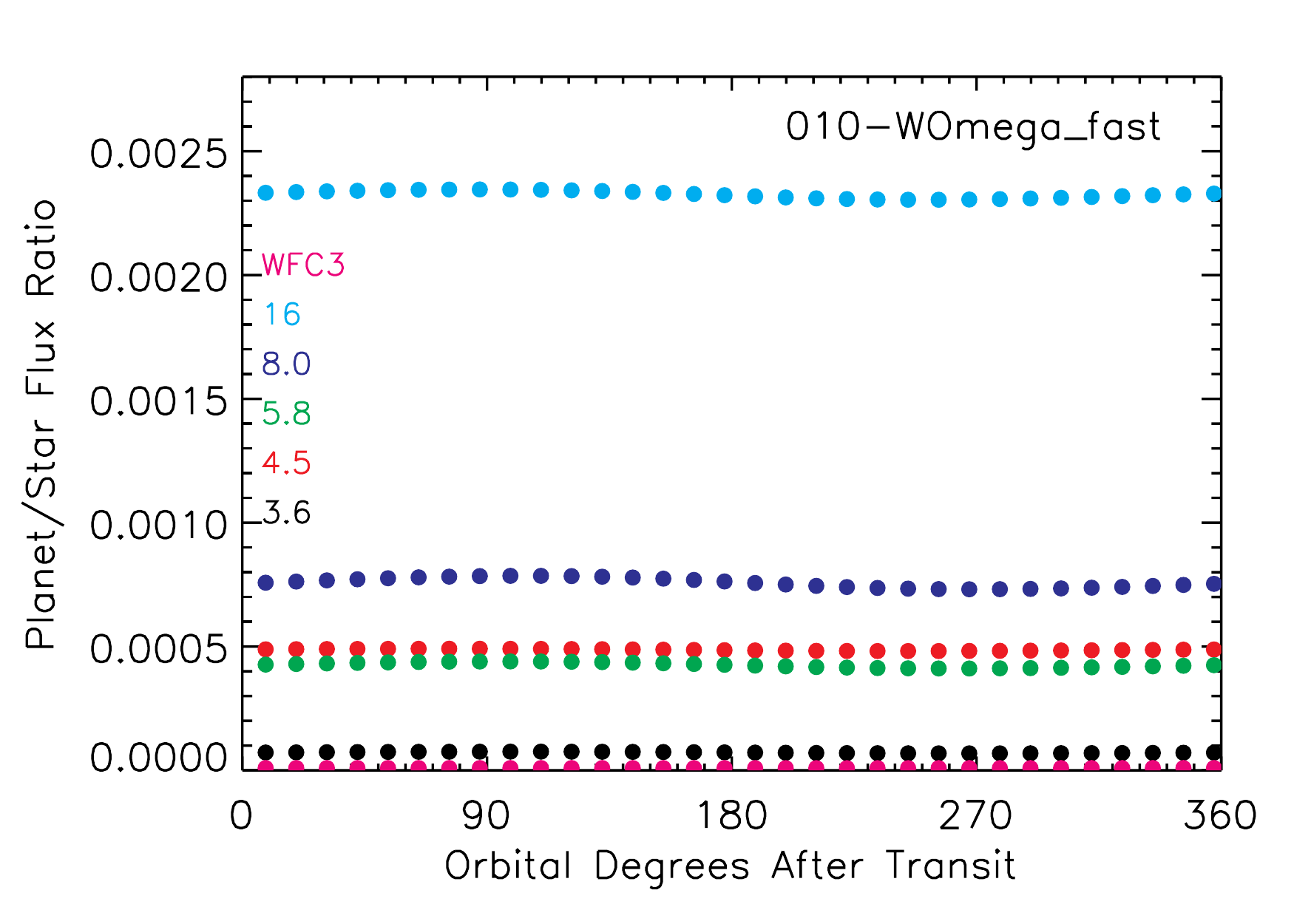}
\includegraphics[scale=0.36, angle=0]{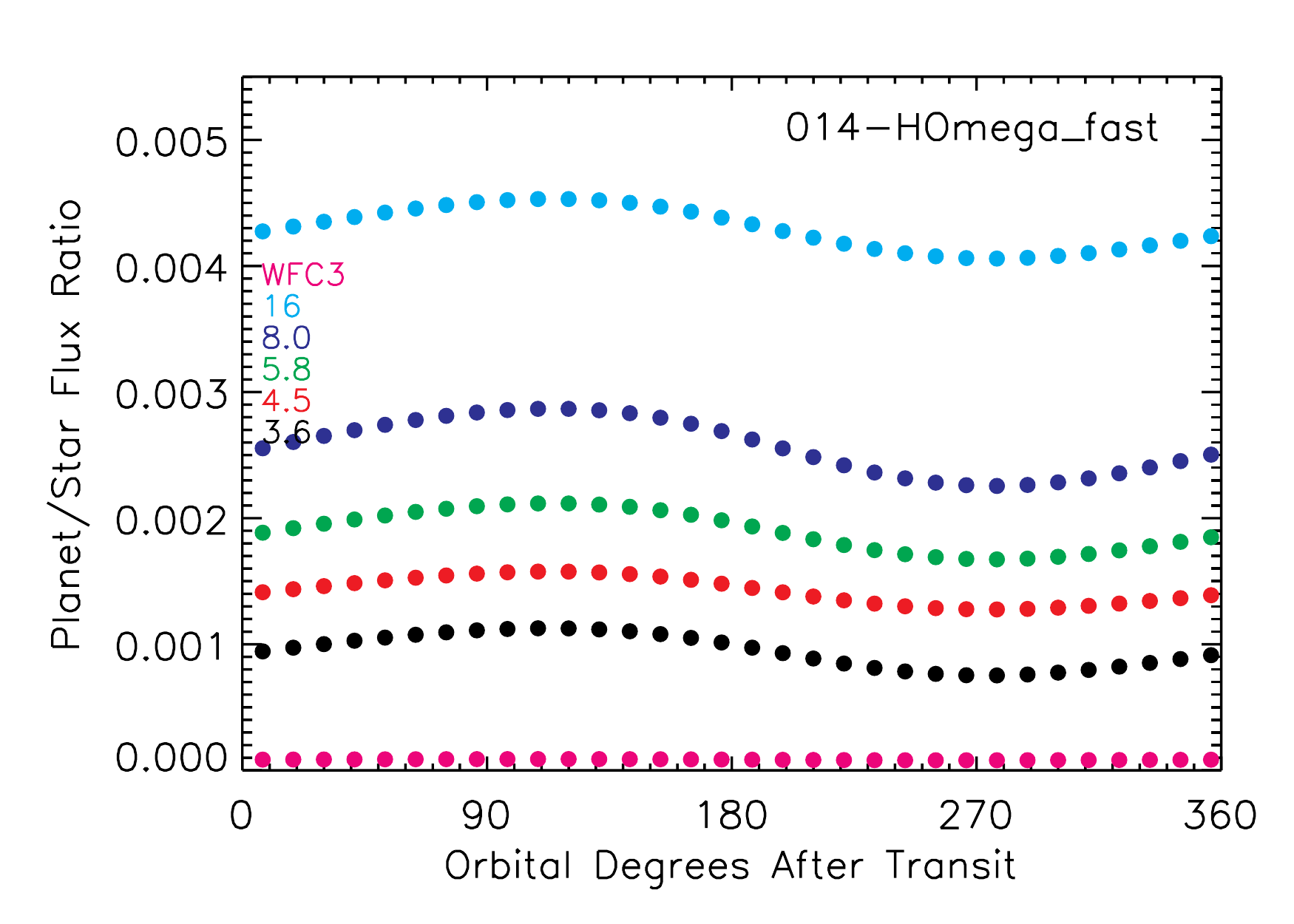}
\end{minipage}
\begin{minipage}[c]{0.3\textwidth}
\includegraphics[scale=0.36, angle=0]{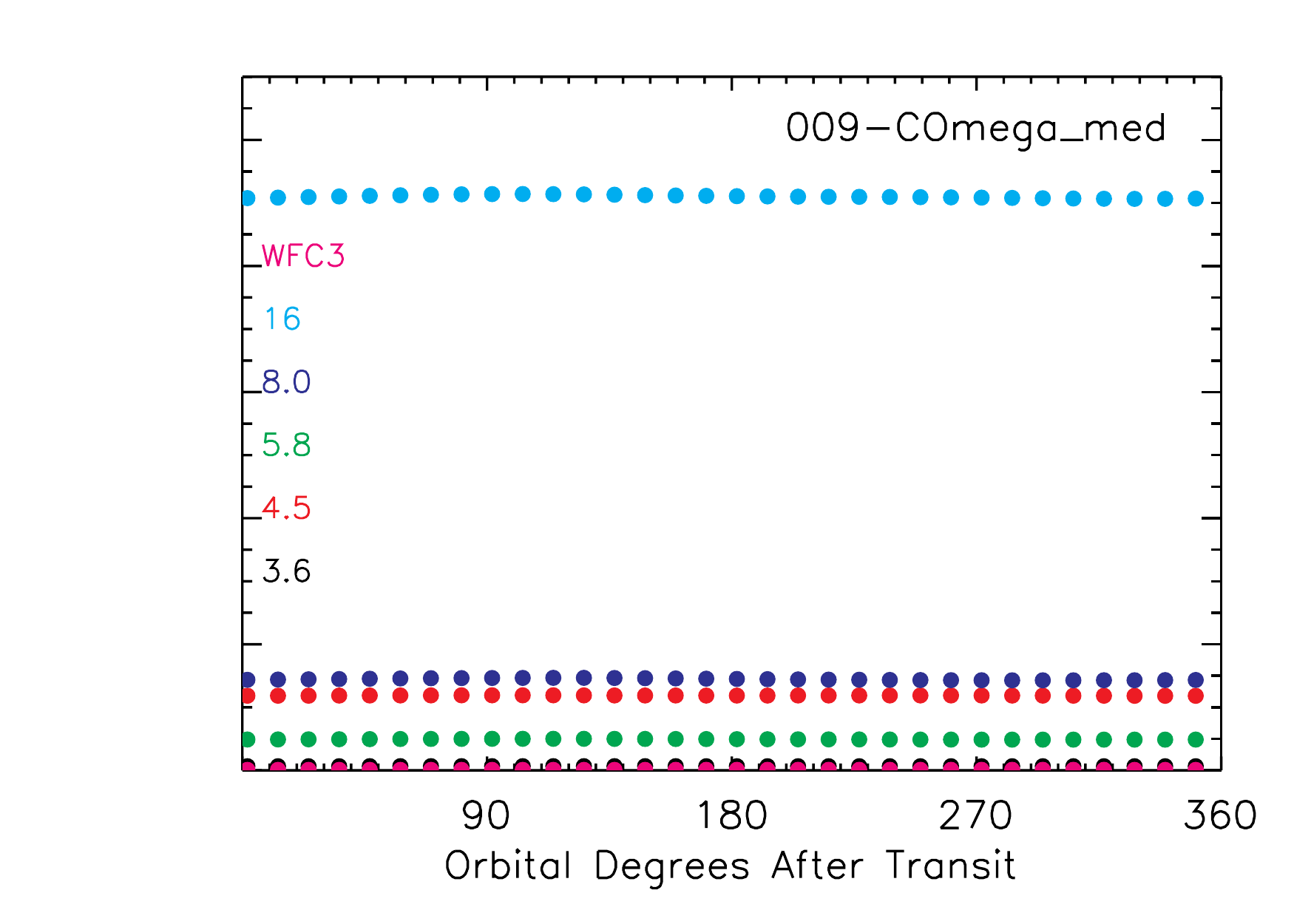}
\includegraphics[scale=0.36, angle=0]{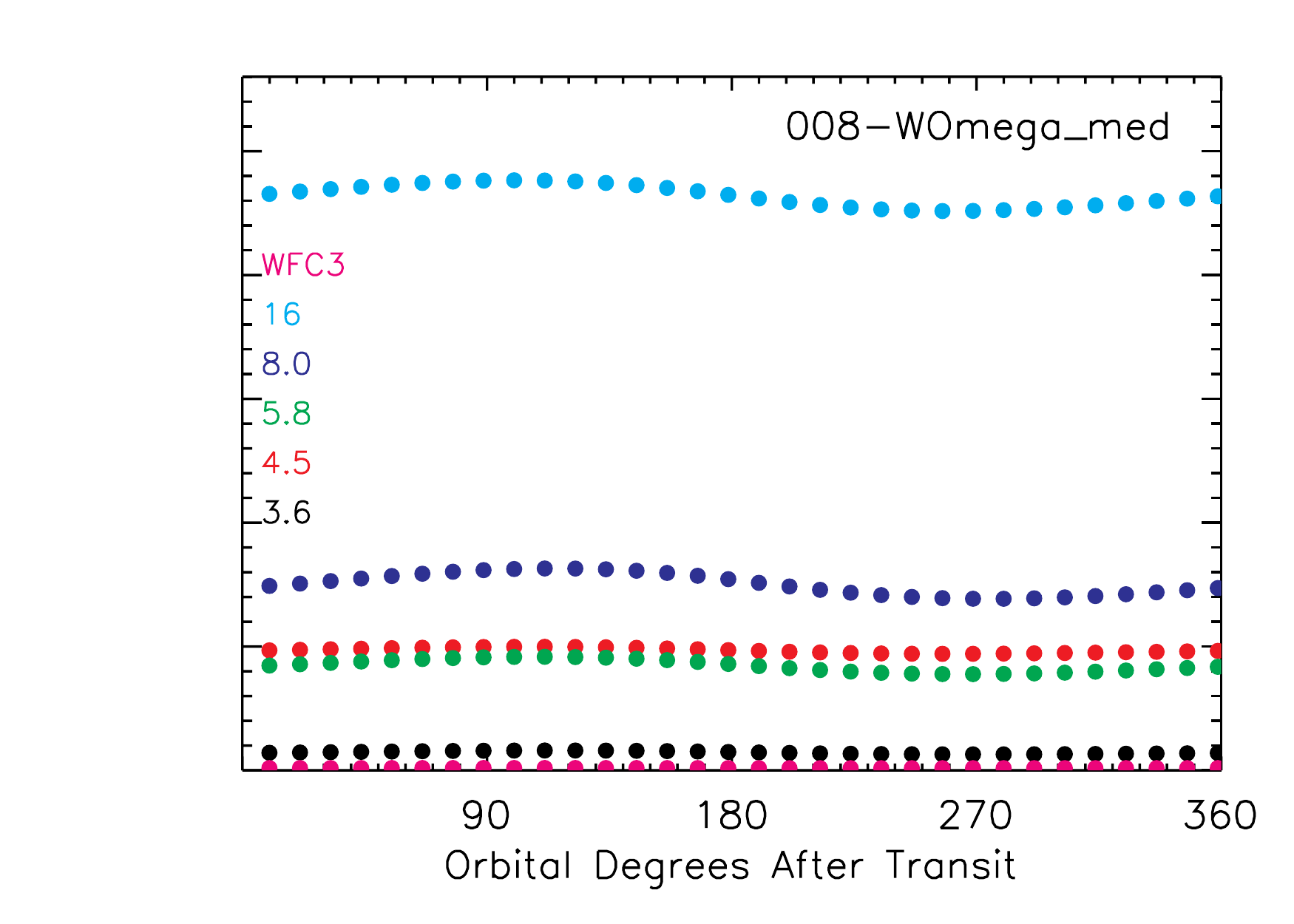}
\includegraphics[scale=0.36, angle=0]{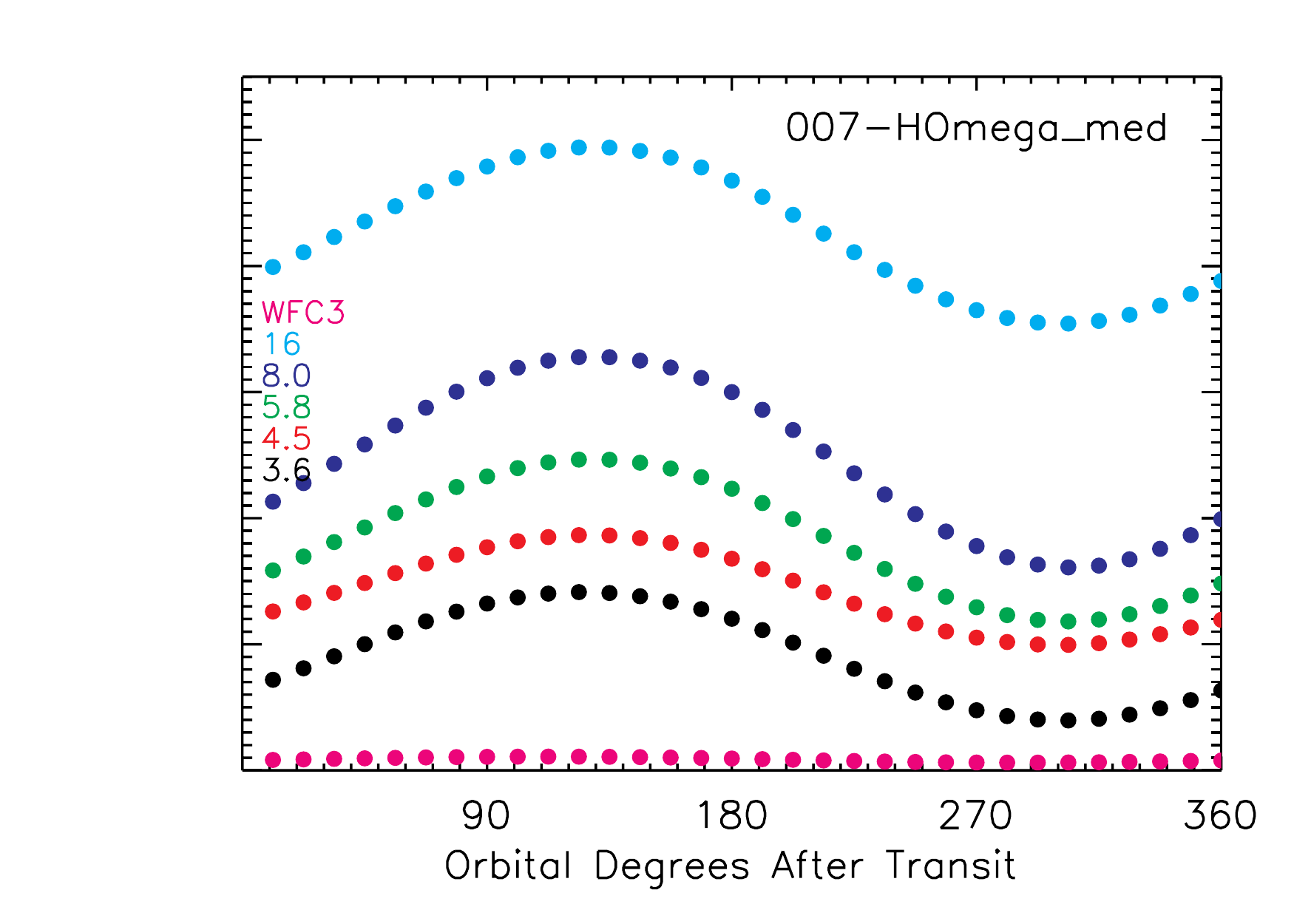}
\end{minipage}
\begin{minipage}[c]{0.3\textwidth}
\includegraphics[scale=0.36, angle=0]{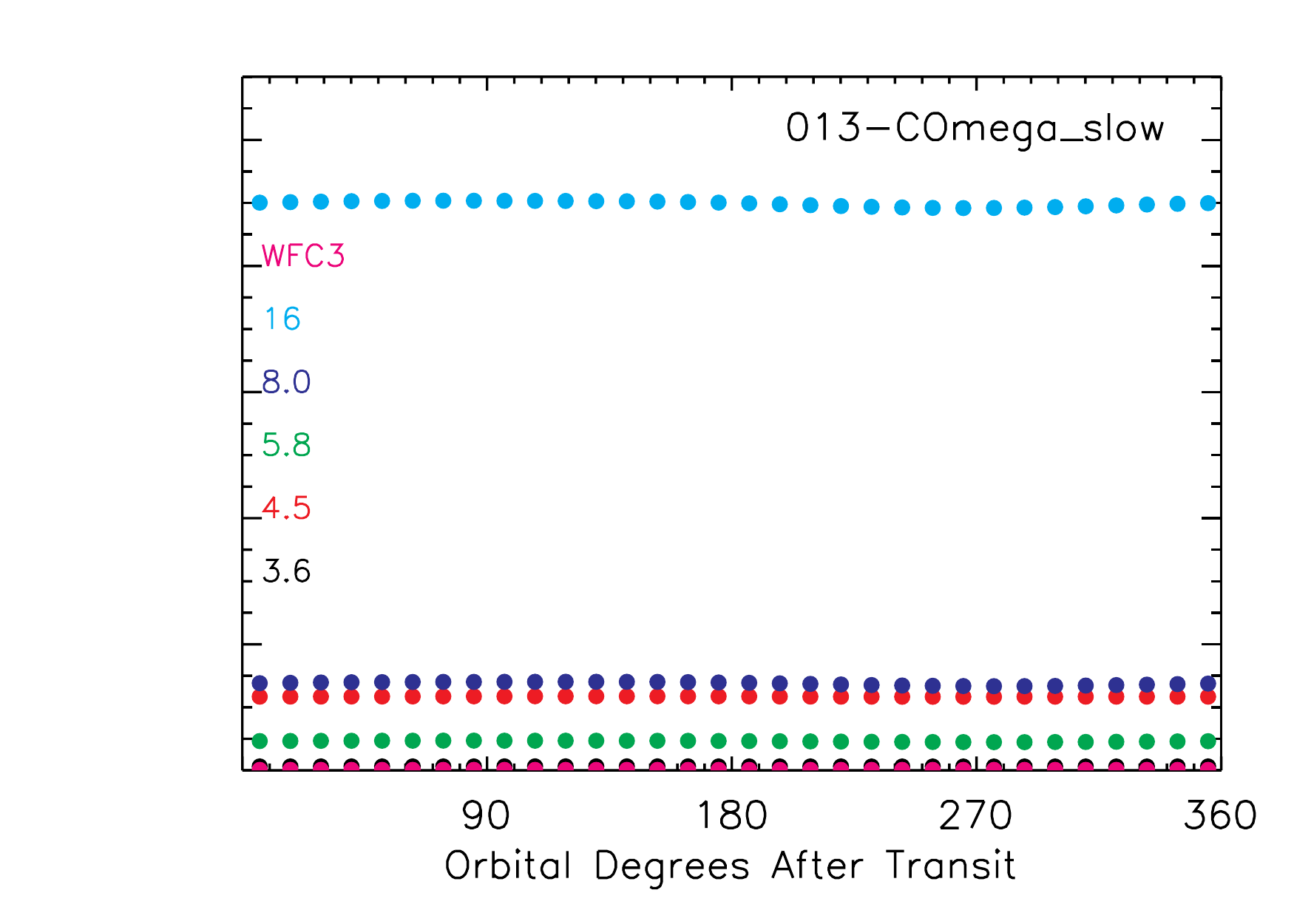}
\includegraphics[scale=0.36, angle=0]{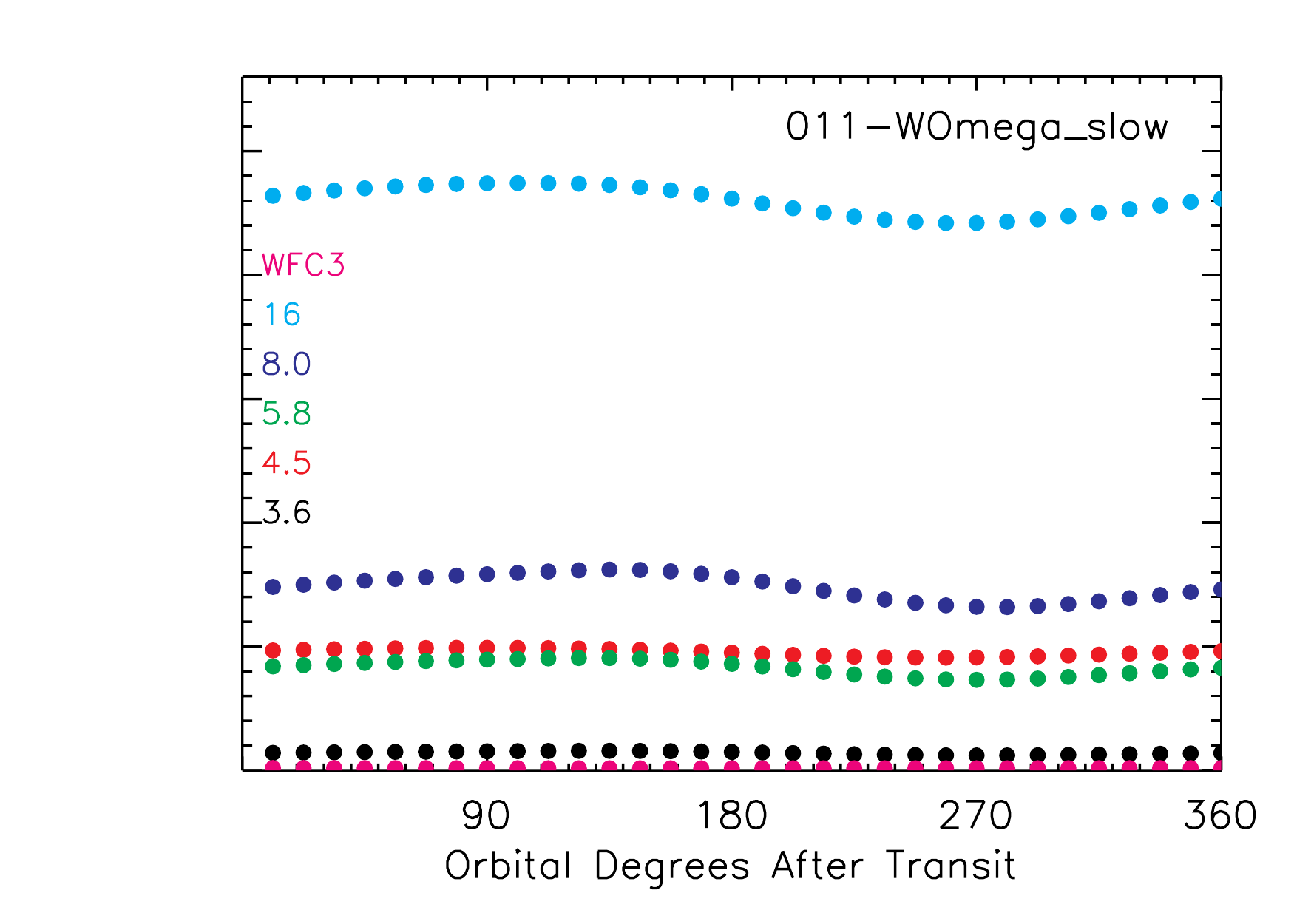}
\includegraphics[scale=0.36, angle=0]{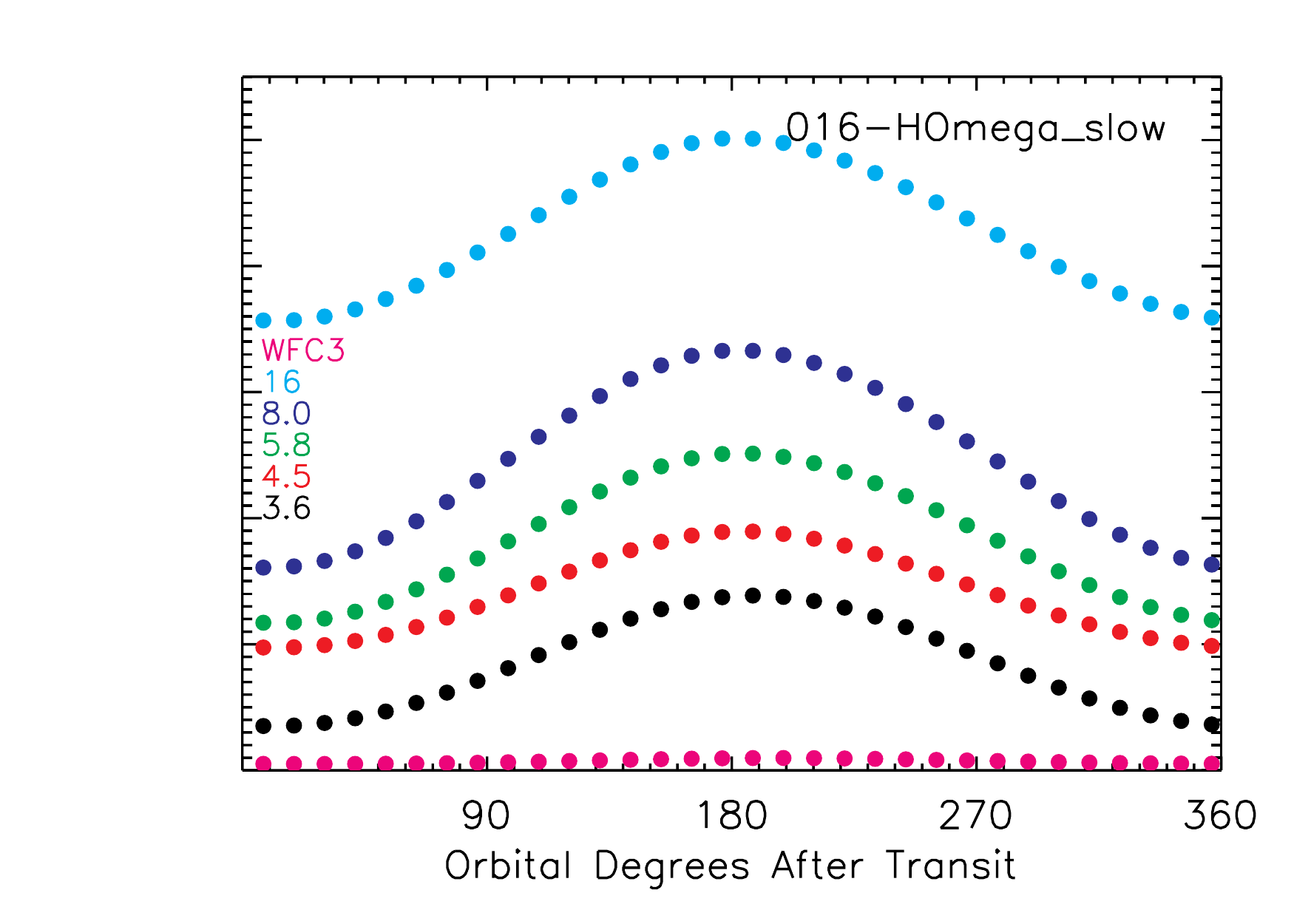}
\end{minipage}
\caption{Light curves in Spitzer IRAC and HST WFC3 bandpasses for the
  nine runs in the nominal grid.  Each panel plots planet-star flux
  ratio versus orbital phase relative to transit.  The left, middle,
  and right columns adopt rotation periods of 0.55, 2.2, and 8.8 days,
  respectively.  The top, middle, and bottom rows adopt orbital
  semimajor axes of 0.2, 0.08, and 0.03 AU, respectively.}
\label{lightcurves}
\end{figure*}

The resulting light curves are shown in Figure~\ref{lightcurves}.  Our
cool models exhibit relatively flat light curves, a result of the
minimal hemispheric-scale longitudinal temperature differences in
these models.  In constrast, phase variations are larger in the warm
and hot models, a result of the larger longitudinal temperature
variations there.  For a given incident stellar flux, slowly rotating
models exhibit larger phase variations than rapidly rotating models.
In the W series, for example, the variations are $\lesssim 5$--10\% in
the rapidly rotating model but reach 10--20\% for the slowly rotating
model.  Unsurprisingly, the hot models exhibit the largest phase
variations, reaching or exceeding a factor-of-two variation at many
wavelengths in H\oslow.  In most of the models with prominent phase
variations---including all three W models, H\ofast, and H\omed---the
IR flux peaks precede secondary eclipse, with offets of
$\sim$40--$80^{\circ}$.  This is the result of advection from the
equatorial superrotation (and, in fast rotation cases, the rotation
itself) displacing the hot regions to the east.  Interestingly, the
phase offset is nearly zero in H\oslow.  This results from the fact
that, in a synchronously rotating reference frame, the planetary
motion is from east-to-west, while the equatorial jet flows from
west-to-east; these phenomena lead to rather weak net flow at low
latitudes in the synchronously rotating frame, and thus a small
offset.

\begin{figure*}
\begin{minipage}[c]{0.3\textwidth}
\includegraphics[scale=0.36, angle=0]{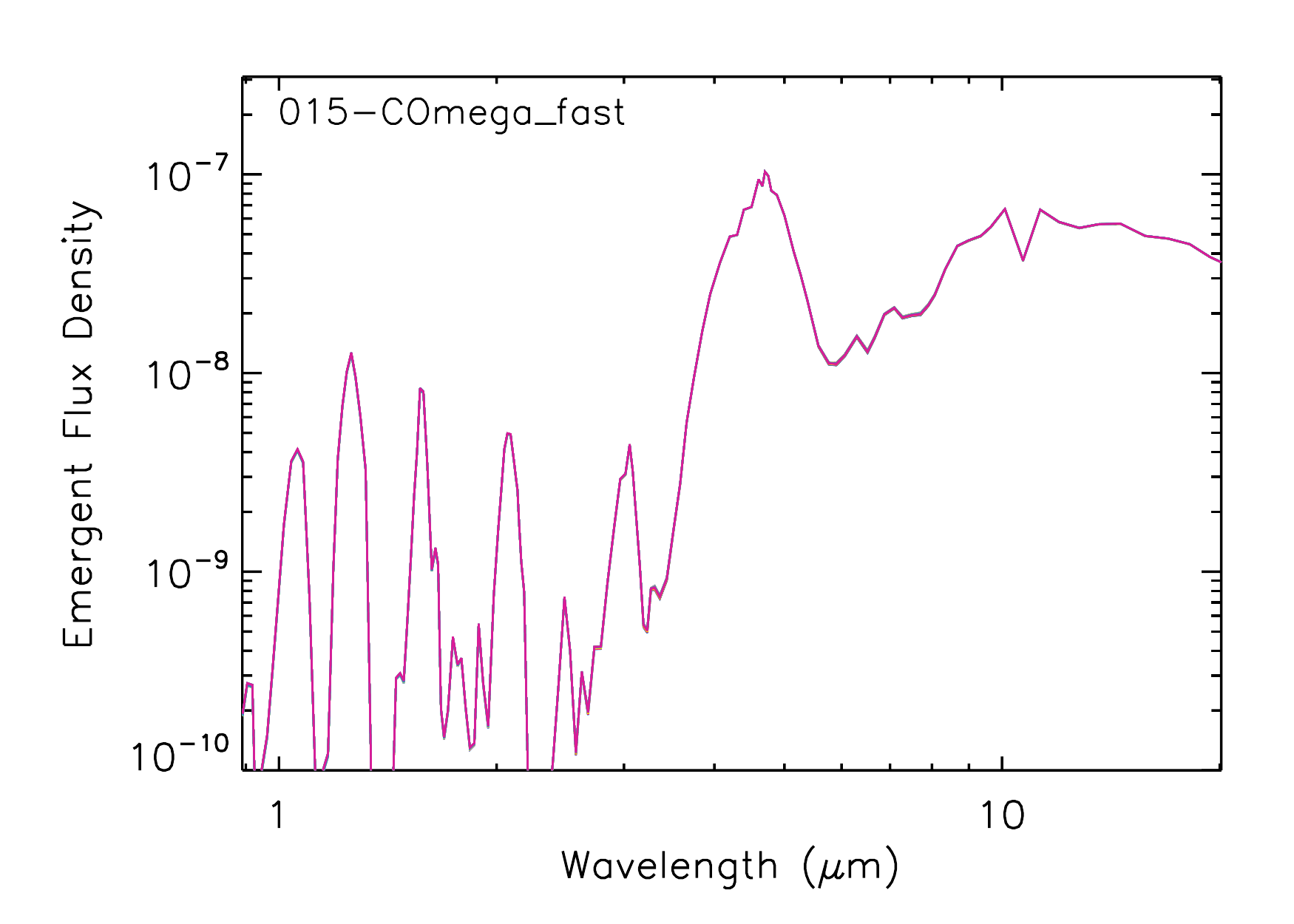}
\includegraphics[scale=0.36, angle=0]{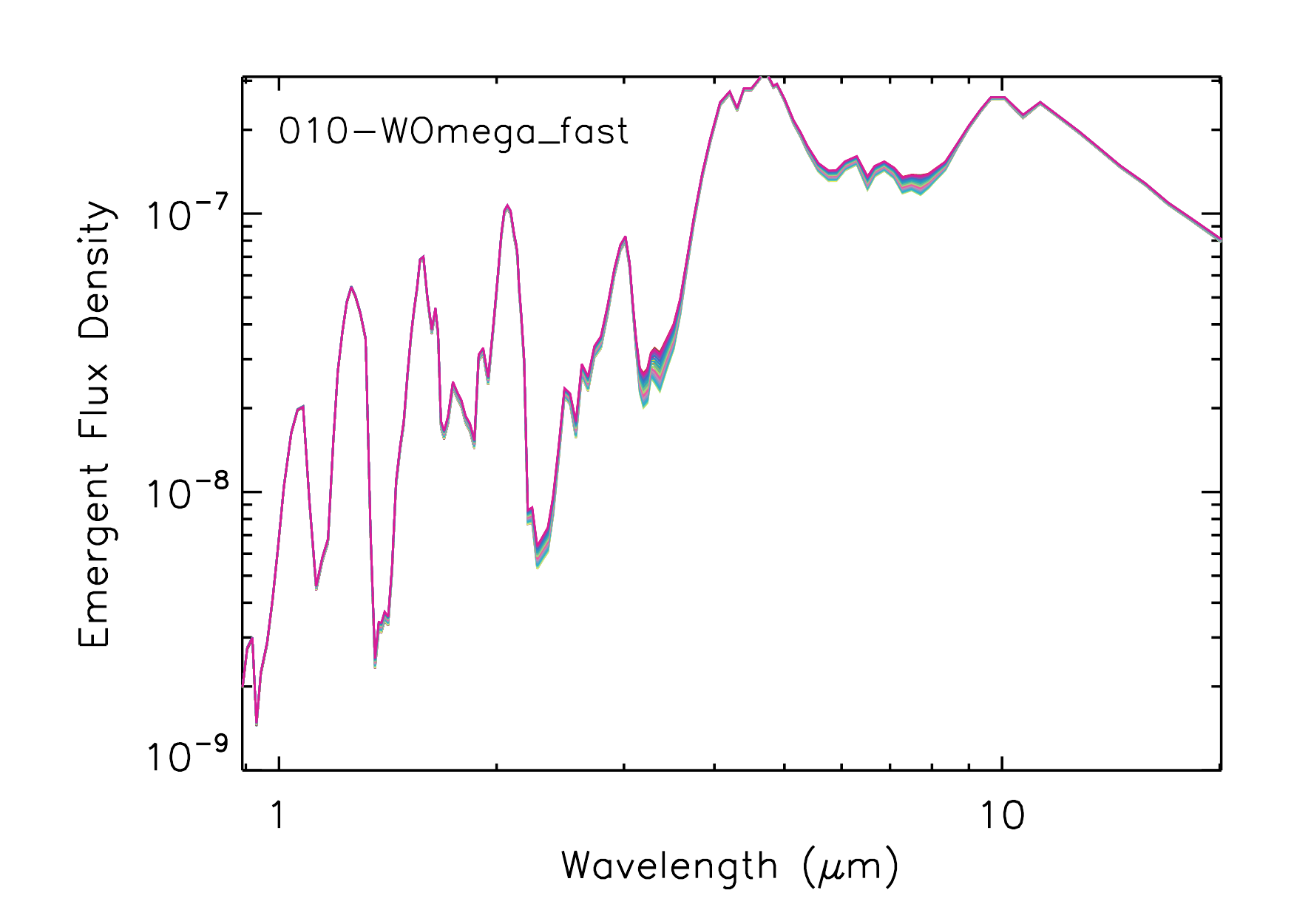}
\includegraphics[scale=0.36, angle=0]{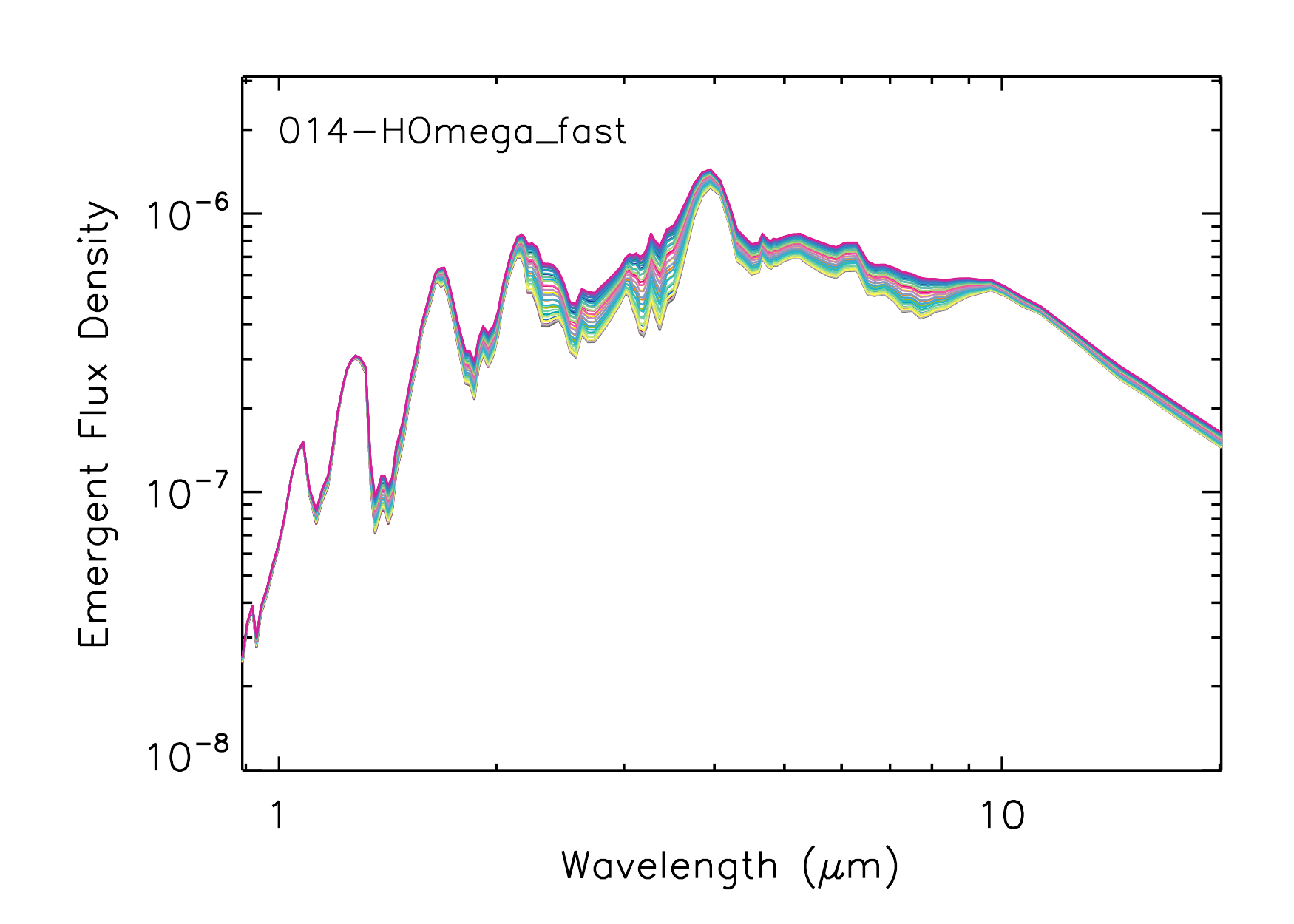}
\end{minipage}
\begin{minipage}[c]{0.3\textwidth}
\includegraphics[scale=0.36, angle=0]{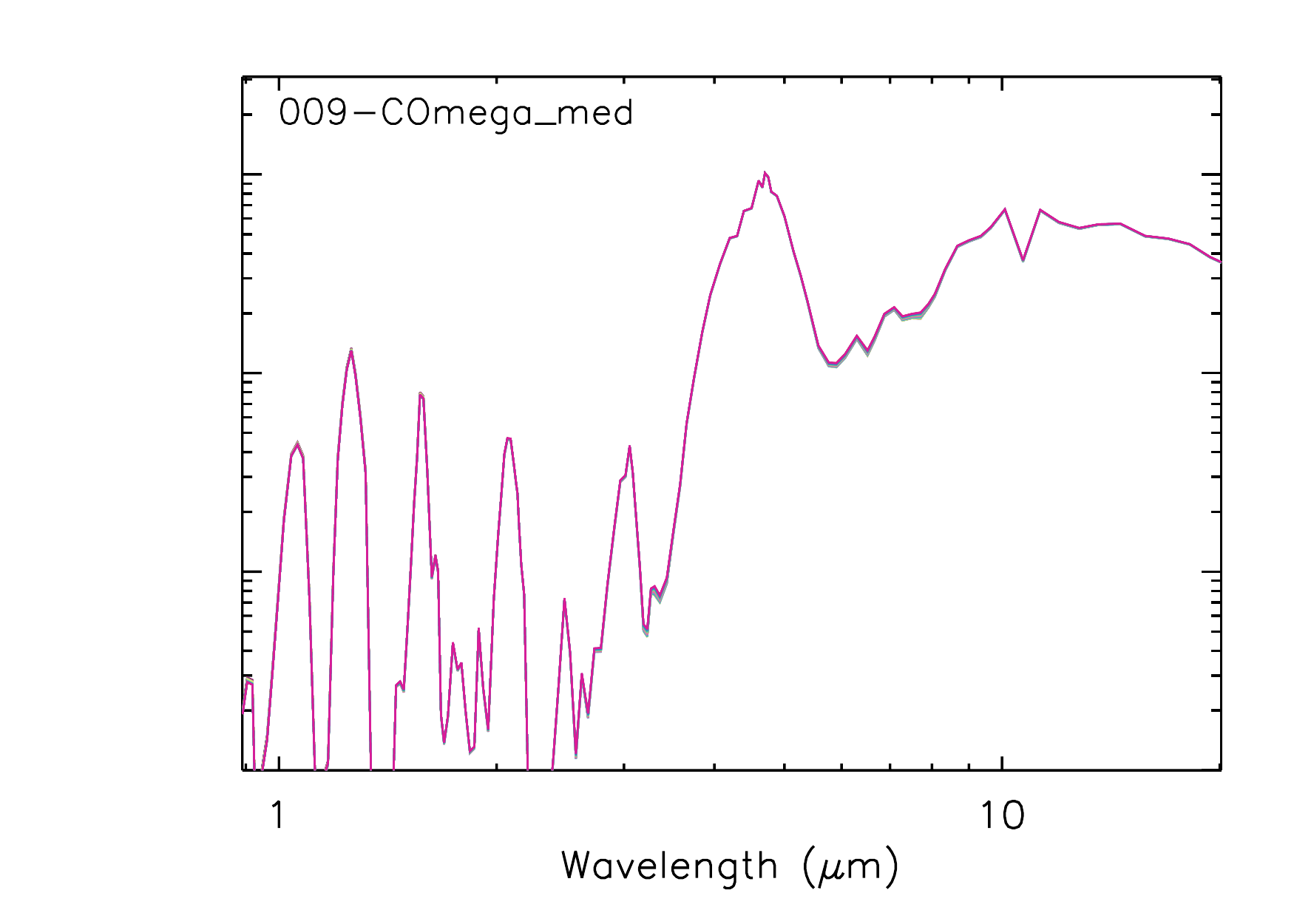}
\includegraphics[scale=0.36, angle=0]{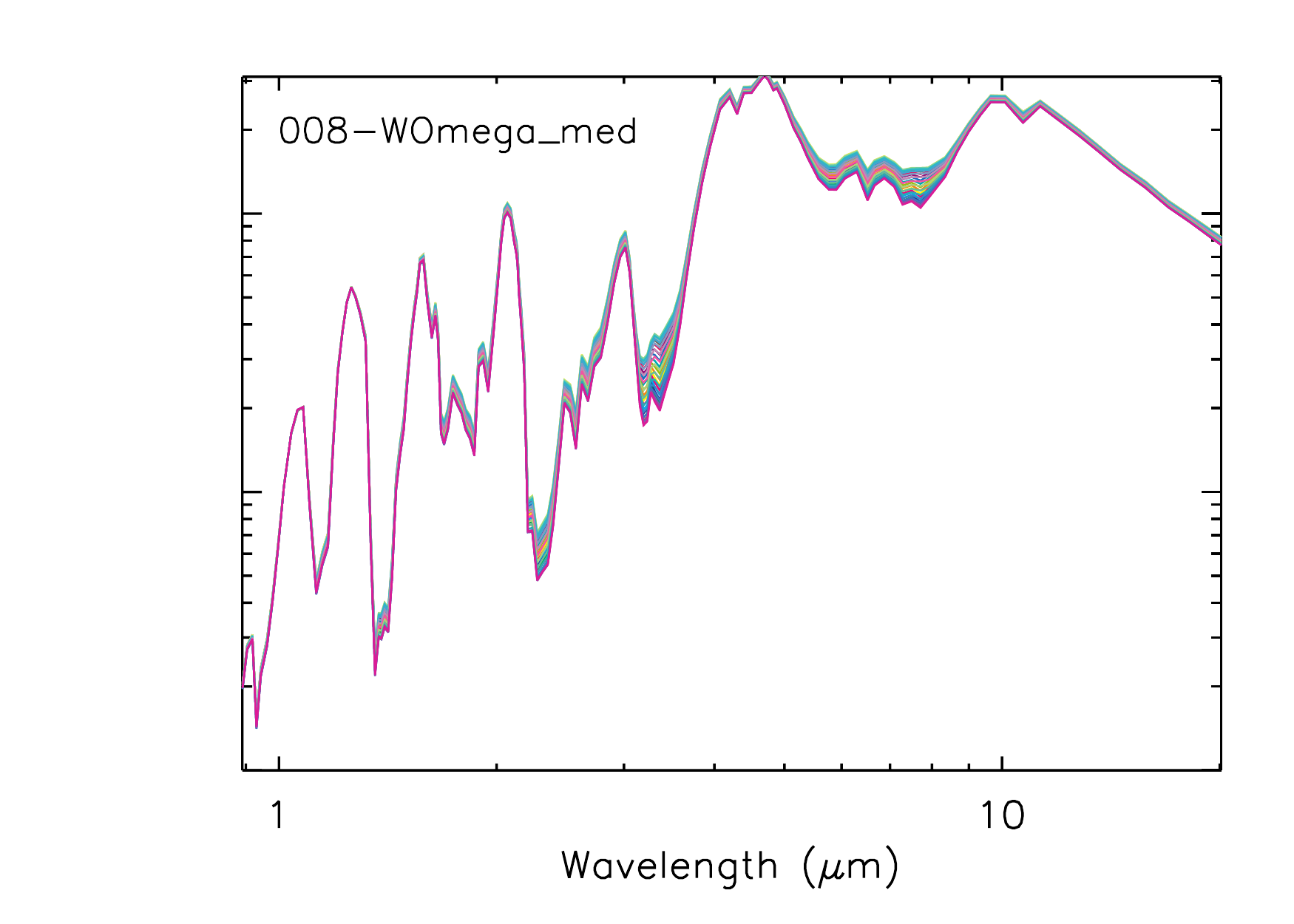}
\includegraphics[scale=0.36, angle=0]{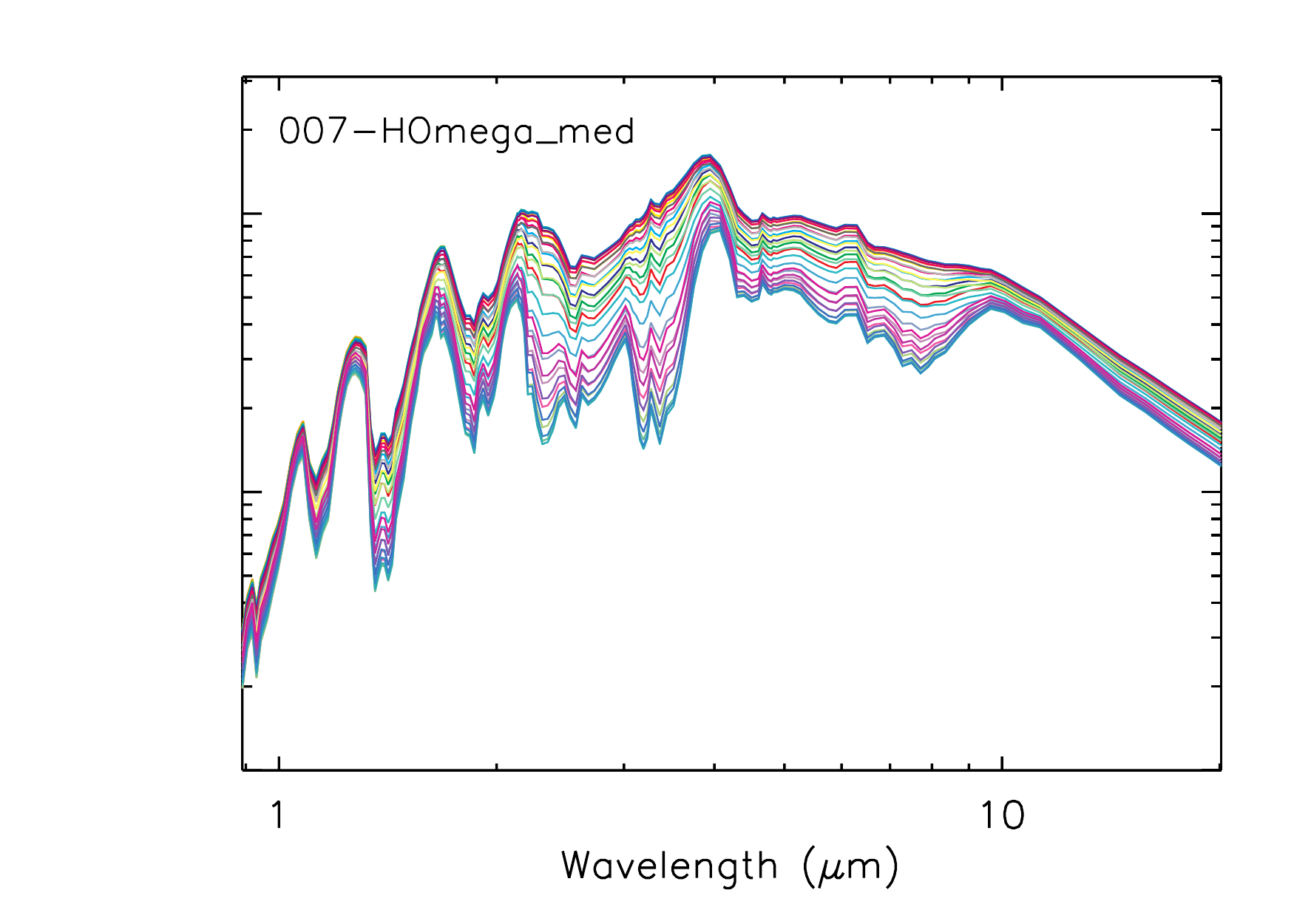}
\end{minipage}
\begin{minipage}[c]{0.3\textwidth}
\includegraphics[scale=0.36, angle=0]{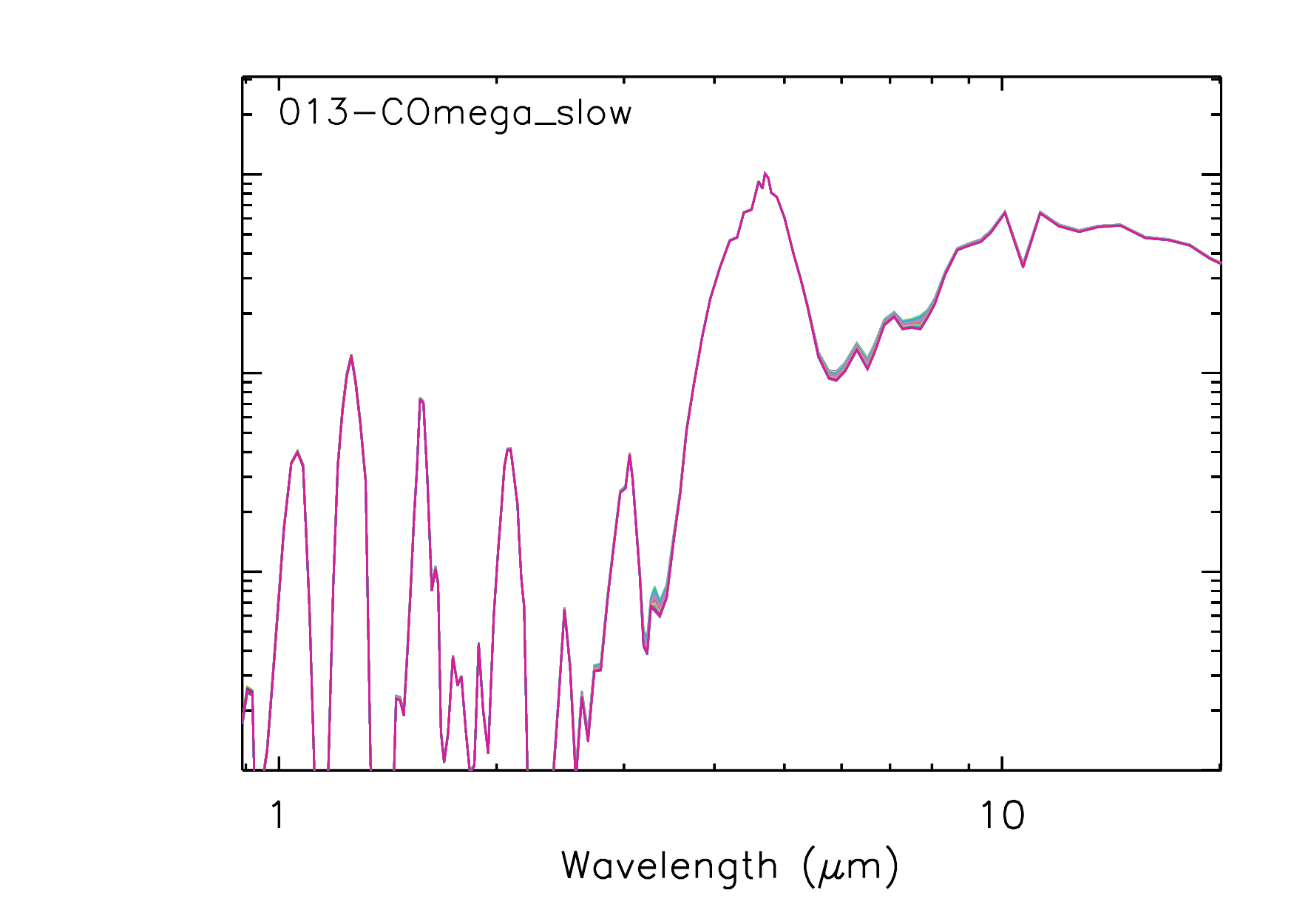}
\includegraphics[scale=0.36, angle=0]{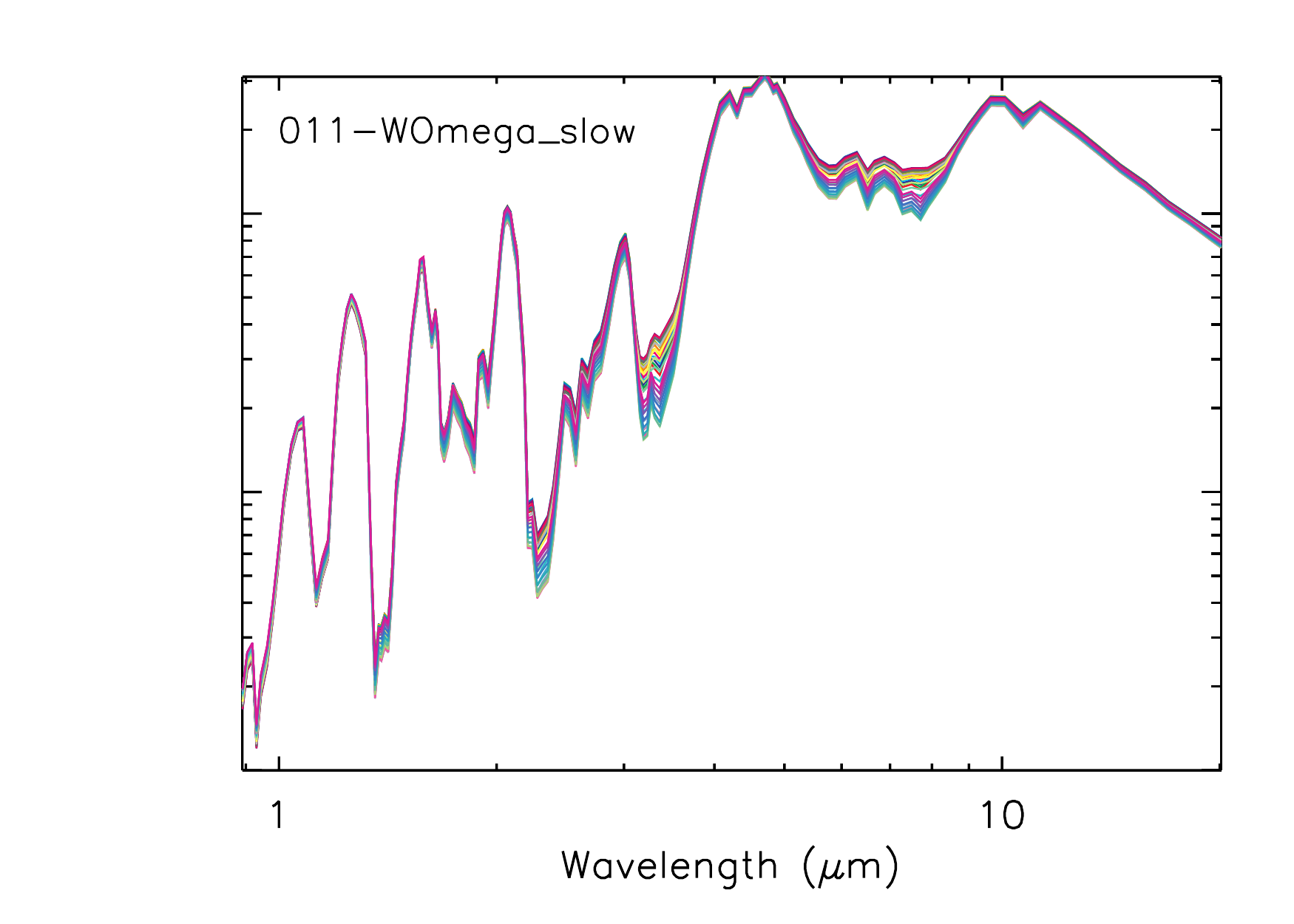}
\includegraphics[scale=0.36, angle=0]{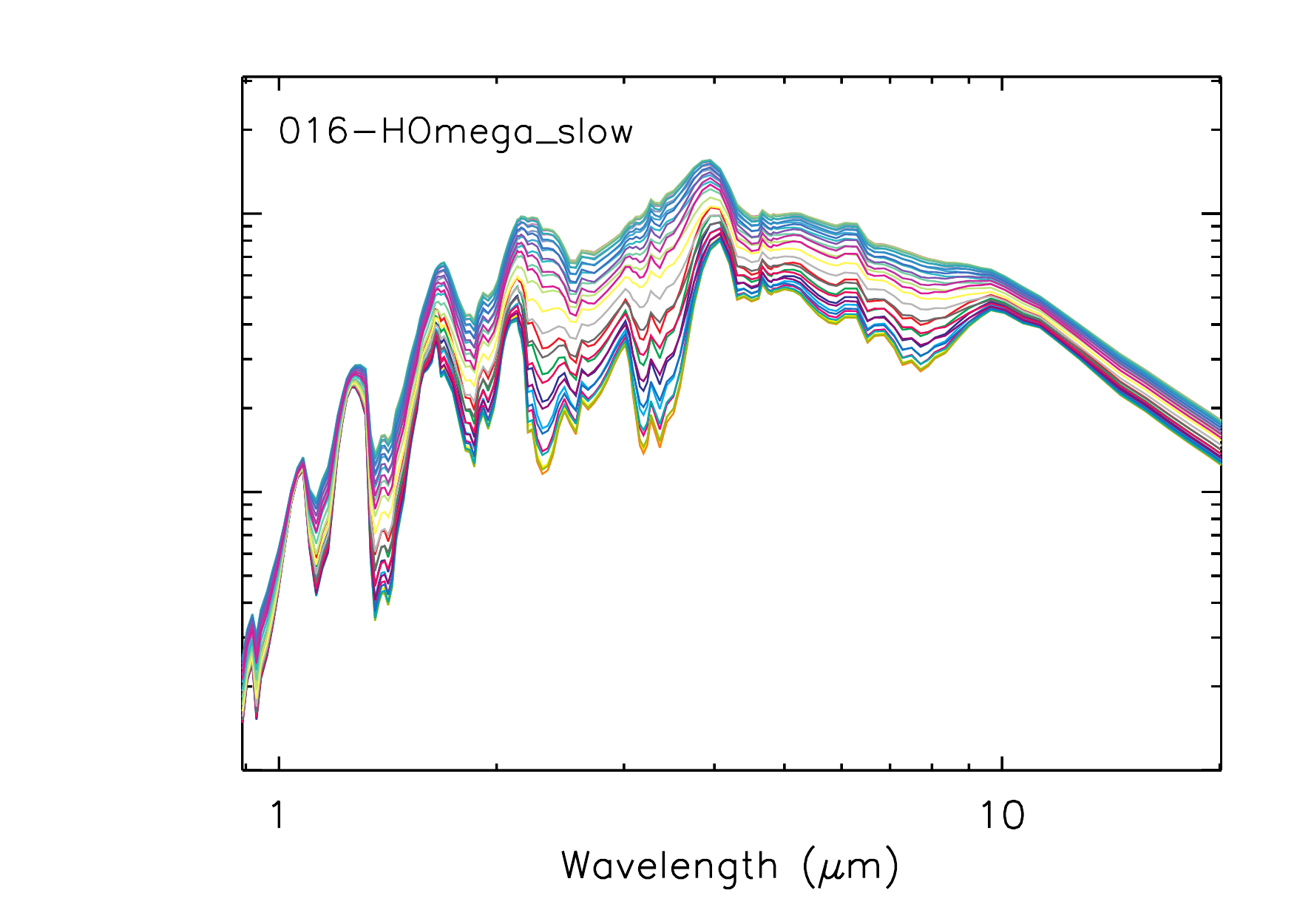}
\end{minipage}
\caption{Infrared spectra for the
  nine runs in the nominal grid.   Different curves show spectra
  at different orbital phases; this quantifies how the amplitude of
  phase variations depends on wavelength.  The left, middle,
  and right columns adopt rotation periods of 0.55, 2.2, and 8.8 days,
  respectively.  The top, middle, and bottom rows adopt orbital
  semimajor axes of 0.2, 0.08, and 0.03 AU, respectively.}
\label{spectra}
\end{figure*}

Figure~\ref{spectra} shows IR spectra for these same nine models at
many different orbital phases.  The spectra deviate strongly from a
blackbody, especially in the W and C models.  Spectral peaks, which
represent atmospheric opacity windows, sample deeper pressures where
temperatures are high; spectral valleys represent absorption bands
that probe lower-pressure, cooler regions.  Because the day-night
temperature differences tend to increase with decreasing pressure,
spectral windows tend to exhibit smaller phase variations than
spectral absorption bands.  This trend can be prominently seen in
Figure~\ref{spectra} where the spread between curves---when it occurs
at all---is greatest in the spectral valleys.  The C models exhibit
essentially no phase variations at any wavelength, except for slight
variations near 3.3--$3.5\,\mu$m and 6--$8\,\mu$m in C\oslow.  In the
W series, spectral peaks exhibit little phase variation, whereas
spectral valleys start to exhibit significant phase variations,
reaching a factor of two at $\sim$$2.2\,\mu$m and $3.3\,\mu$m in
W\omed and W\oslow.  Consistent with Figure~\ref{lightcurves}, the
spectral phase variations are strongest for the hottest models, and
also exhibit significant dependence on rotation rate.  Model H\ofast
exhibits peak phase variations of a factor of two from
$\sim$2--$3.5\,\mu$m, with little phase variations at many other
wavelengths.  In contrast, the variations reach a factor of 5--8 at
these same wavelengths in H\omed and H\oslow.

These light curves and spectra indicate that sufficiently strong
non-synchronous rotation can cause observable variations in flux,
providing a potential observational test of rotation state.
Nevertheless, in practice this may be challenging.  For example, the
amplitude of the phase variations in H\oslow $\,$ is only a factor of
$\sim$2 larger than in H\ofast, despite the 16-fold difference in
rotation rate between these models.  Given that atmospheric
metallicity, the possible influence of clouds or hazes, and other effects
could exert an equal or greater effect on the phase-curve amplitude
and offsets, extracting robust information about rotation rate from
observational phase variations may be difficult.  Our feeling is that
order-of-magnitude variations in rotation rate are observationally
distinguishable in lightcurves, but that factor of two (or smaller)
variations in rotation rate may be degenerate with other unknowns.
The real, but modest, effect of rotation in that case would simply be
difficult to disentangle from other competing, but poorly constrained,
influences on the light curves.

There are several other observational signatures that may help to
constrain the dynamics of these warm-to-hot EGPs, including the regime
transition discussed here.  First, ingress/egress mapping during
secondary eclipse could be used to determine the two-dimensional
thermal pattern across the dayside \citep{majeau-etal-2012,
  de-wit-etal-2012}, which may allow observational discrimination
between planets with dayside hot spots (like our models H\omed,
H\oslow, and W\oslow) versus those with nearly homogeneous
temperatures in longitude (like our models C\ofast, C\omed, and
C\oslow).  Because this technique allows mapping in both longitude
{\it and} latitude, it may allow latitudinal temperature contrasts to
be determined.  Second, vertical mixing will likely cause chemical
disequilibrium between CO and CH$_4$ \citep{cooper-showman-2006}, and
between N$_2$ and NH$_3$ for cooler EGPs.  Thus photometry and spectra
that help to infer the abundances of these species will allow
constraints to be placed on the vertical mixing rates.  Third, cloud
formation would also be expected, particularly in our cooler models
and on the nightsides of our warmer models.  The longitudinal
distribution of cloudiness---which could be constrained via
visible-wavelength light curves---likewise will shed light on the
circulation regime.

\section{Discussion}
\label{discussion}

We presented a systematic analysis of how the atmospheric
circulation regime on warm and hot Jupiters varies with
incident stellar flux and planetary rotation rate.  Basic
theoretical arguments suggest that the ``canonical'' hot Jupiter
regime---of large day-night temperature difference and a fast,
eastward jet with peak zonal-mean zonal winds at the equator---should
give way at smaller stellar flux and/or faster rotation rate 
to a regime with small longitudinal temperature variations
and peak wind speeds occurring in zonal jets at mid-to-high
latitudes.  We argued on theoretical grounds that the
former regime should occur when the radiative time constant at
the photosphere is shorter than the solar day whereas the latter should obtain
when the radiative time constant is longer than the solar day.
To test these ideas and provide a deeper understanding of 
the dynamical behavior, we performed 3D numerical simulations
of the atmospheric flow using the SPARC/MITgcm model, varying
the rotation rate over a factor of 16 and the incident stellar
flux over a factor of 40. 

In agreement with the predictions, our models show that the dynamics
exhibits a regime transition from a circulation dominated by an
equatorial superrotating jet at high irradiation and slow rotation
rates to a regime consisting of off-equatorial eastward jets, with
weaker eastward or even westward flow at the equator, at lower
irradiation and faster rotation rates.   In the latter regime, 
models at modest-to-slow rotation rate exhibit one eastward jet
in each hemisphere, but our fastest-rotating models exhibit multiple
eastward jets in each hemisphere and illuminate the dynamical continuum
from hot Jupiters to Jupiter and Saturn themselves.

Nevertheless, the apparent simplicity of this transition belies a rich
range of dynamical behavior within each regime and associated with
their transition.  For example, the Rossby number ranges from
$\sim$0.1 in the rapidly rotating models to $\gtrsim 1$ in the slowly
rotating models, implying that the relative roles of Coriolis forces
to momentum advection, the way that eddies act to maintain a mean
flow, and the nature of the meridional heat transport undergo
significant variation across our ensemble.   Several sub-regimes
involving significant transitions in the dynamics may 
may exist within each of the broader regimes emphasized here.

Synthetic light curves and spectra calculated from our models
demonstrate that incident stellar flux and planetary rotation rate
exert a strong influence on planetary emission.  Variations of
emission with orbital phase are weak when the incident stellar flux is
low, but become significant as the stellar flux increases.  At a given
stellar flux, slower rotation tends to cause larger-amplitude phase
variations, though the effect is modest (order-of-magnitude variations
in rotation rate lead to $\sim$factor-of-two changes in phase
amplitude).  These results indicate that non-synchronous rotation
could be inferred from light curves of hot Jupiters if the deviation
from synchronous rotation is sufficiently large ($\gtrsim$ factor of
two).  This is consistent with \citet{showman-etal-2009} and
\citet{rauscher-kempton-2014} and extends their results to a wider
range of conditions.

As emphasized by \citet{showman-etal-2009} and
\citet{kataria-etal-2014}, the amplitude of phase variations depends
significantly on wavelength.  The greatest phase variations tend to
occur in spectral absorption bands (which probe low-pressure regions
that tend to exhibit large day-night temperature differences) and
smaller phase variations occur in spectral windows (which probe
higher-presssure regions that generally exhibit smaller day-night
temperature variations).  This wavelength-dependent phase variation
becomes significant in our hot and warm models.  This signature
provides a potentially important way to observationally determine how
the day-night temperature difference of a hot Jupiter depends on
pressure.

The models presented here pose numerous questions amenable to future
research.  The regime of weak longitudinal temperature variations and
fast mid-to-high latitude jets occurring when $\tau_{\rm rad}> P_{\rm
  solar}$ has been little explored in the hot-Jupiter literature, but
this regime likely dominates the dynamical behavior of EGPs from
orbital distances of $\sim$0.1~AU out to several AU.  Given that
the population of such planets greatly exceeds that of hot Jupiters
themselves, further work to understand this regime would be beneficial.
For example, most work done in the terrestrial literature on baroclinic
instabilities is for the Earth-like application of an atmosphere resting
on a surface with a strong entropy gradient, a property that strongly
affects the instability.  In contrast, it would be useful to
explore how baroclinic instabilities equilibrate with the mean flow in
the case of a gas giant where a lower impermeable surface plays no
role---an area where little work has been done.  The extent to which
Venus or Titan can provide analogues to the regime of slowly rotating,
weakly irradiated EGPs would also be worth investigating.  The relative
roles of equator-pole and day-night forcing could be disentangled by
performing idealized models that vary their ratio.  Moreover,
additional work is needed---both in the hot Jupiter and more weakly
irradiated regime---in the mechanisms controlling the meridional
(equator-pole) heat transport, and how that transport depends on
parameters.  Coupled with improved observations over the coming
decade, such an effort would greatly improve our understanding
of EGP atmospheric circulation over a broad range of conditions.


\acknowledgements We thank Tad Komacek for comments on the manuscript.
This research was supported by NASA Origins and Planetary Atmospheres
grants NNX12AI79G and NNX10AB91G to APS.  It was performed in part
under contract with the Jet Propulsion Laboratory (JPL) funded by NASA
through the Sagan Fellowship Program executed by the NASA Exoplanet
Science Institute.

\def\icarus{Icarus}

\bibliographystyle{apj}
\bibliography{showman-bib}

\begin{thebibliography}{100}
\expandafter\ifx\csname natexlab\endcsname\relax\def\natexlab#1{#1}\fi

\bibitem[{{Adcroft} {et~al.}(2004){Adcroft}, {Campin}, {Hill}, \&
  {Marshall}}]{adcroft-etal-2004}
{Adcroft}, A., {Campin}, J.-M., {Hill}, C., \& {Marshall}, J. 2004, Monthly
  Weather Review, 132, 2845

\bibitem[{{Amundsen} {et~al.}(2014){Amundsen}, {Baraffe}, {Tremblin},
  {Manners}, {Hayek}, {Mayne}, \& {Acreman}}]{amundsen-etal-2014}
{Amundsen}, D.~S., {Baraffe}, I., {Tremblin}, P., {Manners}, J., {Hayek}, W.,
  {Mayne}, N.~J., \& {Acreman}, D.~M. 2014, \aap, 564, A59

\bibitem[{{Arras} \& {Socrates}(2010)}]{arras-socrates-2010}
{Arras}, P., \& {Socrates}, A. 2010, \apj, 714, 1

\bibitem[{{Charney}(1963)}]{charney-1963}
{Charney}, J.~G. 1963, Journal of Atmospheric Sciences, 20, 607

\bibitem[{{Conrath} {et~al.}(1981){Conrath}, {Gierasch}, \&
  {Nath}}]{conrath-etal-1981}
{Conrath}, B.~J., {Gierasch}, P.~J., \& {Nath}, N. 1981, \icarus, 48, 256

\bibitem[{{Cooper} \& {Showman}(2005)}]{cooper-showman-2005}
{Cooper}, C.~S., \& {Showman}, A.~P. 2005, \apjl, 629, L45

\bibitem[{{Cooper} \& {Showman}(2006)}]{cooper-showman-2006}
---. 2006, \apj, 649, 1048

\bibitem[{{Cowan} {et~al.}(2012){Cowan}, {Machalek}, {Croll}, {Shekhtman},
  {Burrows}, {Deming}, {Greene}, \& {Hora}}]{cowan-etal-2012}
{Cowan}, N.~B., {Machalek}, P., {Croll}, B., {Shekhtman}, L.~M., {Burrows}, A.,
  {Deming}, D., {Greene}, T., \& {Hora}, J.~L. 2012, \apj, 747, 82

\bibitem[{{Crossfield} {et~al.}(2010){Crossfield}, {Hansen}, {Harrington},
  {Cho}, {Deming}, {Menou}, \& {Seager}}]{crossfield-etal-2010}
{Crossfield}, I.~J.~M., {Hansen}, B.~M.~S., {Harrington}, J., {Cho}, J.,
  {Deming}, D., {Menou}, K., \& {Seager}, S. 2010, \apj, 723, 1436

\bibitem[{{de Wit} {et~al.}(2012){de Wit}, {Gillon}, {Demory}, \&
  {Seager}}]{de-wit-etal-2012}
{de Wit}, J., {Gillon}, M., {Demory}, B.-O., \& {Seager}, S. 2012, \aap, 548,
  A128

\bibitem[{{Del Genio} \& {Barbara}(2012)}]{delgenio-barbara-2012}
{Del Genio}, A.~D., \& {Barbara}, J.~M. 2012, \icarus, 219, 689

\bibitem[{{Del Genio} {et~al.}(2007){Del Genio}, {Barbara}, {Ferrier},
  {Ingersoll}, {West}, {Vasavada}, {Spitale}, \& {Porco}}]{delgenio-etal-2007}
{Del Genio}, A.~D., {Barbara}, J.~M., {Ferrier}, J., {Ingersoll}, A.~P.,
  {West}, R.~A., {Vasavada}, A.~R., {Spitale}, J., \& {Porco}, C.~C. 2007,
  Icarus, 189, 479

\bibitem[{{Del Genio} \& {Suozzo}(1987)}]{delgenio-suozzo-1987}
{Del Genio}, A.~D., \& {Suozzo}, R.~J. 1987, Journal of Atmospheric Sciences,
  44, 973

\bibitem[{{Del Genio} \& {Zhou}(1996)}]{delgenio-zhou-1996}
{Del Genio}, A.~D., \& {Zhou}, W. 1996, Icarus, 120, 332

\bibitem[{{Dobbs-Dixon} \& {Agol}(2013)}]{dobbs-dixon-agol-2013}
{Dobbs-Dixon}, I., \& {Agol}, E. 2013, \mnras, 435, 3159

\bibitem[{{Dobbs-Dixon} \& {Lin}(2008)}]{dobbs-dixon-lin-2008}
{Dobbs-Dixon}, I., \& {Lin}, D.~N.~C. 2008, \apj, 673, 513

\bibitem[{{Dritschel} \& {McIntyre}(2008)}]{dritschel-mcintyre-2008}
{Dritschel}, D.~G., \& {McIntyre}, M.~E. 2008, Journal of Atmospheric Sciences,
  65, 855

\bibitem[{{Fortney} {et~al.}(2006){Fortney}, {Cooper}, {Showman}, {Marley}, \&
  {Freedman}}]{fortney-etal-2006b}
{Fortney}, J.~J., {Cooper}, C.~S., {Showman}, A.~P., {Marley}, M.~S., \&
  {Freedman}, R.~S. 2006, \apj, 652, 746

\bibitem[{{Fortney} {et~al.}(2007){Fortney}, {Marley}, \&
  {Barnes}}]{fortney-etal-2007}
{Fortney}, J.~J., {Marley}, M.~S., \& {Barnes}, J.~W. 2007, \apj, 659, 1661

\bibitem[{{Frierson} {et~al.}(2007){Frierson}, {Lu}, \&
  {Chen}}]{frierson-etal-2007}
{Frierson}, D.~M.~W., {Lu}, J., \& {Chen}, G. 2007, Geophys. Res. Lett., 34,
  L18804

\bibitem[{{Guillot} {et~al.}(1996){Guillot}, {Burrows}, {Hubbard}, {Lunine}, \&
  {Saumon}}]{guillot-etal-1996}
{Guillot}, T., {Burrows}, A., {Hubbard}, W.~B., {Lunine}, J.~I., \& {Saumon},
  D. 1996, \apjl, 459, L35

\bibitem[{{Guillot} \& {Showman}(2002)}]{guillot-showman-2002}
{Guillot}, T., \& {Showman}, A.~P. 2002, \aap, 385, 156

\bibitem[{{Hartmann}(2007)}]{hartmann-2007}
{Hartmann}, D.~L. 2007, J. Meteor. Soc. Japan, 85B, 123

\bibitem[{{Heimpel} {et~al.}(2005){Heimpel}, {Aurnou}, \&
  {Wicht}}]{heimpel-etal-2005}
{Heimpel}, M., {Aurnou}, J., \& {Wicht}, J. 2005, \nat, 438, 193

\bibitem[{{Held}(1975)}]{held-1975}
{Held}, I.~M. 1975, Journal of the Atmospheric Sciences, 32, 1494

\bibitem[{{Held}(2000)}]{held-2000}
---. 2000, Paper presented at 2000 Woods Hole Oceanographic Institute
  Geophysical Fluid Dynamics Program, Woods Hole Oceanographic Institute, Woods
  Hole, MA (available at http://www.whoi.edu/page.do?pid=13076)

\bibitem[{{Held} \& {Andrews}(1983)}]{held-andrews-1983}
{Held}, I.~M., \& {Andrews}, D.~G. 1983, Journal of Atmospheric Sciences, 40,
  2220

\bibitem[{{Held} \& {Hou}(1980)}]{held-hou-1980}
{Held}, I.~M., \& {Hou}, A.~Y. 1980, Journal of Atmospheric Sciences, 37, 515

\bibitem[{{Heng} {et~al.}(2011{\natexlab{a}}){Heng}, {Frierson}, \&
  {Phillipps}}]{heng-etal-2011b}
{Heng}, K., {Frierson}, D.~M.~W., \& {Phillipps}, P.~J. 2011{\natexlab{a}},
  \mnras, 418, 2669

\bibitem[{{Heng} {et~al.}(2011{\natexlab{b}}){Heng}, {Menou}, \&
  {Phillipps}}]{heng-etal-2011}
{Heng}, K., {Menou}, K., \& {Phillipps}, P.~J. 2011{\natexlab{b}}, \mnras, 413,
  2380

\bibitem[{{Holton}(2004)}]{holton-2004}
{Holton}, J.~R. 2004, An Introduction to Dynamic Meteorology, 4th Ed. (Academic
  Press, San Diego)

\bibitem[{{James}(1987)}]{james-1987}
{James}, I.~N. 1987, Journal of Atmospheric Sciences, 44, 3710

\bibitem[{{Karoly} {et~al.}(1998){Karoly}, {Vincent}, \&
  {Shrage}}]{karoly-etal-1998}
{Karoly}, D.~J., {Vincent}, D.~G., \& {Shrage}, J.~M. 1998, {\rm General
  circulation}. {\rm In} {\it Meteorology of the Southern Hemisphere}, American
  Meteorological Society Monographs, vol. 27 (American Meterological Society),
  47--85

\bibitem[{{Kaspi} {et~al.}(2009){Kaspi}, {Flierl}, \&
  {Showman}}]{kaspi-etal-2009}
{Kaspi}, Y., {Flierl}, G.~R., \& {Showman}, A.~P. 2009, Icarus, 202, 525

\bibitem[{{Kaspi} \& {Showman}(2014)}]{kaspi-showman-2014}
{Kaspi}, Y., \& {Showman}, A. 2014, ArXiv e-prints

\bibitem[{{Kataria} {et~al.}(2014{\natexlab{a}}){Kataria}, {Showman},
  {Fortney}, {Marley}, \& {Freedman}}]{kataria-etal-2014}
{Kataria}, T., {Showman}, A.~P., {Fortney}, J.~J., {Marley}, M.~S., \&
  {Freedman}, R.~S. 2014{\natexlab{a}}, \apj, 785, 92

\bibitem[{{Kataria} {et~al.}(2014{\natexlab{b}}){Kataria}, {Showman},
  {Fortney}, {Stevenson}, {Line}, {Kreidberg}, {Bean}, \&
  {D{\'e}sert}}]{kataria-etal-2014b}
{Kataria}, T., {Showman}, A.~P., {Fortney}, J.~J., {Stevenson}, K.~B., {Line},
  M.~R., {Kreidberg}, L., {Bean}, J.~L., \& {D{\'e}sert}, J.-M.
  2014{\natexlab{b}}, ArXiv e-prints

\bibitem[{{Kataria} {et~al.}(2013){Kataria}, {Showman}, {Lewis}, {Fortney},
  {Marley}, \& {Freedman}}]{kataria-etal-2013}
{Kataria}, T., {Showman}, A.~P., {Lewis}, N.~K., {Fortney}, J.~J., {Marley},
  M.~S., \& {Freedman}, R.~S. 2013, \apj, 767, 76

\bibitem[{{Kirk} \& {Stevenson}(1987)}]{kirk-stevenson-1987}
{Kirk}, R.~L., \& {Stevenson}, D.~J. 1987, \apj, 316, 836

\bibitem[{{Knutson} {et~al.}(2012){Knutson}, {Lewis}, {Fortney}, {Burrows},
  {Showman}, {Cowan}, {Agol}, {Aigrain}, {Charbonneau}, {Deming}, {D{\'e}sert},
  {Henry}, {Langton}, \& {Laughlin}}]{knutson-etal-2012}
{Knutson}, H.~A., {Lewis}, N., {Fortney}, J.~J., {Burrows}, A., {Showman},
  A.~P., {Cowan}, N.~B., {Agol}, E., {Aigrain}, S., {Charbonneau}, D.,
  {Deming}, D., {D{\'e}sert}, J.-M., {Henry}, G.~W., {Langton}, J., \&
  {Laughlin}, G. 2012, \apj, 754, 22

\bibitem[{{Kraucunas} \& {Hartmann}(2005)}]{kraucunas-hartmann-2005}
{Kraucunas}, I., \& {Hartmann}, D.~L. 2005, J. Atmos. Sci., 62, 371

\bibitem[{{Lewis} {et~al.}(2013){Lewis}, {Knutson}, {Showman}, {Cowan},
  {Laughlin}, {Burrows}, {Deming}, {Crepp}, {Mighell}, {Agol}, {Bakos},
  {Charbonneau}, {D{\'e}sert}, {Fischer}, {Fortney}, {Hartman}, {Hinkley},
  {Howard}, {Johnson}, {Kao}, {Langton}, \& {Marcy}}]{lewis-etal-2013}
{Lewis}, N.~K., {Knutson}, H.~A., {Showman}, A.~P., {Cowan}, N.~B., {Laughlin},
  G., {Burrows}, A., {Deming}, D., {Crepp}, J.~R., {Mighell}, K.~J., {Agol},
  E., {Bakos}, G.~{\'A}., {Charbonneau}, D., {D{\'e}sert}, J.-M., {Fischer},
  D.~A., {Fortney}, J.~J., {Hartman}, J.~D., {Hinkley}, S., {Howard}, A.~W.,
  {Johnson}, J.~A., {Kao}, M., {Langton}, J., \& {Marcy}, G.~W. 2013, \apj,
  766, 95

\bibitem[{{Lewis} {et~al.}(2014){Lewis}, {Showman}, {Fortney}, {Knutson}, \&
  {Marley}}]{lewis-etal-2014}
{Lewis}, N.~K., {Showman}, A.~P., {Fortney}, J.~J., {Knutson}, H.~A., \&
  {Marley}, M.~S. 2014, ArXiv e-prints

\bibitem[{{Lewis} {et~al.}(2010){Lewis}, {Showman}, {Fortney}, {Marley},
  {Freedman}, \& {Lodders}}]{lewis-etal-2010}
{Lewis}, N.~K., {Showman}, A.~P., {Fortney}, J.~J., {Marley}, M.~S.,
  {Freedman}, R.~S., \& {Lodders}, K. 2010, \apj, 720, 344

\bibitem[{{Li} \& {Goodman}(2010)}]{li-goodman-2010}
{Li}, J., \& {Goodman}, J. 2010, \apj, 725, 1146

\bibitem[{{Lian} \& {Showman}(2008)}]{lian-showman-2008}
{Lian}, Y., \& {Showman}, A.~P. 2008, Icarus, 194, 597

\bibitem[{{Lian} \& {Showman}(2010)}]{lian-showman-2010}
---. 2010, Icarus, 207, 373

\bibitem[{{Lindzen}(1981)}]{lindzen-1981}
{Lindzen}, R.~S. 1981, \jgr, 86, 9707

\bibitem[{{Liu} \& {Showman}(2013)}]{liu-showman-2013}
{Liu}, B., \& {Showman}, A.~P. 2013, \apj, 770, 42

\bibitem[{{Liu} {et~al.}(2008){Liu}, {Goldreich}, \&
  {Stevenson}}]{liu-etal-2008}
{Liu}, J., {Goldreich}, P.~M., \& {Stevenson}, D.~J. 2008, Icarus, 196, 653

\bibitem[{{Liu} \& {Schneider}(2010)}]{liu-schneider-2010}
{Liu}, J., \& {Schneider}, T. 2010, Journal of Atmospheric Sciences, 67, 3652

\bibitem[{{Liu} \& {Schneider}(2011)}]{liu-schneider-2011}
---. 2011, Journal of Atmospheric Sciences, 68, 2742

\bibitem[{{Lu} {et~al.}(2007){Lu}, {Vecchi}, \& {Reichler}}]{lu-etal-2007}
{Lu}, J., {Vecchi}, G.~A., \& {Reichler}, T. 2007, \grl, 34 (L06805)

\bibitem[{{Majeau} {et~al.}(2012){Majeau}, {Agol}, \&
  {Cowan}}]{majeau-etal-2012}
{Majeau}, C., {Agol}, E., \& {Cowan}, N.~B. 2012, \apjl, 747, L20

\bibitem[{{Marley} \& {McKay}(1999)}]{marley-mckay-1999}
{Marley}, M.~S., \& {McKay}, C.~P. 1999, Icarus, 138, 268

\bibitem[{{Mayne} {et~al.}(2014){Mayne}, {Baraffe}, {Acreman}, {Smith},
  {Browning}, {Sk{\aa}lid Amundsen}, {Wood}, {Thuburn}, \&
  {Jackson}}]{mayne-etal-2014}
{Mayne}, N.~J., {Baraffe}, I., {Acreman}, D.~M., {Smith}, C., {Browning},
  M.~K., {Sk{\aa}lid Amundsen}, D., {Wood}, N., {Thuburn}, J., \& {Jackson},
  D.~R. 2014, \aap, 561, A1

\bibitem[{{Menou} \& {Rauscher}(2009)}]{menou-rauscher-2009}
{Menou}, K., \& {Rauscher}, E. 2009, \apj, 700, 887

\bibitem[{{Menou} \& {Rauscher}(2010)}]{menou-rauscher-2010}
---. 2010, \apj, 713, 1174

\bibitem[{{Mitchell} {et~al.}(2006){Mitchell}, {Pierrehumbert}, {Frierson}, \&
  {Caballero}}]{mitchell-etal-2006}
{Mitchell}, J.~L., {Pierrehumbert}, R.~T., {Frierson}, D.~M.~W., \&
  {Caballero}, R. 2006, Proceedings of the National Academy of Science, 103,
  18421

\bibitem[{{Mitchell} \& {Vallis}(2010)}]{mitchell-vallis-2010}
{Mitchell}, J.~L., \& {Vallis}, G.~K. 2010, Journal of Geophysical Research
  (Planets), 115, 12008

\bibitem[{{O'Gorman} \& {Schneider}(2008)}]{ogorman-schneider-2008}
{O'Gorman}, P.~A., \& {Schneider}, T. 2008, Journal of Atmospheric Sciences,
  65, 524

\bibitem[{{Panetta}(1993)}]{panetta-1993}
{Panetta}, R.~L. 1993, Journal of Atmospheric Sciences, 50, 2073

\bibitem[{{Perez-Becker} \& {Showman}(2013)}]{perez-becker-showman-2013}
{Perez-Becker}, D., \& {Showman}, A.~P. 2013, \apj, 776, 134

\bibitem[{{Perna} {et~al.}(2012){Perna}, {Heng}, \& {Pont}}]{perna-etal-2012}
{Perna}, R., {Heng}, K., \& {Pont}, F. 2012, \apj, 751, 59

\bibitem[{{Perna} {et~al.}(2010){Perna}, {Menou}, \&
  {Rauscher}}]{perna-etal-2010}
{Perna}, R., {Menou}, K., \& {Rauscher}, E. 2010, \apj, 719, 1421

\bibitem[{Pierrehumbert \& Swanson(1995)}]{pierrehumbert-swanson-1995}
Pierrehumbert, R.~T., \& Swanson, K.~L. 1995, Annu. Rev. Fluid Mech., 27, 419

\bibitem[{{Polichtchouk} \& {Cho}(2012)}]{polichtchouk-cho-2012}
{Polichtchouk}, I., \& {Cho}, J.~Y.-K. 2012, \mnras, 424, 1307

\bibitem[{{Randel} \& {Held}(1991)}]{randel-held-1991}
{Randel}, W.~J., \& {Held}, I.~M. 1991, Journal of Atmospheric Sciences, 48,
  688

\bibitem[{{Rasio} {et~al.}(1996){Rasio}, {Tout}, {Lubow}, \&
  {Livio}}]{rasio-etal-1996}
{Rasio}, F.~A., {Tout}, C.~A., {Lubow}, S.~H., \& {Livio}, M. 1996, \apj, 470,
  1187

\bibitem[{{Rauscher} \& {Kempton}(2014)}]{rauscher-kempton-2014}
{Rauscher}, E., \& {Kempton}, E.~M.~R. 2014, \apj, 790, 79

\bibitem[{{Rauscher} \& {Menou}(2010)}]{rauscher-menou-2010}
{Rauscher}, E., \& {Menou}, K. 2010, \apj, 714, 1334

\bibitem[{{Rauscher} \& {Menou}(2012{\natexlab{a}})}]{rauscher-menou-2012b}
---. 2012{\natexlab{a}}, \apj, 750, 96

\bibitem[{{Rauscher} \& {Menou}(2012{\natexlab{b}})}]{rauscher-menou-2012}
---. 2012{\natexlab{b}}, \apj, 745, 78

\bibitem[{{Rauscher} \& {Menou}(2013)}]{rauscher-menou-2013}
---. 2013, \apj, 764, 103

\bibitem[{{Read}(1988)}]{read-1988}
{Read}, P.~L. 1988, Quarterly Journal of the Royal Meteorological Society, 114,
  421

\bibitem[{{Rogers} \& {Showman}(2014)}]{rogers-showman-2014}
{Rogers}, T.~M., \& {Showman}, A.~P. 2014, \apjl, 782, L4

\bibitem[{{Saravanan}(1993)}]{saravanan-1993}
{Saravanan}, R. 1993, J. Atmos. Sci., 50, 1211

\bibitem[{{Schneider} \& {Bordoni}(2008)}]{schneider-bordoni-2008}
{Schneider}, T., \& {Bordoni}, S. 2008, Journal of Atmospheric Sciences, 65,
  915

\bibitem[{{Schneider} \& {Liu}(2009)}]{schneider-liu-2009}
{Schneider}, T., \& {Liu}, J. 2009, J. Atmos. Sci., 66, 579

\bibitem[{{Showman}(2007)}]{showman-2007}
{Showman}, A.~P. 2007, J. Atmos. Sci., 64, 3132

\bibitem[{{Showman} {et~al.}(2008){Showman}, {Cooper}, {Fortney}, \&
  {Marley}}]{showman-etal-2008a}
{Showman}, A.~P., {Cooper}, C.~S., {Fortney}, J.~J., \& {Marley}, M.~S. 2008,
  \apj, 682, 559

\bibitem[{{Showman} {et~al.}(2013{\natexlab{a}}){Showman}, {Fortney}, {Lewis},
  \& {Shabram}}]{showman-etal-2013}
{Showman}, A.~P., {Fortney}, J.~J., {Lewis}, N.~K., \& {Shabram}, M.
  2013{\natexlab{a}}, \apj, 762, 24

\bibitem[{{Showman} {et~al.}(2009){Showman}, {Fortney}, {Lian}, {Marley},
  {Freedman}, {Knutson}, \& {Charbonneau}}]{showman-etal-2009}
{Showman}, A.~P., {Fortney}, J.~J., {Lian}, Y., {Marley}, M.~S., {Freedman},
  R.~S., {Knutson}, H.~A., \& {Charbonneau}, D. 2009, \apj, 699, 564

\bibitem[{{Showman} \& {Guillot}(2002)}]{showman-guillot-2002}
{Showman}, A.~P., \& {Guillot}, T. 2002, \aap, 385, 166

\bibitem[{{Showman} \& {Polvani}(2010)}]{showman-polvani-2010}
{Showman}, A.~P., \& {Polvani}, L.~M. 2010, \grl, 37, 18811

\bibitem[{{Showman} \& {Polvani}(2011)}]{showman-polvani-2011}
---. 2011, \apj, 738, 71

\bibitem[{{Showman} {et~al.}(2013{\natexlab{b}}){Showman}, {Wordsworth},
  {Merlis}, \& {Kaspi}}]{showman-etal-2013b}
{Showman}, A.~P., {Wordsworth}, R.~D., {Merlis}, T.~M., \& {Kaspi}, Y.
  2013{\natexlab{b}}, {\rm Atmospheric Circulation of Terrestrial Exoplanets}.
  {\rm In} {\it Comparative Climatology of Terrestrial Planets} (Mackwell,
  S.J., Simon-Miller, A.A., Harder, J.W., and Bullock, M.A., Eds.) (Univ.
  Arizona Press), 277--326

\bibitem[{{Stevenson} {et~al.}(2014){Stevenson}, {coauthors}, \&
  {morecoauthors}}]{stevenson-etal-2014}
{Stevenson}, K.~B., {coauthors}, X.~X., \& {morecoauthors}, X.~X. 2014,
  Science, submitted

\bibitem[{{Suarez} \& {Duffy}(1992)}]{suarez-duffy-1992}
{Suarez}, M.~J., \& {Duffy}, D.~G. 1992, J. Atmos. Sci., 49, 1541

\bibitem[{{Thompson}(1971)}]{thompson-1971}
{Thompson}, R.~O.~R.~Y. 1971, Journal of Physical Oceanography, 1, 235

\bibitem[{{Thrastarson} \& {Cho}(2010)}]{thrastarson-cho-2010}
{Thrastarson}, H.~T., \& {Cho}, J. 2010, \apj, 716, 144

\bibitem[{{Tsai} {et~al.}(2014){Tsai}, {Dobbs-Dixon}, \& {Gu}}]{tsai-etal-2014}
{Tsai}, S.-M., {Dobbs-Dixon}, I., \& {Gu}, P.-G. 2014, ArXiv e-prints

\bibitem[{{Vallis}(2006)}]{vallis-2006}
{Vallis}, G.~K. 2006, Atmospheric and Oceanic Fluid Dynamics: Fundamentals and
  Large-Scale Circulation (Cambridge Univ. Press, Cambridge, UK)

\bibitem[{{Vasavada} \& {Showman}(2005)}]{vasavada-showman-2005}
{Vasavada}, A.~R., \& {Showman}, A.~P. 2005, Reports of Progress in Physics,
  68, 1935

\bibitem[{{Walker} \& {Schneider}(2006)}]{walker-schneider-2006}
{Walker}, C.~C., \& {Schneider}, T. 2006, Journal of Atmospheric Sciences, 63,
  3333

\bibitem[{{Wallace} \& {Hobbs}(2004)}]{wallace-hobbs-2006}
{Wallace}, J.~M., \& {Hobbs}, P.~V. 2004, Atmospheric Science: An Introductory
  Survey, 2nd Ed. (Academic Press, San Diego)

\bibitem[{{Wang}(1990)}]{wang-1990}
{Wang}, B. 1990, Tellus A, Vol.~42, issue 4, p.463, 42, 463

\bibitem[{{Williams}(1979)}]{williams-1979}
{Williams}, G.~P. 1979, Journal of Atmospheric Sciences, 36, 932

\bibitem[{{Williams}(2003)}]{williams-2003a}
---. 2003, Journal of Atmospheric Sciences, 60, 1270

\bibitem[{{Zellem} {et~al.}(2014){Zellem}, {Lewis}, {Knutson}, {Griffith},
  {Showman}, {Fortney}, {Cowan}, {Agol}, {Burrows}, {Charbonneau}, {Deming},
  {Laughlin}, \& {Langton}}]{zellem-etal-2014}
{Zellem}, R.~T., {Lewis}, N.~K., {Knutson}, H.~A., {Griffith}, C.~A.,
  {Showman}, A.~P., {Fortney}, J.~J., {Cowan}, N.~B., {Agol}, E., {Burrows},
  A., {Charbonneau}, D., {Deming}, D., {Laughlin}, G., \& {Langton}, J. 2014,
  \apj, 790, 53

\end{thebibliography}



\end{document}